\documentclass[usenatbib]{mn2e}

\pdfoutput=1

\usepackage{natbib}
\usepackage{graphicx}
\usepackage{apjfonts}
\usepackage{deluxetable}
\usepackage{multicol}
\usepackage{grant_defs}
\usepackage[backref,breaklinks,colorlinks,citecolor=blue]{hyperref} 
\usepackage{verbatim}
\usepackage{array}

\usepackage{aastex_hack}

\usepackage[backref,breaklinks,colorlinks,citecolor=blue]{hyperref} 
\usepackage[all]{hypcap} 

\title[FUV Morphology of Cool Core BCGs]{Far Ultraviolet Morphology of Star Forming Filaments \\ in Cool Core Brightest Cluster Galaxies}

\author[G.~R.~Tremblay et al.]{G.~R.~Tremblay,$^{1,2,}\dagger$ C.~P.~O'Dea,$^{3,4}$ S.~A.~Baum,$^{3,4,5,6}$ R.~Mittal,$^{6,7}$ M.~A.~McDonald,$^{8}\ddagger$ 
\newauthor F.~Combes,$^{9}$ Y.~Li,$^{10}$ B.~R.~McNamara,$^{11, 12}$ M.~N.~Bremer,$^{13}$  T.~E.~Clarke,$^{14}$ M.~Donahue,$^{15}$ \newauthor A.~C.~Edge,$^{16}$  A.~C.~Fabian,$^{17}$ S.~L.~Hamer,$^{9}$ M.~T.~Hogan,$^{11}$
 J.~B.~R.~Oonk,$^{18}$  A.~C.~Quillen,$^{19}$ \newauthor  J.~S.~Sanders,$^{20}$ P.~Salom\'{e},$^{9}$ and G.~M.~Voit$^{15}$\\
\\$^{1}$ Department of Physics and Yale Center for Astronomy \& Astrophysics, Yale University, 217 Prospect Street, New Haven, 
CT 06511, USA
\\$^{2}$ European Southern Observatory, 
Karl-Schwarzschild-Str.~2, 85748 Garching bei M\"{u}nchen, Germany
\\$^{3}$ Department of Physics \& Astronomy, University of Manitoba, Winnipeg, MB R3T 2N2, Canada
\\$^{4}$ School of Physics \& Astronomy, Rochester Institute of Technology, 84 Lomb Memorial Drive, Rochester, NY 14623, USA
\\$^{5}$ Faculty of Science, University of Manitoba, Winnipeg, MB R3T 2N2, Canada
\\$^{6}$ Chester F.~Carlson Center for Imaging Science, Rochester Institute of Technology,  54 Lomb Memorial Drive, Rochester, NY 14623, USA
\\$^{7}$ Max Planck Institut f\"{u}r Gravitationsphysik (Albert Einstein Institut), D-30167 Hannover, Germany
\\$^{8}$ Kavli Institute for Astrophysics and Space Research, Massachusetts Institute of Technology, 77 Massachusetts Avenue, Cambridge, MA 02139, USA
\\$^{9}$ Observatoire de Paris, LERMA, CNRS, 61 Av.~de l'Observatoire, 75014 Paris, France
\\$^{10}$ Department of Astronomy, University of Michigan, 1085 S.~University Ave., Ann Arbor, MI 48109, USA
\\$^{11}$ Physics \& Astronomy Deptartment, Waterloo University, 200 University Ave.~W., Waterloo, ON, N2L, 2G1, Canada
\\$^{12}$ Harvard-Smithsonian Center for Astrophysics, 60 Garden St., Cambridge, MA 02138, USA
\\$^{13}$ H.~W.~Wills Physics Laboratory, University of Bristol, Tyndall Avenue, Bristol, BS8 1TL, UK 
\\$^{14}$ Naval Research Laboratory Remote Sensing Division, Code 7213 4555 Overlook Ave SW, Washington, DC 20375, USA
\\$^{15}$ Michigan State University, Physics and Astronomy Dept., East Lansing, MI 48824-2320, USA
\\$^{16}$ Department of Physics, Durham University, Durham, DH1 3LE, UK
\\$^{17}$ Institute of Astronomy, Madingley Rd., Cambridge, CB3 0HA, UK
\\$^{18}$ ASTRON, Netherlands Institute for Radio Astronomy, P.O. Box 2, 7990 AA Dwingeloo, The Netherlands
\\$^{19}$ Department of Physics and Astronomy, University of Rochester, Rochester, NY 14627, USA
\\$^{20}$ Max Planck Institut f\"{u}r Extraterrestrische Physik, 85748 Garching bei M\"{u}nchen, Germany
\\${\dagger}$ Einstein Fellow | ${\ddagger}$ Hubble Fellow \vspace*{-6mm}}

\begin{document}

%
\def\aj{AJ}%
\def\araa{ARA\&A}%
\def\apj{ApJ}%
\def\apjl{ApJ}%
\def\apjs{ApJS}%
\def\ao{Appl.~Opt.}%
\def\apss{Ap\&SS}%
\def\aap{A\&A}%
\def\aapr{A\&A~Rev.}%
\def\aaps{A\&AS}%
\def\azh{AZh}%
\def\baas{BAAS}%
\def\jrasc{JRASC}%
\def\memras{MmRAS}%
\def\mnras{MNRAS}%
\def\pra{Phys.~Rev.~A}%
\def\prb{Phys.~Rev.~B}%
\def\prc{Phys.~Rev.~C}%
\def\prd{Phys.~Rev.~D}%
\def\pre{Phys.~Rev.~E}%
\def\prl{Phys.~Rev.~Lett.}%
\def\pasp{PASP}%
\def\pasj{PASJ}%
\def\qjras{QJRAS}%
\def\skytel{S\&T}%
\def\solphys{Sol.~Phys.}%
\def\sovast{Soviet~Ast.}%
\def\ssr{Space~Sci.~Rev.}%
\def\zap{ZAp}%
\def\nat{Nature}%
\def\iaucirc{IAU~Circ.}%
\def\aplett{Astrophys.~Lett.}%
\def\apspr{Astrophys.~Space~Phys.~Res.}%
\def\bain{Bull.~Astron.~Inst.~Netherlands}%
\def\fcp{Fund.~Cosmic~Phys.}%
\def\gca{Geochim.~Cosmochim.~Acta}%
\def\grl{Geophys.~Res.~Lett.}%
\def\jcp{J.~Chem.~Phys.}%
\def\jgr{J.~Geophys.~Res.}%
\def\jqsrt{J.~Quant.~Spec.~Radiat.~Transf.}%
\def\memsai{Mem.~Soc.~Astron.~Italiana}%
\def\nphysa{Nucl.~Phys.~A}%
\def\physrep{Phys.~Rep.}%
\def\physscr{Phys.~Scr}%
\def\planss{Planet.~Space~Sci.}%
\def\procspie{Proc.~SPIE}%


\maketitle

\label{firstpage}

\begin{abstract}  
We present  a multiwavelength  morphological analysis of
star forming clouds and filaments in the central  ($\lae 50$ kpc) regions of
16 low redshift ($z<0.3$)  cool core  brightest cluster galaxies  (BCGs). The
sample  spans decades-wide ranges  of X-ray  mass deposition  and star
formation rates  as well as  active galactic nucleus  (AGN) mechanical power,
encompassing  both  high  and  low extremes  of  the  supposed intracluster
medium (ICM) cooling and AGN heating feedback cycle.  New {\it Hubble Space
Telescope} (HST) imaging  of far ultraviolet continuum emission from young
($\lae 10$ Myr), massive ($\gae 5$ \Msol) stars  reveals filamentary and
clumpy morphologies, which we quantify by means of structural indices. The FUV
data are compared with  X-ray,  Ly$\alpha$, narrowband  H$\alpha$,  broadband
optical/IR,  and radio maps, providing a high  spatial resolution atlas of
star formation locales  relative  to the  ambient  hot  ($\sim10^{7-8}$  K)
and  warm ionised  ($\sim  10^4$ K)  gas  phases, as  well  as  the old
stellar population  and  radio-bright  AGN   outflows.   Nearly half of the
sample possesses kpc-scale  filaments  that, in  projection,  extend toward
and  around radio lobes and/or X-ray  cavities.  These filaments may have been
uplifted by the propagating jet or buoyant X-ray  bubble, or may have formed
{\it in   situ}  by cloud collapse at the interface of a radio lobe or rapid
cooling in a cavity's compressed shell.  Many other extended filaments,
however, show no such spatial correlation, and the dominant driver of their
morphology remains unclear.  We nevertheless show that the morphological
diversity of nearly the entire FUV sample   is reproduced by recent
hydrodynamical simulations in which the AGN powers  a self-regulating rain of
thermally unstable star forming clouds that precipitate from the hot
atmosphere.  In this model, precipitation triggers where the  cooling-to-
freefall time ratio is $t_{\mathrm{cool}}/t_{\mathrm{ff}}\sim 10$.  This
condition is roughly met at the maxmial projected FUV radius for more than
half of our sample,  and clustering about this ratio is stronger for sources
with higher star formation rates.

 \end{abstract}


\section{Introduction}
\label{section:introduction}

Many  giant elliptical, groups,  and clusters  of galaxies  inhabit an
X-ray bright halo of $\gae10^7$ K plasma whose core radiative lifetime
is  much shorter  than its  age.  Absent  a heating  mechanism, simple
models predict  that the rapid cooling  of gas within  this $\sim 100$
kpc  ``cool core''  (CC)  should  result in  a  long-lived cascade  of
multiphase clouds  collapsing into the galaxy at  its center, fueling
extreme  star formation  rates ($10^{2}-10^{3}$  \Msol\  yr\mone) amid
massive reservoirs ($\sim 10^{12}$ \Msol) of cold molecular gas (e.g.,
review  by \citealt{fabian94}).   Although brightest  cluster galaxies
(BCGs)  embedded  in  CC  clusters  do  preferentially  harbour  these
supposed cooling  flow mass sinks,  the observed star  formation rates
and cold gas  masses are often orders of  magnitude below predictions,
and  high resolution  X-ray  spectroscopy of  the intracluster  medium
(ICM,  e.g.,  \citealt{sarazin86})  is  only consistent  with  reduced
cooling at $\sim 10 \%$  of the expected classical rates (e.g., review
by \citealt{peterson06}).

The mechanical  dissipation of active galactic nucleus  (AGN) power is
now routinely invoked by theorists  and observers as a solution to the
problem,  as  the average  associated  energy  budget  for groups  and
clusters  is  large  enough  to  inhibit  or  replenish  cooling  flow
radiative    losses     not    only    at     late    epochs    (e.g.,
\citealt{birzan04,birzan08,rafferty06,dunn06,best06,best07,mittal09,dong10}),
but  perhaps  over  a  significant  fraction  of  cosmic  time  (e.g.,
\citealt{hlavacek-larrondo12,simpson13,mcdonald13b}).  The paradigm is
motivated  by  strong circumstantial  evidence,  including nearly
ubiquitous  observations of radio-bright  AGN outflows  driving shocks
and excavating  kpc-scale buoyant cavities  in the ambient  X-ray gas,
acting  as  lower-limit  calorimeters  to  the  often  extreme  ($\lae
10^{46}$  ergs sec\mone) AGN  kinetic energy  input (e.g.,  reviews by
\citealt{mcnamara07,mcnamara12,sun12}).  Yet amid panoramic supporting
evidence (reviewed by \citealt{fabian12}), the physics that govern the
spatial distribution and thermal  coupling of AGN mechanical energy to
the  multiphase  ($10-10^7$   K)  gaseous  environment  remain  poorly
understood, and cooling flow  alternatives invoking (e.g.) wet mergers, thermal conduction, and evaporation
 have been  a persistent matter of debate (e.g.,
\citealt{bregman88,sparks92,sparks97,sparks89,sparks09,sparks12,fabian94sparks,soker03,voit08,voit11b,smith13, canning15, voit15}).

Although  often  invoked exclusively  as  a  star formation  quenching
mechanism,  observations have  long demonstrated  that  AGN mechanical
feedback does  not completely offset radiative losses  or establish an
impermeable  ``entropy floor'',  instead  permitting residual  cooling
either    at    constant    low    ($\sim   10$\%)    rates,    (e.g.,
\citealt{tremblay12a,tremblay12b}), or in elevated episodes as the AGN
varies  in  power  (e.g., \citealt{odea10,tremblay11}).   Relative  to
field galaxies or those in  non-cool core clusters, BCGs in cool cores
preferentially  harbour   radio  sources  and   kpc-scale  filamentary
forbidden  and  Balmer  emission  line nebulae  amid  $10^{9}-10^{11}$
\Msol\ repositories  of vibrationally  excited and cold  molecular gas
\citep{heckman81,hu85,baum87,heckman89,burns90,jaffe97,donahue00,edge01,edge03,salome03,mcnamara04,egami06,salome06,edwards07,vonderlinden07,salome11,wilman09,edge10phot,edge10spec, mcnamara14, russell14}.
Low  to moderate levels  ($\sim 1  - \gae10$  \Msol\ yr\mone)  of star
formation  appear   to  be  ongoing  amid   these  mysteriously  dusty
\citep{quillen08,odea08,edge10phot,edge10spec,mittal11,tremblay12a,rawle12},    PAH-rich
\citep{donahue11} cold  reservoirs on $\lae  50$ kpc scales  in clumpy
and             filamentary            distributions            (e.g.,
\citealt{johnstone87,mcnamara04,odea04,odea08,rafferty06,rafferty08,odea10,mcdonald11a,tremblay14}).
The ionisation  states of  the nebulae have  been a mystery  for three
decades,  and  debate  continues  over  the roles  played  by  stellar
photoionization,  shocks, thermal conduction,  mixing, and  cosmic ray
heating                                                          (e.g.,
\citealt{voit97,ferland09,sparks09,sparks12,odea10,mcdonald10,mcdonald11b,tremblay11,fabian11a,mittal11,oonk11,johnstone12}).
Whatever the case, there is strong evidence that young stars might play an
important (although not exclusive) role
in dictating  the physics of  both the warm  ($\sim 10^4$ K)  and cold
($\sim      100$       K)      phases      of       the      filaments
\citep{voit97,odea10,mcdonald10,mcdonald11b,canning10,canning14}.

Recent work on star formation in CC BCGs has demonstrated its efficacy
as an observable tracer  for otherwise unobservable physical processes
regulating the heating and  cooling balance in hot atmospheres.  While
low in general  and effectively zero in some  cases, the observed star
formation  rates in  CC BCGs  are  sometimes high  enough ($\gae  100$
\Msol\  yr\mone)  to match  condensation  rates  from  the X-ray  halo
\citep{odea08}.   
Emergent work at higher redshift has shown that the long-ago-predicted
{\it classical}  cooling flows may  exist after all, forming  stars at
many  hundreds of  solar masses  per year  (i.e.~the  Phoenix cluster,
\citealt{mcdonald12,mcdonald13a,mcdonald14}; see also work on Abell 1068 by  \citealt{mcnamara04,wise04}).
Cooling  flows may  begin  to  form  stars when  the
central  entropy or  cooling  time drops  below  a critical  threshold
(e.g.,
\citealt{voitdonahue05,rafferty08,cavagnolo08,guo13,voit08,voit15b}),
or when the ratio of  cooling-to-dynamical times permits a self-regulating ``rain'' of
thermally unstable,  spatially inhomogeneous clouds  condensing from the 
hot  atmosphere 
\citep{sharma12,mccourt12,gaspari12, li14a, li14b, li15, brighenti15, voit15, voit15b}.   There  is  also  some
observational  evidence for enhanced cooling  in spatially confined ``cooling channels'' 
where AGN heating
may  be locally  inefficient (see, e.g., evidence for enhanced cooling in regions perpendicular to the projected cavity/radio ``heating axis'' in Perseus and Abell 2597;  \citealt{lim08,tremblay12a}).

\begin{table*}
\centering
    \caption{Basic information on the 16 low-redshift cool core brightest cluster galaxies that make up our sample. 
(1) Source name (note that while we list the most commonly used name for the cluster, the actual target studied in this paper 
is the central brightest cluster galaxy of the named cluster); 
(2) non-exhaustive list of other commonly used names for the cluster, central brightest cluster galaxy, or central radio source;
(3) right ascension and
(4) declination for the J2000.0 epoch; 
(5) redshift  ($z$) as listed in the NASA/IPAC Extragalactic database (NED);
(6) the number of kiloparsecs (kpc) that correspond to one arcsecond at the given redshift in our assumed cosmology ($H_0 = 70$ km  s$^{-1}$ Mpc$^{-1}$);
(7) the associated figure number(s) where the multiwavelength data for the listed BCG can be viewed. The FUV continuum images for the entire sample 
can be viewed at a glance in Fig.~\ref{fig:postagestamps}. }
\begin{tabular}{ccccccc}
\hline
          &
          &
R.A.      & 
Dec. & 
Redshift          &
          &
Shown in           \\
Source Name &
Associated Name(s) &
(J2000)  &
(J2000)   &
($z$) &
kpc/arcsec  &
Figure \# \\
(1) & (2) & (3) & (4) & (5) & (6) & (7) \\
\hline
\hline
{\bf Abell 11}    & \nodata & 00h 12m 44.8s & -16$^\circ$ 26$\arcmin$ 19$\arcsec$   & 0.1660 & 2.81 & \ref{fig:postagestamps}, \ref{fig:lyacompare}, \ref{fig:a11_figure}  \\ 
{\bf Abell 1068}  &  ZwCl 1037.6+4013 & 10h 40m 47.1s  &  +39$^\circ$ 57$\arcmin$ 19$\arcsec$   & 0.1375 & 2.40 & \ref{fig:postagestamps}, \ref{fig:highsfr}, \ref{fig:a1068_figure}   \\
{\bf Abell 1664}  & RX J1303.7-2414 & 13h 03m 41.8s  & -24$^\circ$ 13$\arcmin$ 06$\arcsec$     & 0.1283 & 2.27 & \ref{fig:postagestamps}, \ref{fig:lyacompare}, \ref{fig:midsfr}, \ref{fig:a1664_figure} \\
{\bf Abell 1795}  & ZwCl 1346.9+2655 & 13h 49m 00.5s  & +26$^\circ$ 35$\arcmin$ 07$\arcsec$  &  0.0625 &1.19 & \ref{fig:postagestamps}, \ref{fig:halphacompare}, \ref{fig:jet_triggered_sfr_figure}, \ref{fig:a1795_figure}   \\
{\bf Abell 1835}  & ZwCl 1358.5+0305  & 14h 01m 02.0s  &  +02$^\circ$ 51$\arcmin$ 32$\arcsec$   &  0.2532 & 3.91 & \ref{fig:postagestamps}, \ref{fig:lyacompare}, \ref{fig:highsfr}, \ref{fig:filamentcavity}, \ref{fig:a1835_figure} \\
{\bf Abell 2199}  & NGC 6166, 3C 338  & 16h 28m 38.5s  &  +39$^\circ$ 33$\arcmin$ 06$\arcsec$  &  0.0302 & 0.60  &  \ref{fig:postagestamps}, \ref{fig:lowsfr}, \ref{fig:a2199_figure}  \\
{\bf Abell 2597}  & PKS 2322-122 & 23h 25m 18.0s &  -12$^\circ$ 06$\arcmin$ 30$\arcsec$   & 0.0821  & 1.53 & \ref{fig:postagestamps}, \ref{fig:halphacompare}, \ref{fig:midsfr}, \ref{fig:lyacompare}, \ref{fig:jet_triggered_sfr_figure}, \ref{fig:filamentcavity}, \ref{fig:a2597_figure}  \\
{\bf Centaurus}  &  NGC 4696, Abell 3526, PKS 1245-41  & 12h 48m 49.2s & +41$^\circ$ 18$\arcmin$ 39$\arcsec$ & 0.0099 & 0.20 & \ref{fig:postagestamps}, \ref{fig:lowsfr}, \ref{fig:filamentcavity}, \ref{fig:centaurus_figure}  \\
{\bf Hydra A} &     Abell 780, 3C 218  & 09h 18m 05.7s & -12$^\circ$ 05$\arcmin$ 44$\arcsec$    &  0.0549 & 1.05 & \ref{fig:postagestamps}, \ref{fig:halphacompare}, \ref{fig:midsfr}, \ref{fig:jet_triggered_sfr_figure}, \ref{fig:filamentcavity}, \ref{fig:hydraA_companion}, \ref{fig:hydra_figure}  \\
{\bf Perseus}  &    NGC 1275, Abell 426, 3C 84 & 03h 19m 48.1s & +41$^\circ$ 30$\arcmin$ 42$\arcsec$    & 0.0176 &0.35 & \ref{fig:postagestamps}, \ref{fig:halphacompare}, \ref{fig:filamentcavity}, \ref{fig:perseus_figure}  \\
{\bf PKS 0745-191}  & \nodata &  07h 47m 31.3s &   -19$^\circ$ 17$\arcmin$ 40$\arcsec$  & 0.1028  &1.89 & \ref{fig:postagestamps}, \ref{fig:filamentcavity}, \ref{fig:pks0745_figure}  \\
{\bf RX J1504.1-0248}  &\nodata  &  15h 04m 07.5s &   -02$^\circ$ 48$\arcmin$ 16$\arcsec$   & 0.2153 & 3.46 & \ref{fig:postagestamps}, \ref{fig:highsfr}, \ref{fig:rxj1504_figure}  \\
{\bf RX J2129.6+0005}  & \nodata  & 21h 29m 37.9s  & +00$^\circ$ 05$\arcmin$ 39$\arcsec$   &  0.2350 &3.70  &  \ref{fig:postagestamps}, \ref{fig:lyacompare}, \ref{fig:rxj2129_figure} \\
{\bf ZwCl 0348}  & ZwCl 0104.4+0048   & 01h 06m 58.0s & +01$^\circ$ 04$\arcmin$ 01$\arcsec$    &   0.2545 &3.93  & \ref{fig:postagestamps}, \ref{fig:lyacompare}, \ref{fig:zw0348_figure}  \\
{\bf ZwCl 3146}  &  ZwCl 1021.0+0426 & 10h 23m 39.6s   & +04$^\circ$ 11$\arcmin$ 10$\arcsec$    &  0.2906 &4.32 & \ref{fig:postagestamps}, \ref{fig:lyacompare}, \ref{fig:zw3146_figure} \\
{\bf ZwCl 8193}  & ZwCl 1715.5+4229 & 17h 17m 19.0s &   +42$^\circ$ 26$\arcmin$ 57$\arcsec$    & 0.1829  &3.04  &  \ref{fig:postagestamps}, \ref{fig:lyacompare}, \ref{fig:zw8193_figure}  \\
  \hline
  \end{tabular}
\label{tab:sample}
\end{table*}

Direct observations of young stars in BCGs can test predictions
 of these various models. 
To that  end, this  paper presents  a morphological
analysis of new and archival  {\it Hubble Space Telescope} ({\it HST})
far-ultraviolet (FUV)  continuum images of young, massive  stars in 16
low-redshift  ($z<0.29$)  CC   BCGs.   X-ray,  Ly$\alpha$,  H$\alpha$,
broadband  optical, and  radio data  are also  leveraged to  create an
``atlas'' of star formation locales relative to the ambient hot ($\gae
10^{7}$ K) and warm ionised ($\sim 10^4$ K) gas phases, as well as the
old stellar population and radio-bright AGN outflows.
In Section 2  we discuss the sample selection,  observations, and data
reduction.   Our results  are presented  in  Section 3, discussed  in Section
4, and summarised in Section 5.   An appendix contains additional multiwavelength overlay 
figures for all sources in our sample. We will frequently abbreviate  target
names in  an obvious manner  (i.e. Abell 2597  is written  as A2597, etc.).
Unless otherwise noted,  we use the names of the parent clusters to  refer to
their  central BCGs (i.e. ``Perseus''  refers to its brightest cluster galaxy, NGC 1275).
Cosmology dependent physical quantities quoted in  this paper assume a flat
$\Lambda$CDM model wherein  $H_0 =  70$ km  s$^{-1}$ Mpc$^{-1}$,
$\Omega_M =  0.3$, and $\Omega_{\Lambda} =  0.7$. Errors are quoted at  the
$1\sigma$ level, unless otherwise noted.

\section{Sample, Observations, and Data Reduction}

\subsection{Sample selection}

The 16  low redshift  ($z<0.3$) CC  BCGs that make  up our  sample are listed
in  Table  \ref{tab:sample}.   All  are well  studied  in  the literature,
and enjoy nearly  complete cross-spectrum  (radio through X-ray)  data
coverage  from many  ground- and  space-based facilities, including  the {\it
Chandra  X-ray Observatory},  {\it Hubble Space Telescope},  {\it   Spitzer
Space Telescope}, and  {\it Herschel Space Observatory}. Eleven  of  these
targets  constitute  the  {\it  Herschel} cool  core clusters  Open Time Key
Project sample  of A.~Edge  and collaborators (\citealt{edge10phot,edge10spec,
mittal11,mittal12,tremblay12a,tremblay12b,rawle12,hamer14}),
selected to span a  wide range of both  cooling flow and  BCG physical
properties. The  remaining five  targets are  from the  non-overlapping
sample of \citet{odea10},  selected  from the  \citet{quillen08}  sample on
the basis of elevated infrared-estimated star formation rates.

Although biased,  our sample spans  decades-wide ranges of  X-ray mass
deposition  and  star  formation  rates,  Balmer  and  forbidden  line
luminosities, as  well as  AGN, radio source,  and X-ray  cavity power
(including sources that lack  detected cavities). Its constituent galaxies
therefore occupy unique milestones in the supposed ICM cooling and AGN
heating feedback  cycle over the last  $\sim 3$ Gyr  of cosmic history
(redshifts $0.0099 \le z \le 0.2906$), including sources with high and
low star formation rates, strong  and weak AGN feedback signatures, as
well  as   many  intermediate   locales  between  these   extremes.  A
non-exhaustive  summary of these  various properties  can be  found in
Table \ref{tab:properties}.

\begin{table*}
\setlength{\extrarowheight}{1.2pt}
\setlength{\tabcolsep}{3pt}
\centering
    \caption{A summary of physical properties of the BCGs in our sample, including 
    their surrounding $\sim 100$ kpc-scale environment. References for the quantities presented here can be found below this table.  
(1) Target name; 
(2) Infrared-estimated star formation rate (SFR), typically from {\it Spitzer} or {\it Herschel}. SFRs flagged with a $\dagger$ symbol may suffer from a non-negligible Type II AGN contribution 
to the IR luminosity, in which case the IR-estimated SFR may be somewhat overestimated. 
Note that SFR estimates vary greatly depending on the method, model, and waveband used. 
We demonstrate this in column (3), which shows an average of all published star formation 
rates for each source (see \citealt{mittal15} for the specific values used in these 
averages);  
(4) Lowest and highest X-ray mass deposition rates that have been published for the listed source. Most of 
these come from {\it Chandra} upper limits, rather than more reliable {\it XMM Newton} Reflection Grating Spectrometer (RGS) data; 
(5) Cold molecular hydrogen gas mass in units of $\times10^{9}$ \Msol;  
(6) Central ICM entropy in units of keV cm$^2$;  
(7) 1.4 GHz radio luminosity, based on NVSS or Parkes $L$-band flux densities; 
(8) Jet mechanical power roughly estimated from (7) using the scaling relation of \citet{cavagnolo10};  
(9) X-ray cavity power estimated from {\it Chandra} observations. }
\begin{tabular}{c|ccccc|ccc}
\hline
\multicolumn{9}{r}{{\sc icm cooling proxies ~~~~~~~~~~~~~~~~~~~~~~~~~~~~~~~~~~~~~~~~~~~~~~~~~~~~~~~~~~~~~~~~~~~~~ agn heating proxies~~~~~~~~~~~~~~~~~~~~~~~~~~~~~}} \\
\hline
          &
SFR (IR est.) &
Published SFRs &
$\dot{M}_{\mathrm{cool}}$   &
$M_{\mathrm{H}_2}$  &
$K_0$           & 
$P_{\mathrm{1.4~GHz}}$            &
$\sim P_{\mathrm{jet}}$ (C10 scaling)        &
$P_{\mathrm{cavity}}$ (X-ray) \\
Source Name &
(\Msol\ yr\mone)  &
(\Msol\ yr\mone)  &
(\Msol\ yr\mone)  &
($\times 10^9$ \Msol)  &
(keV cm$^2$)          & 
($\times 10^{40}$ ergs sec\mone)    &
($\times 10^{44}$ ergs sec\mone)  &
($\times 10^{44}$ ergs sec\mone)  \\
(1) & (2) & (3) & (4) & (5) & (6) & (7) & (8) & (9) \\
\hline
\hline
{\bf Abell 11}   & $35^{(\mathrm{a})}$  & 35 &  \nodata & 1.1  & \nodata                                            &  $9.57\pm0.23$          & 4.42   &   \nodata              \\ 
{\bf Abell 1068}   & $188^{(\mathrm{a})\dagger}$  & $80\pm60$ & $40-150$ & 43 &   72                                         &  $1.34\pm0.03$          & 1.01  &   \nodata              \\
{\bf Abell 1664} & $15^{(\mathrm{a})}$ &  $16\pm7$ & \nodata & 23 &  14.4                                                   &  $2.13\pm0.06$          & 1.43  & $0.7\pm0.3$       \\
{\bf Abell 1795}   & $8^{(\mathrm{b})}$ & $8\pm8$ & $1-21$ & 6 &  19                                                      &  $11.73\pm0.32$         & 5.15   & $16.0^{+2.3}_{-0.5}$     \\
{\bf Abell 1835}  & $138^{(\mathrm{c})}$ & $119\pm83$ & \nodata & 90  &  11.4                                                &  $7.85\pm0.16$          & 3.81 & $18.0^{+19.0}_{-6.0}$  \\
{\bf Abell 2199}  & $0.6^{(\mathrm{b})}$ & $0.2\pm0.1$&  $0-3$  & 1.4 & 13.3                                                  &  $10.45\pm0.33$         & 4.72  & $2.7^{+2.5}_{-0.6}$     \\
{\bf Abell 2597} &  $5^{(\mathrm{d})}$  & $5\pm8$ & $20-40$   &  1.8  &  10.6                                               &  $42.04\pm1.11$         & 13.42    & $1.9^{+1.0}_{-0.8}$     \\
{\bf Centaurus}    &  $0.2^{(\mathrm{b})}$ & $0.2\pm0.1$ & $2.6-2.9$ & $<1$  & 2.25                                           &  $1.16\pm0.03^\ddagger$ & 0.91  & $0.08^{+5.8}_{-1.8}$    \\
{\bf Hydra A}     & $4^{(\mathrm{b})}$ & $8\pm7$ & $11-21$  & 3.2 &  13.3                                                 &  $395.46\pm11.37$       & 72.08 & $6.5\pm0.5$      \\
{\bf Perseus}    & $24^{(\mathrm{e})}$ & $30\pm23$ & $12-29$  & 8.5 & 19.4                                                  &  $21.54\pm0.63$         & 8.13 & $1.5^{+1.0}_{-0.3}$     \\
{\bf PKS 0745-191} &  $17^{(\mathrm{a})}$ & $70\pm94$  & $80-260$  & 4.5 & 12.4                                                &  $85.53\pm2.58$         & 22.86 & $17.0^{+14.0}_{-3.0}$  \\
{\bf RX J1504.1-0248}   & $140^{(\mathrm{f})}$ & $237\pm92$ & \nodata & 10 &  13.1                                           &  $11.05\pm0.34$         & 4.93   & \nodata                \\
{\bf RX J2129.6+0005}  & $13^{(\mathrm{a})}$ &  $9\pm5$ & \nodata & \nodata  &  21                                         &  $5.70\pm0.19$          & 3.00   & \nodata                \\
{\bf ZwCl 0348}   & $52^{(\mathrm{a})}$ & $52$ & \nodata  & \nodata &\nodata                                           &  $0.44\pm0.01$          & 0.44  & \nodata                \\
{\bf ZwCl 3146}   &  $67^{(\mathrm{b})}$ & $67\pm59$  & $420-780$  & 80  &   11.4                                            &  $2.60\pm0.12$          & 1.66 & $58.0^{+68.0}_{-15.0}$ \\
{\bf ZwCl 8193}   &  $59^{(\mathrm{a})}$ & 59 & \nodata & \nodata &\nodata                                           &  $16.75\pm0.38$         & 6.73  & \nodata                \\
  \hline
\end{tabular}
\vspace*{-8mm}
\tablerefs{(1) IR-estimated star formation rates are adopted from: 
(a) \citet{odea08}; 
(b) \citet{hoffer12}; 
(c) \citet{mcnamara06}; 
(d) \citet{donahue07}; 
(e) \citet{mittal12}; and  
(f) \citet{ogrean10}.
(3) \citet{mittal15}; 
(4 \& 9) X-ray Mass deposition rates and cavity powers are collected from \citet{birzan04,dunn06,rafferty06,wise07,tremblay12a,kirkpatrick09}.  
(5) Cold molecular gas masses are adopted from \citet{edge01,edge03,salome03,tremblay12a}.
(6) Central ICM entropy $K_0$ is adopted from the main table of the ACCEPT sample, see e.g. \citet{cavagnolo09}; 
(7) $K$-corrected 1.4 GHz luminosities are based on flux densities from NVSS \citep{condon98}, except in the case of Centaurus, which uses the 1.4 GHz flux density from the Parkes Radio Telescope. The flux-to-luminosity conversion is given in Section 2.3; 
(8) \citet{cavagnolo10}. 
} 
\label{tab:properties}
\end{table*}

\begin{table*}
\setlength{\extrarowheight}{-0.3pt}
\scriptsize
\centering
    \caption{A summary of the new and archival observations used in this analysis. Those targets 
for which new FUV continuum or optical data are presented are highlighted in boldface. Where applicable, a reference is given to the earliest  
publication in which the data were first directly analysed.}  
\begin{tabular}{cccccccc}
\hline
{\bf Source Name} &
{\bf $\lambda$ Regime} &
{\bf Facility} &
{\bf Inst. / Mode} &
{\bf Exp. Time / RMS noise} &
{\bf Obs. / Prog. ID}  &
{\bf Reference / Comment} \\
\hline
\hline
Abell 11    & FUV Cont. & {\it HST}  & ACS/SBC F150LP  & 1170 sec & 11230  &   \citet{odea10}      \\ 
            & Ly$\alpha$& {\it HST}  & ACS/SBC F125LP  & 1170 sec  & 11230  &   \citet{odea10}     \\
            & Optical   & {\it HST}  & WFPC2 F606W & 800 sec   &  8719  &  \citet{odea10}    \\
            & X-ray     & {\it ROSAT} & \nodata    & \nodata   &   \nodata &    All Sky Survey \\
            & 8.46+1.46 GHz radio & VLA     & A,B array    &   $74 \mu$Jy    &     AB0878, AL0578   &       \citet{odea10}    \\
\hline
{\bf Abell 1068}  & FUV Cont. & {\it HST}  & ACS/SBC F150LP  &  2766 sec & 12220  &  {\bf New}; \citet{mittal15}   \\
            & Optical   & {\it HST}  & WFPC2 F606W  &  600 sec  & 8301  &   \nodata     \\
            & X-ray     & {\it Chandra}  &  ACIS-S   & 26.8 ksec   & 1652  &    \citet{mcnamara04,wise04}  \\
            & 8.46 GHz radio &  VLA    &  A array    &   \nodata         &  AE0117  &         \nodata        \\
\hline
Abell 1664  & FUV Cont.  & {\it HST}  & ACS/SBC F150LP  & 1170 sec &  11230   &   \citet{odea10}      \\
            & Ly$\alpha$ & {\it HST}  & ACS/SBC F125LP & 1170 sec  &  11230   &  \citet{odea10}  \\
            & Optical    & {\it HST}  & WFPC2 F606W   &  1800 sec   &  11230   &  \citet{odea10} \\
            & X-ray      & {\it Chandra} & ACIS-S    & 36.6 ksec    & 7901    & \citet{kirkpatrick09}  \\
            & 4.86 GHz radio & VLA       &  C array    &   $100 \mu$Jy           &   AE0099      &        \citet{odea10}                 \\

\hline
Abell 1795  & FUV Cont.  & {\it HST}  & ACS/SBC F140LP  & 2394 sec   & 11980  &     \nodata     \\
            & Optical    & {\it HST}  & WFPC2 F555W     & 1600 sec   & 5212   &   \nodata   \\
            & H$\alpha$ Narrow Band&  Baade 6.5m &  IMACS / MMTF      &   1200 sec  &    \nodata      & \citet{mcdonald09} \\
            & X-ray     &  {\it Chandra}  & ACIS-S    & 30 ksec      & 10900 etc.    &  \nodata \\
            & 8.4 GHz radio & VLA      &   A,C,A/D arrays  &   18 hrs      &   AG0273      &      \citet{ge93}     \\
\hline
Abell 1835  & FUV Cont.  & {\it HST}  & ACS/SBC F165LP   &  1170 sec  & 11230  & \citet{odea10}      \\
            & Ly$\alpha$ & {\it HST}  & ACS/SBC F140LP   &  1170 sec  & 11230  &  \citet{odea10} \\
            & Optical    & {\it HST}  & WFPC2 F702W     &  7500 sec   & 8249   &  \nodata  \\
            & X-ray      & {\it Chandra} & ACIS-I      &  117.9 ksec  & 6880   &  \nodata   \\
            &  \nodata   &  \nodata    &  \nodata     & 36.3 ksec    & 6881   &  \nodata   \\
            &  \nodata   &  \nodata    &  \nodata     & 39.5 ksec   & 7370    &   \nodata  \\
            & 4.76 GHz radio & VLA        &  A,C arrays     &  2 hrs / $47 \mu$Jy      &  AT0211   &   \citet{govoni09}  \\
\hline
{\bf Abell 2199}  & FUV Cont. & {\it HST}  & ACS/SBC F140LP   & 2767 sec   & 12220    &  {\bf New}; \citet{mittal15}    \\
              & Optical   & {\it HST} & WFPC2 F555W    &  5200 sec       & 7265      &  \nodata    \\ 
            & X-ray      & {\it Chandra}  &  ACIS-I    &  120 ksec   &  10748 etc.  & \citet{nulsen13}\\
            & 5 GHz radio & VLA       &  B,C,D arrays  &   6 hrs     &    AG0269            &  \citet{ge94}   \\
\hline
Abell 2597  & FUV Cont. & {\it HST}  & ACS/SBC F150LP   & 8141 sec  &  11131    & \citet{oonk10,tremblay12a}      \\
            & Ly$\alpha$ & {\it HST}  & STIS F25SRF2    & 1000 sec  & 8107     & \citet{odea04,tremblay12a} \\
            & Optical \& H$\alpha$  & {\it HST}  & WFPC2 F702W  & 2100 sec   & 6228  &  \citet{holtzman96} \\
            & H$\alpha$ Narrow Band&  Baade 6.5m &  IMACS / MMTF      &   1200 sec  &    \nodata       & \citet{mcdonald11b,mcdonald11a} \\
            & X-ray    &  {\it Chandra} &  ACIS-S  &   39.8 ksec  & 922     &  \citet{mcnamara01,tremblay12a,tremblay12b} \\
            & \nodata  & \nodata    &  \nodata   &   112 ksec  & 6934, 7329    & \citet{tremblay12a,tremblay12b} \\
            & 8.4 GHz radio &  VLA     &   A array &   15 min     & AR279    & \citet{sarazin95}   \\
            & 330 MHz radio & VLA      &   A array  &  3 hrs    & AC674     & \citet{clarke05}  \\
\hline
Centaurus  & FUV Cont.  & {\it HST}  & ACS/SBC F150LP   &  1780 sec  & 11681    &  \citet{mittal11}      \\
           & Optical    &  {\it HST}  & ACS/WFC F814W  &  8060 sec  & 9427    &  \citet{harris06}          \\
           & \nodata    & \nodata    &  ACS/WFC F435W  & 8654 sec   & 9427      &  \citet{harris06}        \\
           & \nodata    & \nodata    &  WFC3 F160W   &  392 sec        & 11219  & \citet{baldi10}        \\ 
           & X-ray      & {\it Chandra} & ACIS-S     &  200 ksec    &  504,5310,4954,4955   & \citet{fabian05}       \\
           & 1.46 GHz radio & VLA     &  A,B/A      &   1.5 hr    &      AT211      &    \citet{taylor02}          \\
\hline
{\bf Hydra A}  & FUV Cont.    & {\it HST}  & ACS/SBC F140LP    & 2709 sec   & 12220    &  {\bf New}; \citet{mittal15}   \\
           & Optical    & {\it HST}   & ACS/WFC F814W   & 2367 sec       & 12220      &{\bf New}; \citet{mittal15}  \\
           & H$\alpha$ Narrow Band&  Baade 6.5m &  IMACS / MMTF      &   1200 sec  &    \nodata       & \citet{mcdonald10,mcdonald11a} \\
          & X-ray       & {\it Chandra}    &  ACIS-S  &  196 ksec   & 4969,4970   &  \citet{nulsen05}          \\
          &  4.6 GHz radio  &     VLA    &  A,A/B,B,C,D     &       9 hrs      &    AL0032           &     \citet{taylor90}             \\
\hline
Perseus  &  FUV Cont.   & {\it HST}  & ACS/SBC F140LP    & 2552 sec   &  11207   &  \citet{fabian08}     \\
         &  \nodata      &  \nodata  & STIS F25SRF2      &  1000 sec  & 8107     &  \citet{baum05} \\
         &  Optical \& H$\alpha$     & {\it HST}  & ACS/WFC F625W    & 4962 sec    & 10546     & \citet{fabian08} \\
         &  H$\alpha$ Narrow Band  & KPNO WIYN 3.5m & S2kB CCD / KP1495 & 3200 sec &         & \citet{conselice01} \\
         & X-ray      &  {\it Chandra} &  ACIS-S \& -I &   1.4 Msec    & 11713 etc.   &  \citet{fabian11b}  \\
         &   1.4 GHz radio   &    VLA   &    C,D,C/D arrays            &    3 min       &      AT149A            &     \citet{condon96}             \\
\hline
{\bf PKS 0745-191}  &  FUV Cont. & {\it HST}  & ACS/SBC F140LP  &  2715 sec  & 12220      & {\bf New}; \citet{mittal15}      \\
            & Optical   & {\it HST}    & WFPC2 F814W      &  1200 sec      &  7337       & \citet{sand05}   \\
        & X-ray     & {\it Chandra}  &  ACIS-S    &   50 ksec    &  508,2427       &   \citet{fabian99}   \\
          &  8.4 GHz radio     & VLA   &  A array   &     2 hr        &       BT024             &      \citet{taylor94}                 \\
\hline
{\bf RX J1504.1-0248}  & FUV Cont. & {\it HST}  & ACS/SBC F165LP  &  2700 sec  & 12220     & {\bf New}; \citet{mittal15}        \\
                      & Optical    & {\it HST}  & WFC3/UVIS F689M & 2637 sec   & 12220     &  {\bf New}; \citet{mittal15}   \\ 
         &  X-ray     &  {\it Chandra}   &   ACIS-I   &  40 ksec    & 5793    & \citet{bohringer05,ogrean10}    \\

        &  8.46 GHz radio    &   VLA   &    A array     &     3 hr          &      AB1161         &       \nodata    \\
\hline
RX J2129.6+0005  & FUV Cont.  & {\it HST}  &  ACS/SBC F165LP   &  1170 sec  & 11230  &  \citet{odea10}      \\
              & Ly$\alpha$  & {\it HST}  &  ACS/SBC F140LP   & 1170 sec   & 11230   & \citet{odea10} \\
               &  Optical    & {\it HST}   & WFPC2 F606W   &  1000 sec   & 8301   &   \citet{donahue07}     \\
           &  X-ray      &  {\it Chandra}  &  ACIS-I      &  30 ksec    &   9370   &    \nodata              \\
           &  8.46 GHz radio  &  VLA      &   A array    &   $50 \mu$Jy  &  AE117  &    \citet{odea10}    \\
\hline
ZwCl 0348  & FUV Cont.   & {\it HST}  &  ACS/SBC F165LP   & 1170 sec   & 11230   &    \citet{odea10}     \\
           & Ly$\alpha$  & {\it HST}  &  ACS/SBC F140LP   & 1170 sec   & 11230   & \citet{odea10} \\
           & Optical     & {\it HST}  &  WFPC2 F606W      &  700 sec   & 11312   & \citet{smith10} \\
           & X-ray       &  {\it Chandra} & ACIS-S        &  50 ksec   &  10465   &   \nodata             \\
           &  4.86 GHz radio  &   VLA   &  A/B array     &   $66 \mu$Jy &   AK359 &    \citet{odea10}    \\
\hline
ZwCl 3146  & FUV Cont.    & {\it HST}  & ACS/SBC F165LP    & 1170 sec   & 11230   &   \citet{odea10}     \\
           & Ly$\alpha$   & {\it HST}  & ACS/SBC F140LP    & 1170 sec   & 11230   &   \citet{odea10}     \\
           & Optical     & {\it HST}  & WFPC2 F606W       &  1000 sec   & 8301    &  \citet{donahue07} \\ 
           & X-ray       &  {\it Chandra}  &  ACIS-I      &  90 ksec    & 909,9371   &  \citet{boschin02}     \\
           & 4.86 GHz radio   &   VLA       &  A/B array   &   $50 \mu$Jy &  ACTST    &  \citet{odea10}                    \\
\hline
ZwCl 8193  & FUV Cont.   & {\it HST}  & ACS/SBC F150LP    & 1170 sec   & 11230   &    \citet{odea10}     \\
           & Ly$\alpha$   & {\it HST}  & ACS/SBC F140LP    & 1170 sec   & 11230   &   \citet{odea10}     \\
           & Optical     & {\it HST}   & WFPC2 F606W      & 1900 sec   & 11230   &   \citet{odea10}    \\
           & X-ray       &  {\it ROSAT} &  \nodata        & \nodata    & \nodata   &  All Sky Survey \\
           & 8.46 GHz radio  &  VLA    &   A array     &  $180 \mu$Jy  &  AE117    &   \citet{odea10}             \\
  \hline
  \end{tabular}
\label{tab:observations}
\end{table*}

\subsection{New Observations}

Along with  more than  50 multiwavelength archival  observations, this
paper presents  five new FUV  continuum and two new  broadband optical
{\it  HST}  observations,  all   of  which  are  summarised  in  Table
\ref{tab:observations}.    Although  much  of   this  data   has  been
re-reduced   in   a   homogeneous   way   for   this   analysis   (see
\S\ref{section:datareduction}), we refer  the reader to the references
listed  in  the rightmost column of  Table  \ref{tab:observations}  for  more
observational details pertaining to the archival data.

The new {\it  HST} FUV and optical images we  present were obtained in
Cycle 19  as part of  General Observer program 12220  (PI: R.~Mittal).
The line-free  FUV continuum data  were obtained with the  Solar Blind
Channel (SBC) MAMA  detector of the Advanced Camera  for Surveys (ACS,
\citealt{clampin04}).   Total  exposure  times  for each  target  were
roughly $\sim 2700$ sec (roughly one {\it HST} orbit minus overheads),
and  the observations  were carried  out with  a  standard three-point
dither  pattern.  Depending on  target redshift,  we used  the F140LP,
F150LP, and  F165LP long-pass filters  with pivot wavelengths  of 1527
\AA,  1611 \AA,  and  1758 \AA,  respectively.   These filter  choices
ensured that Ly$\alpha$ emission did not fall within their bandpasses,
which  have similar  red cutoff  wavelengths of  $\sim 2000$  \AA, but
different blue cutoff (or minimum)  wavelengths of 1370 \AA, 1470 \AA,
and 1650 \AA, respectively. The plate scale for the SBC is $0\farcs034
\times  0\farcs030$ pixel\mone,  and  the detector  field  of view  is
$34\farcs6  \times  30\farcs8$.   Although  we will  not  discuss  the
archival  FUV observations  for all  other sources,  we note  that the
observing strategy  employed for those  datasets was very  similar (if
not identical) to that used for our new FUV observations.

To  fill a  data  coverage gap,  we  also obtained  two new  line-free
optical images of Hydra A  and RX J1504.1-0248 (hereafter R1504) using
the  UVIS  channel   of  {\it  HST}'s  Wide  Field   Camera  3  (WFC3,
\citealt{dressel12}).   As for  the FUV  images, a  three-point dither
pattern was used  over a $\sim 2500$ sec total  exposure time for each
target.  The  F814W and F689M  filters with pivot wavelengths  of 8024
\AA\ and 6876 \AA\ were used for Hydra A and R1504, respectively.  The
widths of these passbands (1536  \AA\ and 683 \AA) forbid optical line
contamination (though  note that some  archival optical images  we use
for other sources do contain  optical line emission like H$\alpha$+[N~{\sc ii}]).   
More details on these new {\it
  HST}  observations  can be
found in Mittal et al.~(2015, submitted).

\begin{figure*}
\begin{center}
\includegraphics[scale=1.12]{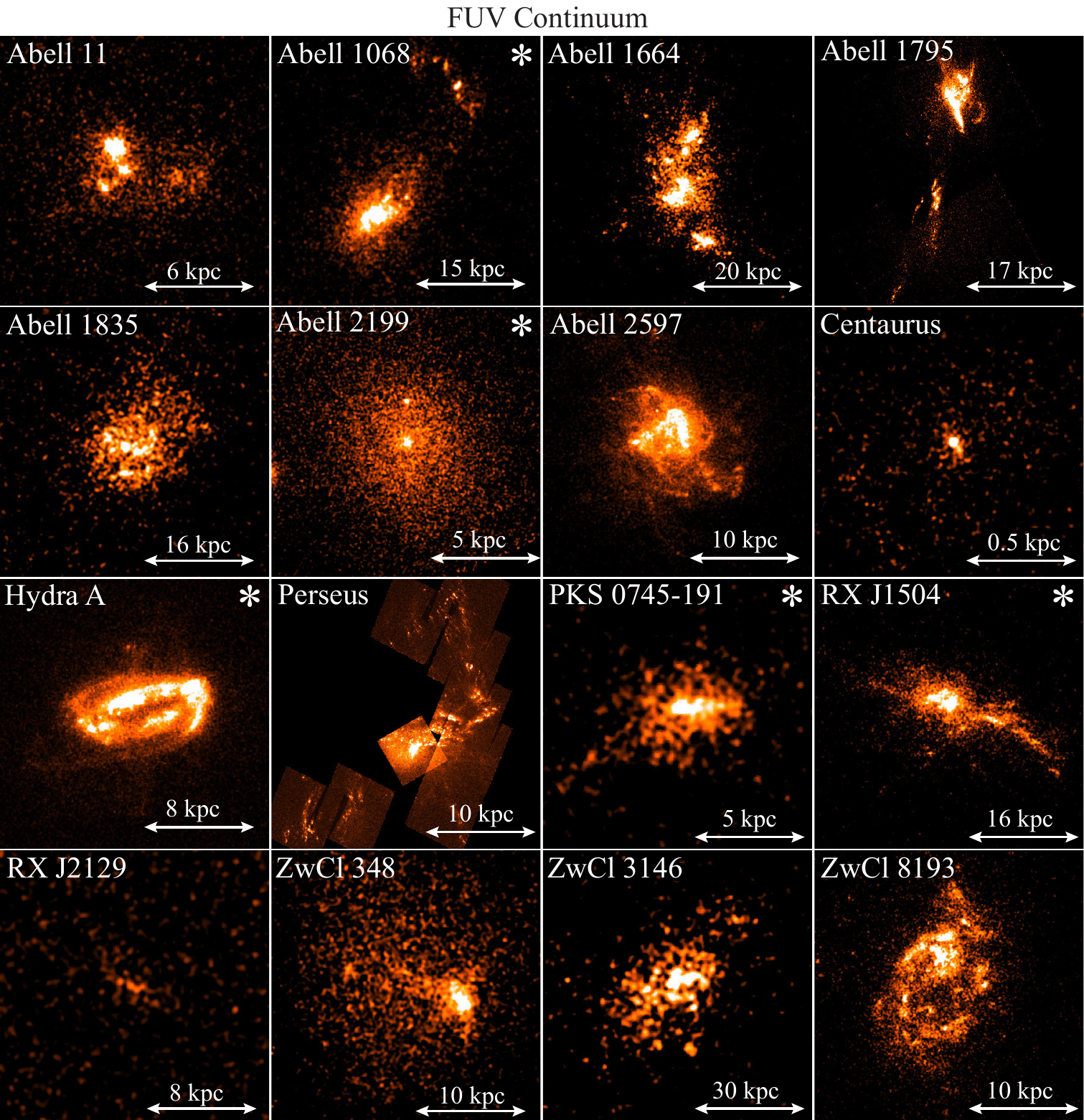}
\end{center}
\caption{{\it HST}/ACS SBC images of FUV continuum emission associated
  with young,  massive stars in  the central $\sim  50$ kpc of  the 16
  low-redshift CC BCGs in our  sample.  It is immediately obvious that
  the star  formation in  these systems does  not occur  in monolithic
  slabs of  gas, but rather  in highly complex filamentary  and clumpy
  distributions.   These   images   are   discussed  in   general   in
  \S\ref{section:general_morphology}.  The five  sources for  which we
  are showing  newly obtained  FUV data are  marked with  an asterisk.
  All panels are rotated such that east is left and north is up.  This
  paper presents multiwavelength data for each of these sources, which
  can be  seen by  referring to  the Figures listed  in column  (7) of
  Table \ref{tab:sample}. }
\label{fig:postagestamps}
\end{figure*}

\subsection{Data Reduction}
\label{section:datareduction}

All  new or  archival {\it  HST}  FUV and  optical data  used in  this
analysis  were retrieved from  either the  Mikulski Archive  for Space
Telescopes  (MAST\footnote{\texttt{http://archive.stsci.edu/}}) or the
Hubble  Legacy Archive (HLA\footnote{\texttt{http://hla.stsci.edu/}}),
and  MAST   products  were  reduced  using   the  standard  on-the-fly
recalibration (OTFR) pipelines.  {\it Chandra} X-ray observations were
obtained  as   level  one  products   from  the  {\it   Chandra}  Data
Archive\footnote{\texttt{http://asc.harvard.edu/cda/}}.       Exposures
were reduced, reprojected,  exposure-corrected, and merged using using
the   standard    {\sc   CIAO}   \citep{fruscione06}    v4.5   scripts
(\texttt{chandra\_repro}, \texttt{reproject\_obs}, \texttt{flux\_obs})
with v4.5.5.1  of the calibration database.  Finally,  while many high
resolution Very Large  Array (VLA) radio maps were  kindly provided by
colleagues, some raw datasets (for  A1068, RX J1504, and PKS 0745) had
to          be         obtained         from          the         NRAO
Archive\footnote{\texttt{https://archive.nrao.edu/}}.      The    NRAO
AIPS\footnote{\texttt{http://www.aips.nrao.edu/}} package was used for
(self-) calibration, imaging, and deconvolution of these data.

\begin{figure*}
\begin{center}
\includegraphics[scale=1.1]{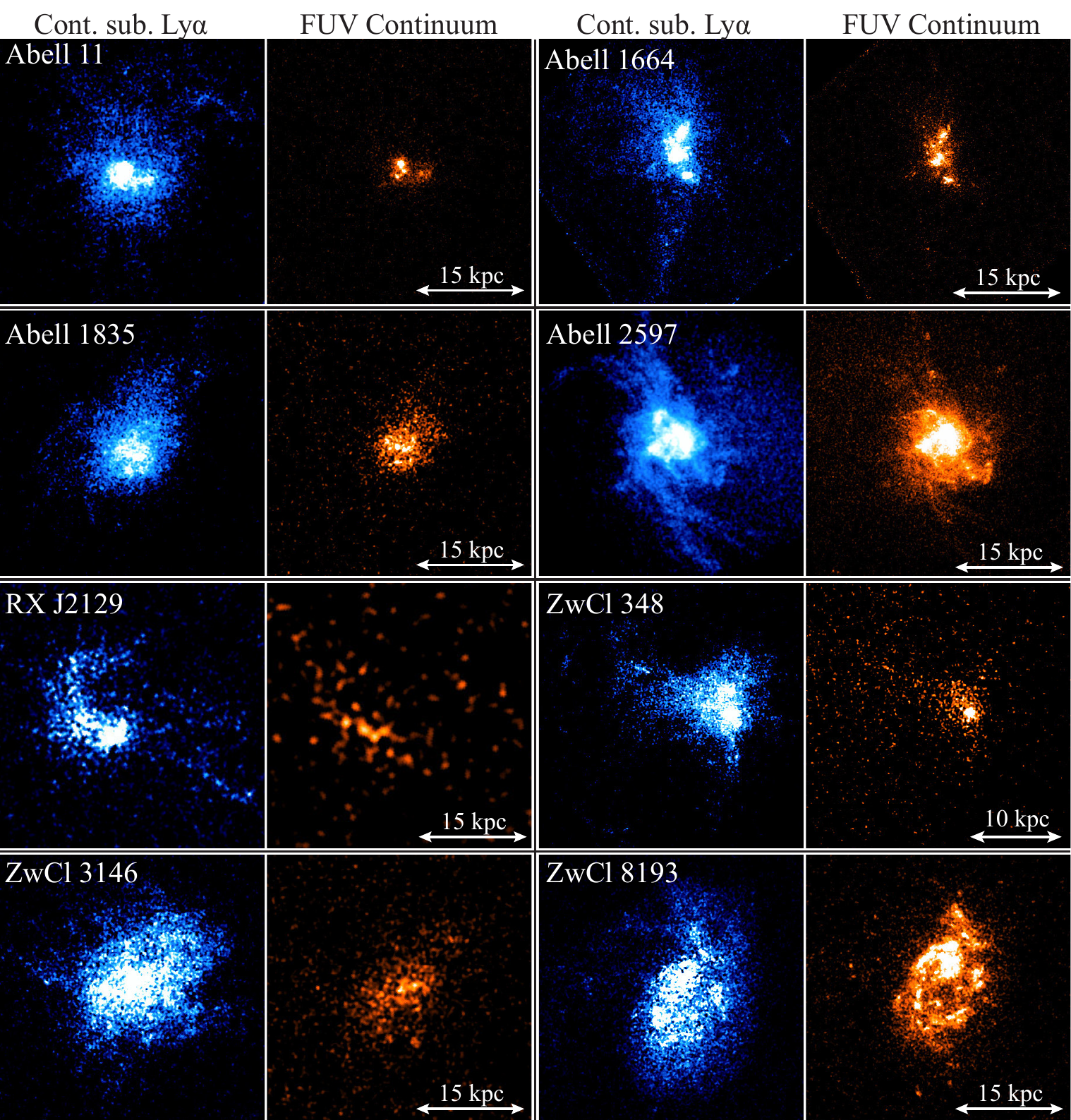}
\end{center}
\caption{A comparison of Ly$\alpha$ and FUV continuum morphologies for
  a  subset of  our sample.  In  blue we  show the  {\it HST}/ACS  SBC
  continuum-subtracted Ly$\alpha$ images from \citealt{odea10}, and in
  orange  we  show  the   FUV  continuum  images.  While  the  general
  morphologies are  very similar, the  Ly$\alpha$ emission is  far more
  extended  than   the  underlying  FUV  continuum.   This  result  is
  unsurprising  given  the   sensitivity  of  Ly$\alpha$  emission  to
  resonant scattering. \citet{odea10} demonstrated that the underlying
  FUV  continuum   strength  was  sufficient  to   fully  account  for
  production   of  the  observed   Ly$\alpha$  emission   via  stellar
  photoionization. All image  pairs are aligned and shown  on a common
  spatial scale, with east left and north up.}
\label{fig:lyacompare}
\end{figure*}

\begin{figure*}
\begin{center}
\includegraphics[scale=0.9]{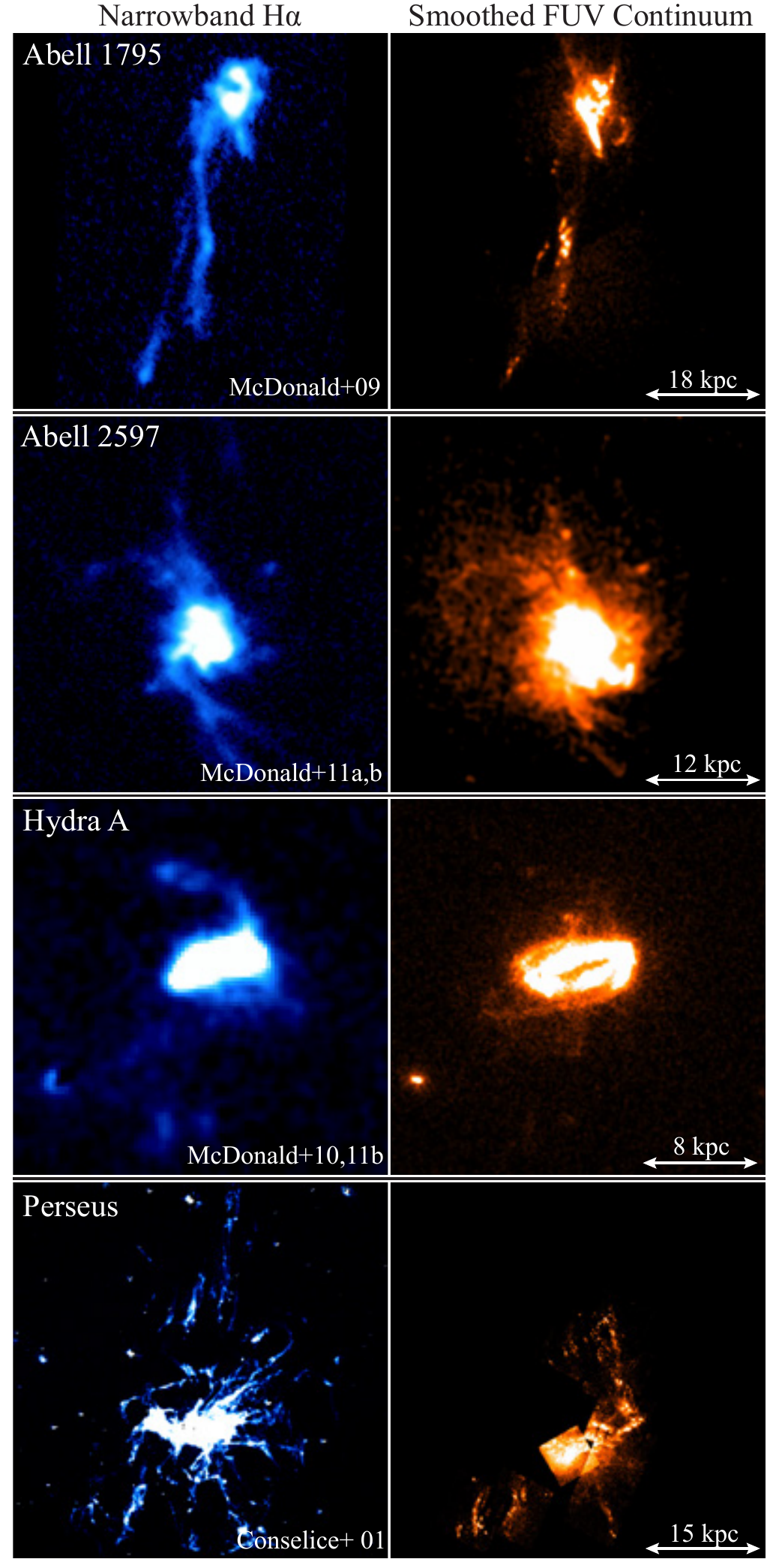}
\end{center}
\vspace*{-2mm}
\caption{A comparison of H$\alpha$  and FUV continuum morphologies for
  a subset  of our  sample. In blue  we show the  narrowband H$\alpha$
  images    from    \citealt{mcdonald10,mcdonald11b,mcdonald11a}   and
  \citealt{conselice01},  and  in orange  we  show  the FUV  continuum
  images after gaussian smoothing to (approximately) match the spatial
  resolution of the  H$\alpha$ maps.  All image pairs  are aligned and
  shown on a  common spatial scale, with east left  and north up.  The
  morphologies  are very similar  overall, though  it is  important to
  note  that  some H$\alpha$  filaments  lack  detected cospatial  FUV
  continuum, and some FUV  filaments lack detected cospatial H$\alpha$
  emission (though non-detection of  course does not necessarily imply
  absence).  Many works have demonstrated that continuum emission from
  young  stars can  account for  a dominant  fraction of  the ionizing
  photons    needed   to   power    the   H$\alpha$    nebula   (e.g.,
  \citealt{odea10,mcdonald11b}).  However,  an  additional  ionization
  mechanism (e.g.,  shocks, cosmic rays, thermal  conduction, etc.) is
  needed in all cases (see \S1). }
\label{fig:halphacompare}
\end{figure*}

This paper also  presents some datasets that have  not been re-reduced for
this  analysis.  Data reduction  and continuum-subtraction details for   the
{\it   HST}   ACS/SBC   Ly$\alpha$    images   shown   in
Fig.~\ref{fig:lyacompare}   can  be   found  in   \citet{odea10}.  Reduction
of the  Maryland-Magellan  Tunable  Filter  (MMTF) narrowband  H$\alpha$  maps
shown in Fig.~\ref{fig:halphacompare}  is   described  in \citet{mcdonald10}.
The 1.4  Msec {\it Chandra} X-ray map  of  Perseus (shown  in
Fig.~\ref{fig:perseus_figure}) is discussed at length in \citet{fabian11b}. $L$-band radio luminosities quoted in Table 2 use flux densities 
from the NRAO VLA Sky Survey (NVSS, \citealt{condon98})  
 assuming the relation $P_{\mathrm{1.4~GHz}} = 4\pi D_L^2 S_{\nu_0} \nu_0
\left(1+z\right)^{\alpha-1}$, where $D_L$ is the Luminosity
Distance to the source, $S_{\nu_0}$ is the 1.4 GHz radio flux
density integrated over the source area, $\nu_0=1.4$ GHz is the frequency of
the observation, and $\alpha$ is the radio spectral index  used in the
(negligible) $K$-correction, assumed here to be $\alpha=0.8$  if $S_\nu \propto \nu^{-\alpha}$ (these
luminosities are insensitive to choice of $\alpha$ given the narrow and low
redshift range of our targets).

All images were spatially aligned using IRAF shifting and registration
tasks.   To aid  viewing  of certain  X-ray  or optical  morphological
features, in many  cases we show unsharp masks  wherein the ``smooth''
X-ray or optical light has been subtracted from the surface brightness
map, highlighting residual edge  structures.  X-ray unsharp masks were
made in the CIAO  environment by gaussian smoothing exposure-corrected
maps with both small and  large kernel sizes. The heavily smoothed map
was then  subtracted from the  lightly smoothed map, and  the residual
image was normalized by the  sum of both smoothed maps.  Unsharp masks
of the FUV and  optical {\it  HST} data were  made using essentially  the same
technique in the IRAF environment. \vfill

\subsection{The SBC red leak and other contaminants}
\label{section:redleak}

The ACS  Solar Blind Channel  suffers from a poorly  characterized and
highly variable  red leak (e.g., \citealt{ubeda12}),
wherein the FUV  long pass filters can permit  a substantial amount of
``red''   (i.e.   optical)  interloper   flux  through   the  bandpass,
contaminating what should  otherwise be a pure FUV  image.  The effect
is extremely difficult to correct  for in the absence of multi-band UV
imaging, as it depends on  both time and detector temperature, varying
by as  much as  30\%\ across five  consecutive orbits.  There  is some
evidence  that  the  effect  has  decreased since  2008,  and  current
estimates suggest  that, at worst,  it may artificially boost  the FUV
count rate by about 10-20\%\ (STScI ACS team, private communication).

While one  must therefore be wary  when interpreting SBC  FUV images of
otherwise red,  luminous elliptical galaxies,  solace is found  in the fact
that the contributors of  red leak photons are almost exclusively solar-  and
later type  stars.  This  means that  the effect  can only significantly
contaminate by  means of  a smooth,  diffuse,  and very faint  background
whose  surface  brightness  tracks  the  underlying optical isophotes of the
host  galaxy's old stellar component.  We are therefore able to circumvent the
issue in this paper by quantitatively and  qualitatively interpreting  {\it
only}  high  surface brightness, spatially anisotropic  FUV-bright clumps  and
filaments, to  which the red leak  cannot significantly contribute beyond a
slight increase in count rate  that is effectively  uniform across such
structures.  The same argument applies for highly variable contamination from
the old stellar ``UV upturn''  population (see e.g.,  \citealt{oconnell99},
for  a review). One  can  therefore  be  confident  that the  clumpy  and
filamentary kpc-scale   emission   ubiquitously    seen   in   our   images
(see Fig.~\ref{fig:postagestamps}) is almost  entirely due to young ($\lae 10$
Myr), massive  ($\gae 5$  \Msol) stars.   We  nevertheless caution  against
over-interpretation  of  the smooth, diffuse  emission seen  in a  few of  our
images (Abell 2199, for example\footnote{Abell 2199 and Centaurus are the only
two sources in our sample for which the SBC red leak may make a dominant
contribution to observed morphology. As a test, we have scaled the $V$-band
optical image of A2199 by a factor consistent with the  known range of
reasonable FUV/optical colors  (discussed at length in e.g.~\citealt{oonk11}),
then subtracted this scaled image from the FUV map. We are unable  to rule out
the possibility that {\it all} emission seen in the A2199 ``FUV''  image is a
combination of red leak from the  underlying old stellar  component, plus a
central  FUV point source  associated with  the AGN.   The  star formation
rate in  this source may therefore  be effectively zero. The same is true for
Centaurus,  however all other sources in our sample are dominated by bright
FUV clumps and filaments whose  morphology will be unaffected by the red
leak.}),   and stress  that no quantitative plots or scientific conclusions
presented in this paper are based  in any way on this diffuse emission.

\begin{figure*}
\begin{center}
\includegraphics[scale=0.48]{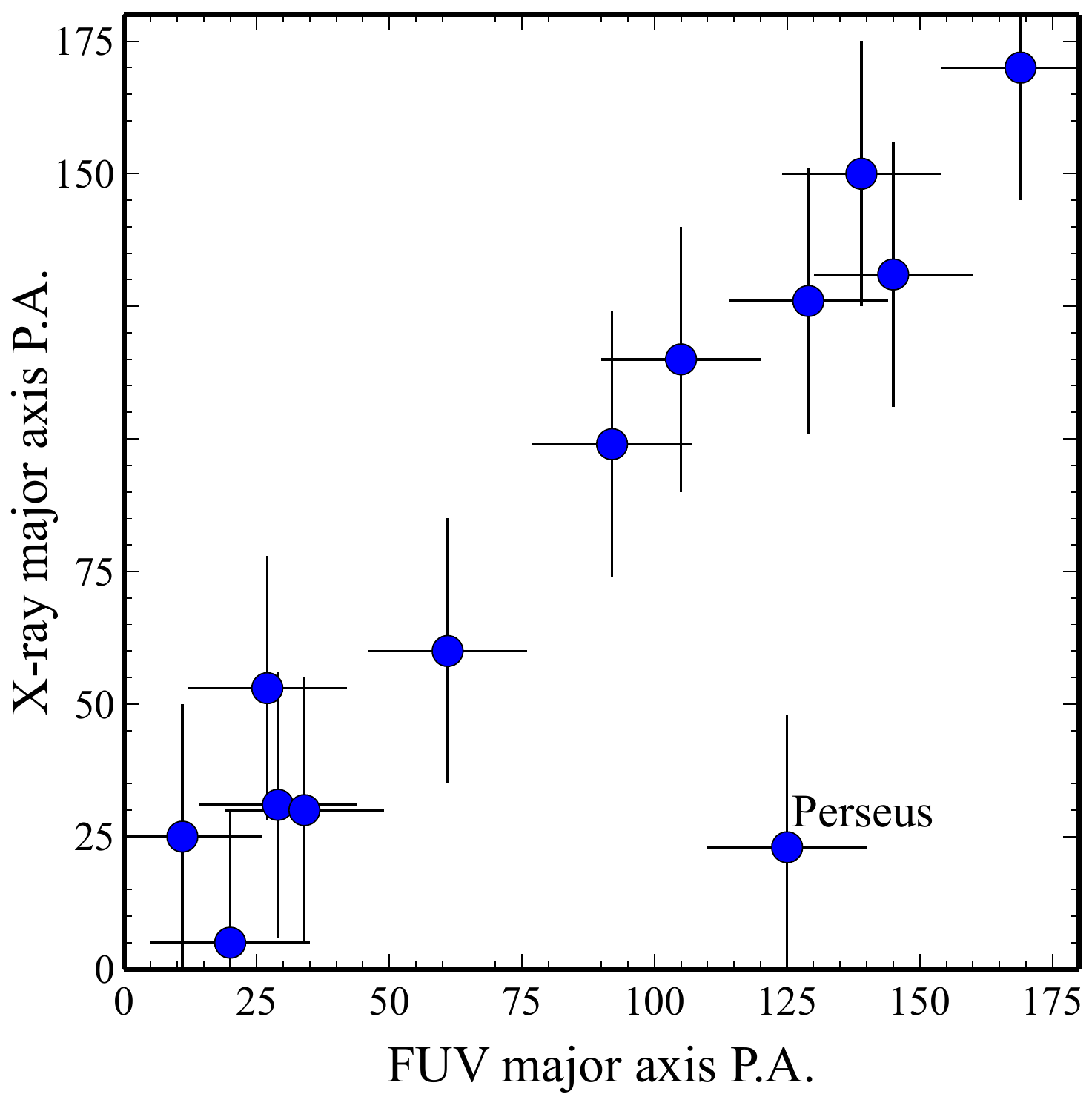}~~~~
\includegraphics[scale=0.48]{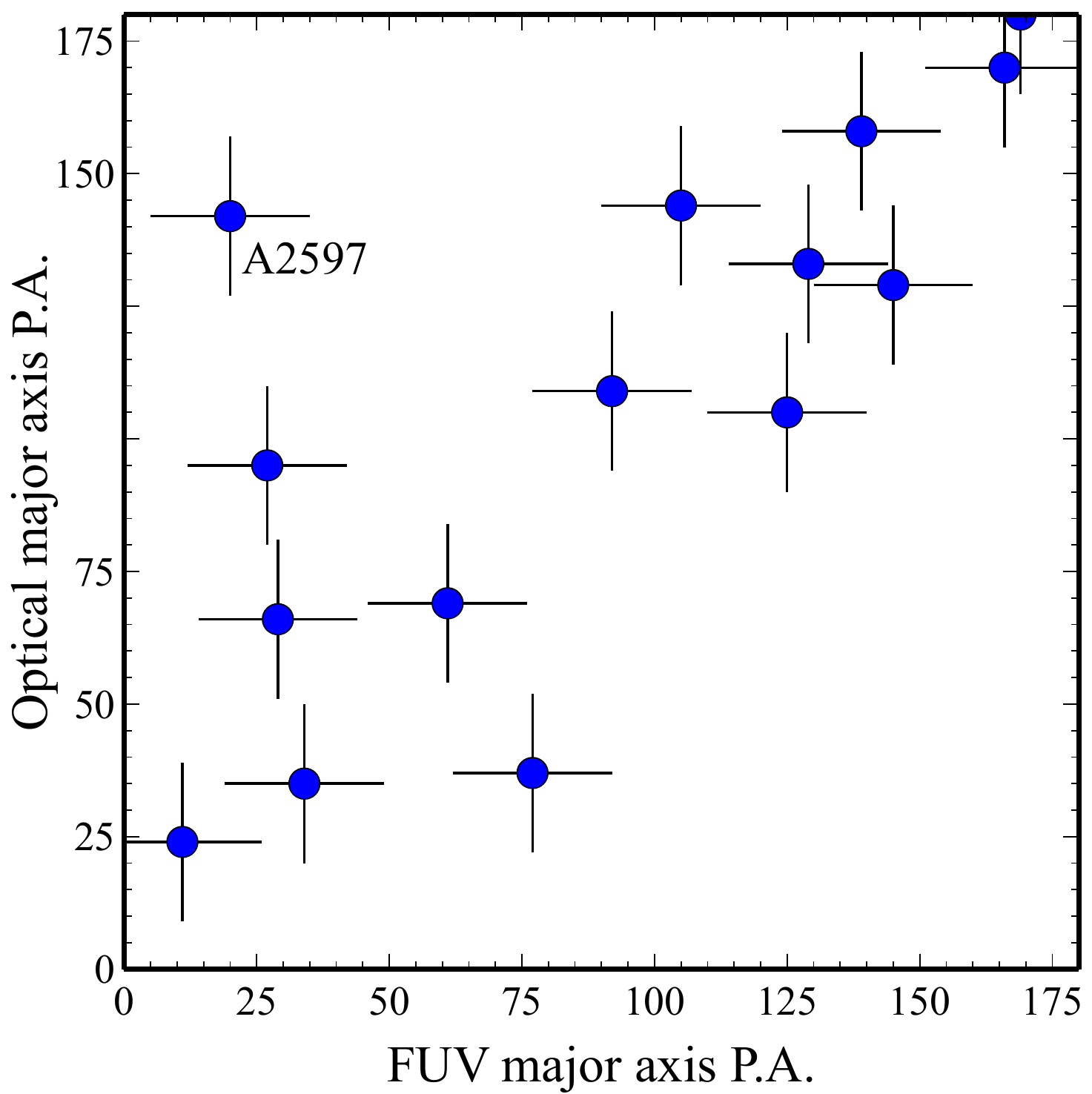}\\
\includegraphics[scale=0.48]{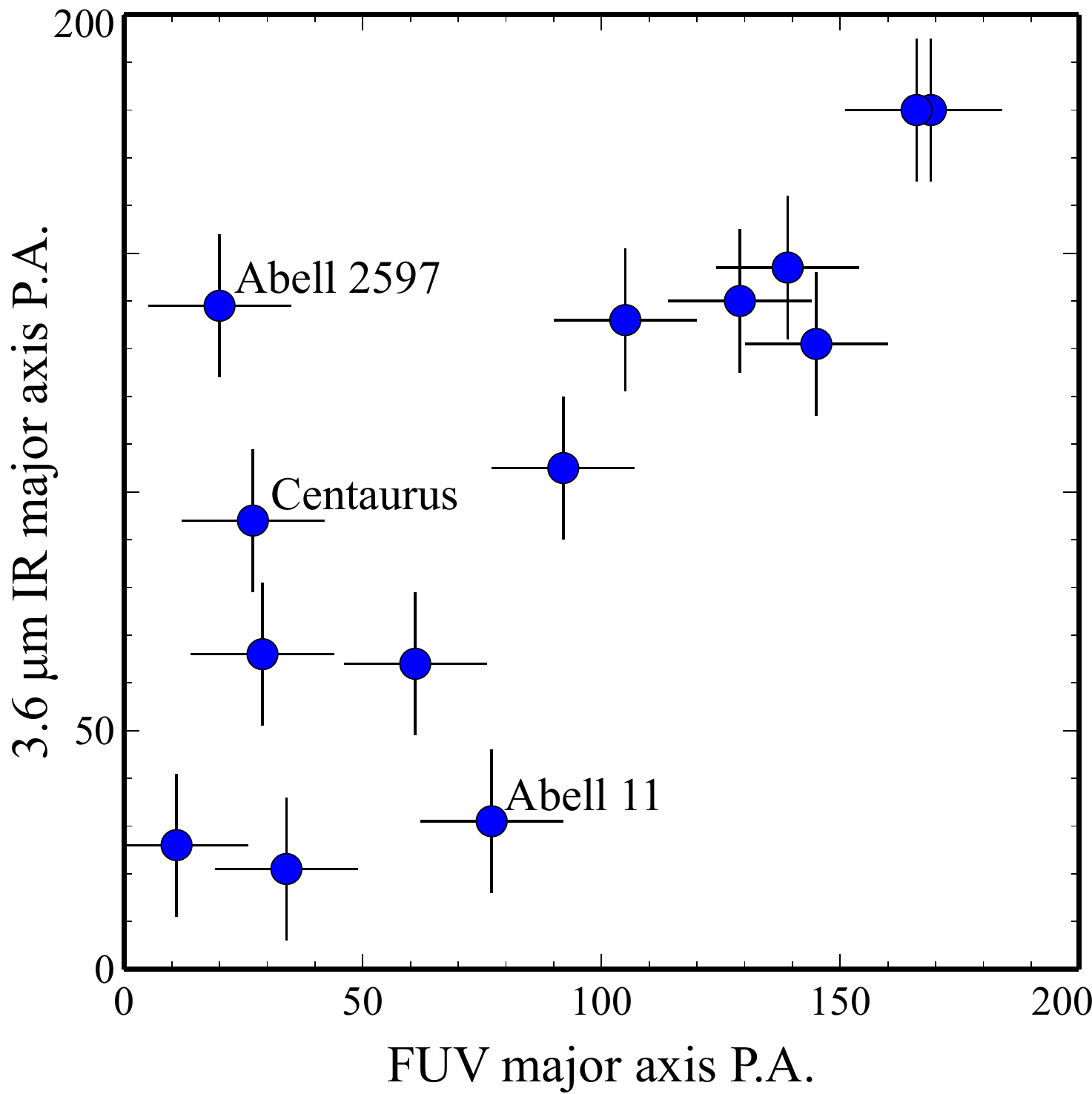}~~~~
\includegraphics[scale=0.48]{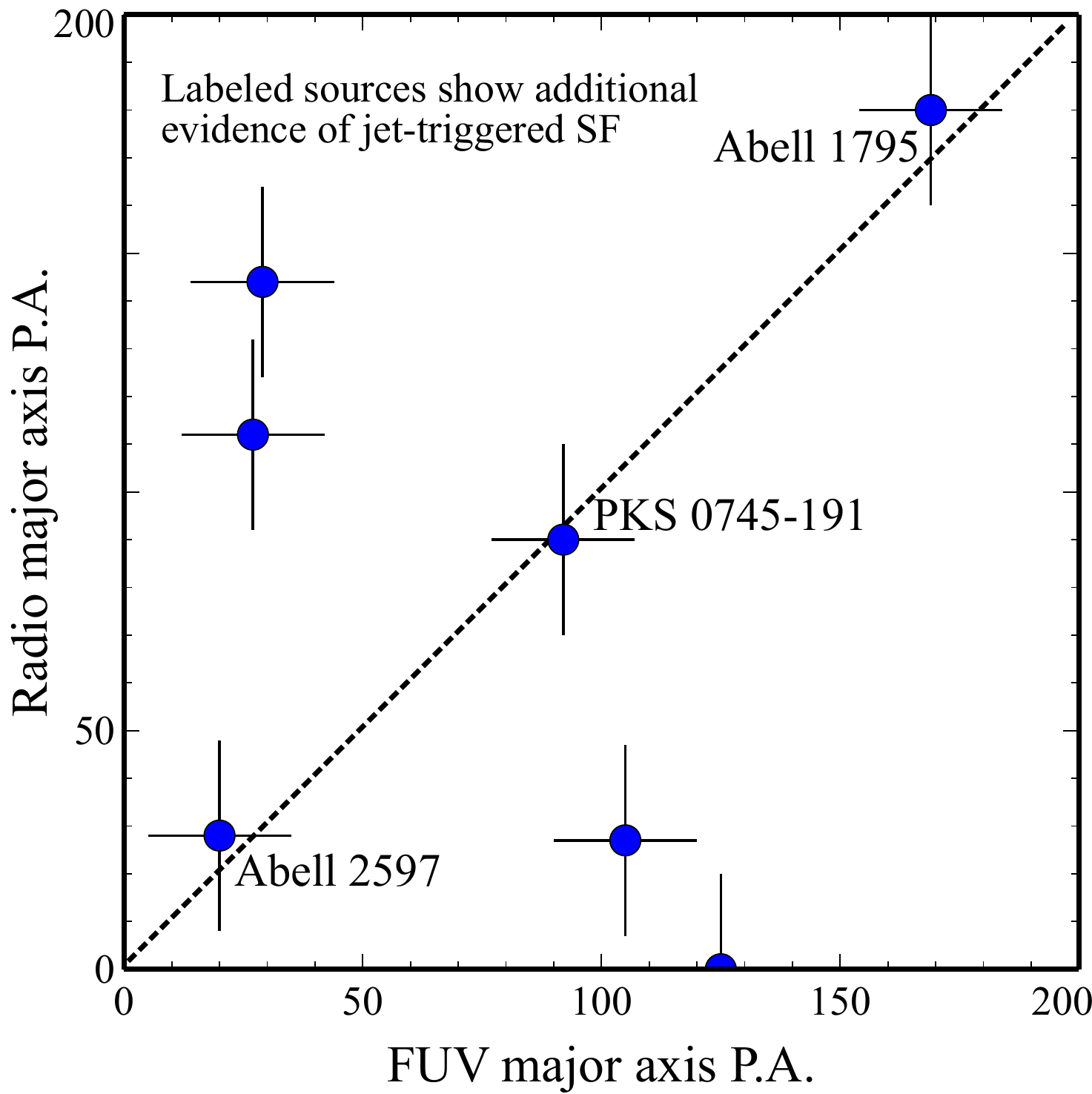}
\end{center}
\caption{A comparison of projected position angles (P.A.) for the FUV,
  X-ray,  optical, IR,  and  radio isophotal  major  axes at  matching
  spatial scales. P.A.~has  been measured N through E.  If the P.A. of
  the X-ray, optical, or IR major axis was found to vary strongly as a
  function  of radius, we  measured the  P.A. of  the major  axis that
  matched  the  largest  angular  size~of  the  FUV  emission.  Strong
  alignment  between  FUV,  X-ray,  optical, and  IR  counterparts  is
  observed. Those  sources that are  outliers to any  particular trend
  are labeled  in the  respective plot. The  radio vs.~FUV  major axis
  comparison  ({\it  bottom right})  shows  alignment  only for  those
  sources  that   exhibit  evidence  for   either  jet-triggered  star
  formation or  strong dynamical interaction between  the radio source
  and star  forming gas. The dashed  line on the bottom  right plot is
  the one-to-one line. Spearman-rank and Pearson correlation coefficients for these
  plots are shown in Table~\ref{tab:spearman}.  }
\label{fig:paplots}
\end{figure*}

\section{Results}

\label{section:general_morphology}

The  FUV  continuum  images  for  our full  sample  are  presented  in
Fig.~\ref{fig:postagestamps}.  The scales over  which FUV  emission is
detected varies  from 500 pc (Centaurus)  to 67 kpc  (A1795). Mean and median
largest  angular  sizes  (L.A.S)   are  30  kpc  and  33  kpc, respectively.
Estimated star formation rates range from
effectively zero or  $\ll 0.1$ \Msol\ yr\mone\ (Centaurus, RX J2129,  A2199)
to  $\sim  150$  \Msol\ yr\mone\  (A1068,  A1835,  RX J1504). These  and other
properties such as  cold molecular  gas mass, X-ray estimated mass deposition
rates, radio source power, and X-ray cavity power are
summarised for all targets in Table \ref{tab:properties}.

The FUV morphological analyses in the sections below come with an important caveat:  FUV
emission is  highly sensitive to extinction  by dust. The  FUV  emission  that
we  do  detect  likely  stems only  from  the outermost  layers  of dense,
dusty  star  forming  clouds, which  are themselves obscured  by intervening
dust  along the line of  sight.   As the FUV is particularly sensitive to
young stars less than $\sim10$ Myr old and more  massive than $\sim 5$ \Msol,
our images should be considered instantaneous ``snapshots''  of ongoing or
very recent {\it unobscured} star formation.   A detailed treatment of
extinction for a majority subset of our sample is  provided by \citet{mittal15},  
and will  not  be  discussed  here beyond  cautioning
against over-interpretation of observed FUV structures. The clumps and
filaments  we do detect are likely ``tips of icebergs'', and smooth,
diffuse emission  may be significantly contaminated by red leak and the UV
upturn population (as discussed  in \S\ref{section:redleak}).

It is nevertheless obvious from Fig.~\ref{fig:postagestamps} that star
formation in our sample is not occurring amid  monolithic slabs of gas. The 
observed FUV
morphologies are instead   highly  clumpy  and  filamentary,
exhibiting   a  variety   of   associations   (and  sometimes
interesting  {\it non}-associations)  with X-ray,  optical,  and radio
features, as well as galaxy properties such as central X-ray entropy 
and the relative strength of AGN feedback signatures. 
These associations are discussed in the following sections.

\subsection{Comparison of FUV, Ly$\alpha$, and H$\alpha$ morphology}

In Fig.~\ref{fig:lyacompare}  we compare a subset of  our targets with
the continuum-subtracted Ly$\alpha$  data of \citet{odea10}.  Although
the morphologies are very similar  overall, the Ly$\alpha$ is far more
extended   than   the  underlying   FUV   continuum.    In  a   simple
photon-counting exercise using  the same FUV data for  a subset of our
sample, \citet{odea10}  demonstrated that the  young stellar component
traced by the FUV continuum can roughly account for the photon budget required to photoionise the
Ly$\alpha$   nebula,  although   there   is  unavoidably   significant
uncertainty in the extinction  correction used in this analysis.

That the Ly$\alpha$ is far more extended than the FUV continuum may be
due to a simple sensitivity issue. Ly$\alpha$ is far brighter than the
local FUV  continuum, as  the  average Ly$\alpha$/FUV flux  density ratio
for our  sample is roughly $\sim  3$. There are several examples in our sample
where outer Ly$\alpha$ filaments are detected 
in a region where this  Ly$\alpha$/FUV ratio would result in an FUV continuum 
flux that is below the sensitivity limit of the observation 
(the  outer  Ly$\alpha$  filaments in  Abell 11  are  one
example). 
We are therefore unable to rule out the possibility that {\it all} Ly$\alpha$ 
emission is cospatial with underlying FUV continuum from young stars. 
Alternatively, it is still possible that the  Ly$\alpha$  is intrinsically 
more  extended than  the  FUV  continuum. This would be similarly unsurprising and 
consistent with many previous  observations, as this
is  attributable  to  Ly$\alpha$'s  very  high  sensitivity  to
resonant  scattering  (e.g.,   \citealt{laursen07}).   This  can  make
Ly$\alpha$  morphology difficult  to  interpret in  a physical  sense,
though it does serve as an excellent (although ``tip-of-the-iceberg'')
tracer for  neutral hydrogen.  We  leverage these Ly$\alpha$  data for
this purpose in many  of the multiwavelength comparison figures listed
in column (7) of Table \ref{tab:sample}.

In Fig.~\ref{fig:halphacompare} we compare  a subset of our FUV sample with
the        narrowband       H$\alpha$       maps       from
\citet{mcdonald09,mcdonald10,mcdonald11b,mcdonald11a}               and
\citet{conselice01}.  We have smoothed the FUV maps with a gaussian in order
to degrade  their spatial resolution to (roughly)  match that of the
H$\alpha$ images.   Although the  FUV and  H$\alpha$ morphologies closely
match  one another  (see  also e.g.,  \citealt{mcdonald11b}), the match is not
nearly one-to-one.  It has been  known for many years that some  of the
H$\alpha$  filaments in CC BCGs are devoid of a detectable FUV counterpart,
with Perseus\footnote{One must  be wary  of   confusing star forming filaments
in Perseus with FUV and blue excess   emission  from the  foreground High
Velocity System  (HVS)  that is   superimposed along the line of sight. This
disrupted galaxy is $\sim   100$ kpc  closer in projection and  is unrelated
to  the BCG; see e.g.,   \citealt{sanders07}.} being the most obvious example
(e.g.,  \citealt{hatch06,canning10}). Moreover, some blue    star  forming
filaments apparently  lack cospatial  H$\alpha$ emission (e.g., the ``blue
loop'' in Perseus, \citealt{fabian08,canning10,canning14}).   H$\alpha$ traces
the contemporary star formation rate via the instantaneous flux of ionizing
photons from the most massive ($M_\star \gae 15 M_\odot$) O and early B-type
stars, while the more heavily extincted UV excess associated  with the
photospheres of less massive ($M_\star \gae 5 M_\odot$) young stars can shine
long after those most massive stars powering the H$\alpha$ flux  are gone.
H$\alpha$ and the FUV therefore sample smaller and larger temporal slices
($\sim10^{6-7}$ yr vs.~$\sim10^8$ yr) of the star formation history,
respectively.
More importantly, many authors have demonstrated that the
H$\alpha$ nebulae cannot be heated by star formation alone (see \S1,
although this issue is  not the  focus  of our paper).
The slight morphological mismatch between H$\alpha$ and FUV is therefore 
not necessarily surprising. Fig.~\ref{fig:halphacompare} only demonstrates 
that the FUV and H$\alpha$ filaments are roughly cospatial in projection, with 
clear important exceptions. While some authors have shown that 
stars can indeed play a very important role in the ionisation states 
of the filaments (e.g., \citealt{odea10,mcdonald11a}), 
we reiterate that another 
heat source (acting either alone or in concert with the young stars) is needed (e.g. \citealt{voit97,ferland09,oonk11,fabian11b,mittal12,johnstone12,canning15})

\begin{table}
  \centering
  \caption{\small: The Spearman-rank and Pearson correlation
  coefficients between the FUV postion angle ($X$) and the X-ray,
  optical, infrared and radio position angles ($Y$). These quantities are plotted 
  against one another in Fig.~\ref{fig:paplots}.}
  \label{cc} 
  \begin{tabular}{ c | c | c | c }
     \hline
     Quantity ($Y$)  &  Spearman-Rank  & Pearson &  Best Fitting Correlation  \\
     \hline \hline                                   
     X-ray         &  0.81           & 0.84    &  $Y$ $\propto X^{1\pm0.2}$   \\ 
     Optical       &  0.73       & 0.74    &  $Y$ $\propto X^{0.9\pm0.1}$ \\
     Infarred      &  0.72       & 0.76    &  $Y$ $\propto X^{1\pm0.2}$   \\
     Radio     &  0          & 0.10    &  ...               \\
     \hline
  \end{tabular}
\label{tab:spearman}
\end{table}

\subsection{Galaxy-scale position angle alignment of FUV with 
X-ray, optical, and IR major axes}

\begin{figure}
\begin{center}
\includegraphics[scale=0.55]{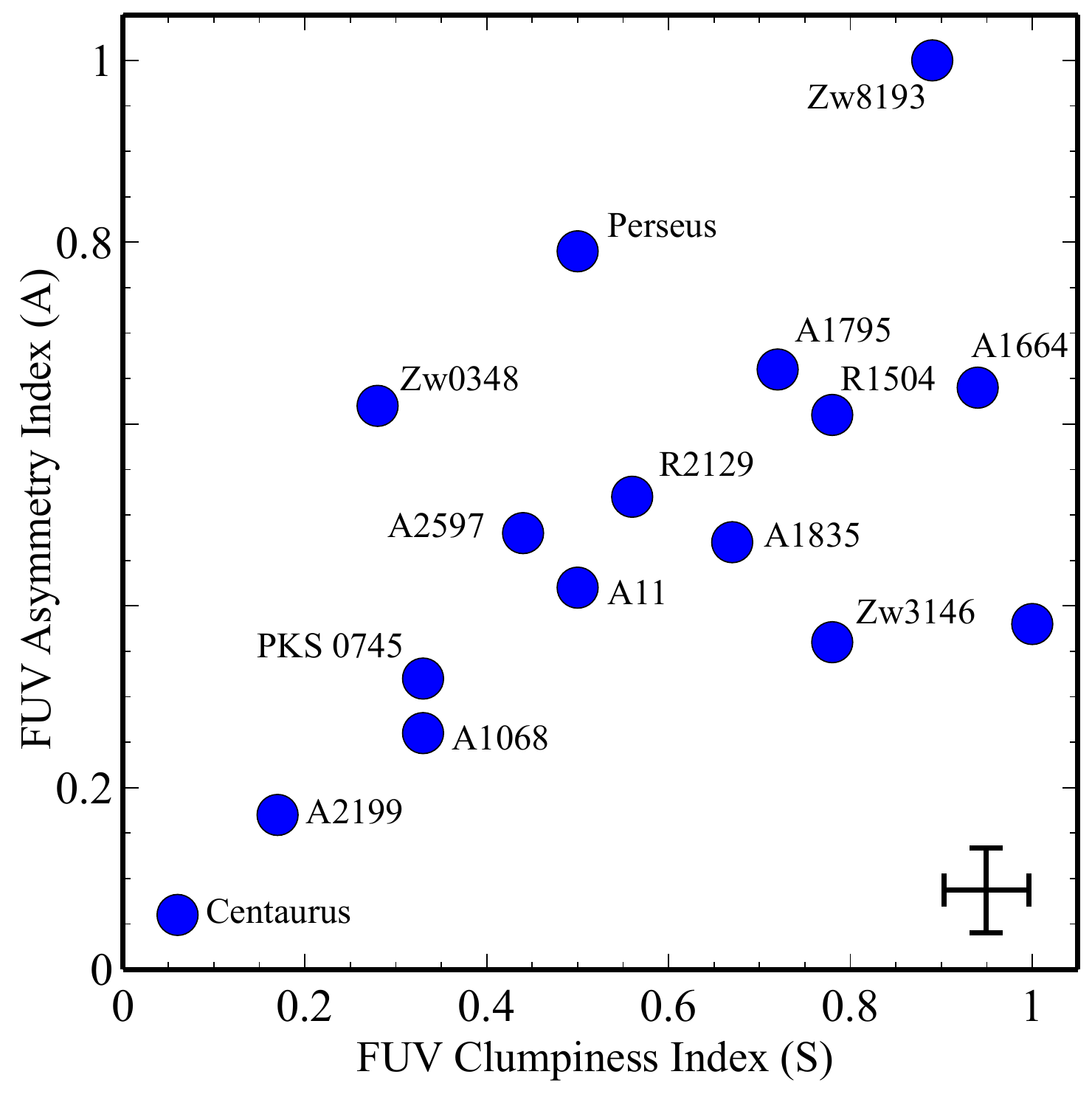}
\end{center}
\caption{The FUV asymmetry index $A$, as defined by Eq.~\ref{equation:asymmetry}, vs. 
the FUV ``clumpiness index'' $S$, defined in Eq.~\ref{equation:clumpiness}. 
Higher values of $A$ means that the FUV surface brightness distribution is more 
azimuthally asymmetric about the galaxy centre. Higher values of $S$ means that the 
FUV emission features more high spatial frequency clumps, and less smooth emission. 
The two indices strongly correlate, such that a galaxy with a high $A+S$ value can be 
considered more spatially anisotropic (in terms of FUV emission) than a galaxy with a low
$A+S$ value. 
These indices
are discussed in \S\ref{section:AS}. }
\label{fig:a_vs_s}
\end{figure}

In Fig.~\ref{fig:paplots} we plot the position angle (P.A., measured N through
E) of the projected X-ray, optical, IR (3.6 $\mu$m), and radio major  axis
vs.~the  projected FUV  major axis. The FUV major axis was taken to be the
position angle of the isophote at roughly twice the FUV half-light radius in lightly smoothed maps.
We then measured the X-ray,  optical, and  IR major
axis within the isophote at roughly the same radius. 
Sources that  are point like  or circular in
any of these  bands have been excluded from that particular plot.

As is  evident in Fig.~\ref{fig:paplots}, we  observe weak-to-strong projected
alignment  between FUV,  X-ray, optical,  and IR  counterparts.  Those
sources that are outliers to any  of these trends have been labeled in
the  respective  plot.   It  is  possible  that  the  alignment  is  a
reflection of  the old and  young stellar components sharing  a common
origin  in the  ambient hot  gas.   Another possibility  
is that the alignment  is merely due  to the fact
that the  various components all reside within  the same gravitational
potential, and  have had sufficient time to  dynamically relax, torque
toward a common axis, etc. We caution against over-interpretation of these 
apparent projected alignments: these are chaotic, messy systems with morphologies 
that probably vary strongly with time.

\subsection{Kiloparsec-scale offsets between FUV and X-ray surface brightness peaks}

While FUV and X-ray  surface brightness peaks are spatially coincident
for the  majority of  our sample, A1664,  A1835, Centaurus,  PKS 0745,
R2129, and Zw3146 show projected offsets  of 9 kpc, 14.8 kpc, 1.3 kpc,
4.65  kpc, 7  kpc,  and 11  kpc,  respectively.  The  mean and  median
offsets  are  are both  roughly  8  kpc.   Offsets between  the  X-ray
emission  and the  optical/IR BCG  peak are  effectively the  same for
these  targets  (as  the  FUV  and optical  peaks  are  almost  always
cospatial, at least within our  sample).  The offset in A1664 has been
previously noted  by \citet{kirkpatrick09} in their  detailed study of
the  source.  The A1835,  R2129,  and  Zw3146  offsets were  noted  by
\citet{odea10}.

The  X-ray surface  brightness  maps for  all  objects with  X-ray/FUV
photocentroid  offsets show  large scale  ($\gae50$  kpc) asymmetries,
suggestive of  complex gas dynamics  in the hot  phase.  Photocentroid offsets
between the BCG and the  cluster X-ray emission is a proxy for how  close  (or
how far)  the  system  is  to  a state  of  dynamical equilibrium,  such that
the offsets  should decrease  as  the cluster evolves  \citep{katayama03}.
Sloshing  motions in  the X-ray  gas can nevertheless remain long-lived even
after the supposed virialisation of the cluster (e.g.
\citealt{markevitch07}).  Large optically selected samples  of both  cool core
and  non-cool core  BCGs frequently  show median      X-ray/BCG      offsets
of     $\sim      15$      kpc \citep{bildfell08,sanderson09,loubser09}.  CC
BCGs  systematically lie below this median  at $\sim 10$ kpc, which is close
to the median for those objects  (with observed X-ray offsets) within  our
sample ($\sim 8$ kpc).  If we include those  galaxies in our sample that do
not show any  measurable projected  offset (ten  out of  sixteen  sources),
the sample-wide  median and  mean projected  offsets are  $\sim0$  kpc and
$\sim3 $ kpc, respectively. The sources exhibiting kpc-scale offsets do not
appear to prefer any particular galaxy property -- instead they inhabit the
full range of FUV morphology, star formation rate, radio power, etc.  that is
spanned by our whole sample.

\subsection{Quantifying morphology by Asymmetry and Clumpiness indices}

\label{section:AS}

Fig.~\ref{fig:postagestamps} shows that our sample spans a 
 diverse range of FUV morphologies, 
including  sources
that can be described as
amorphous  (e.g.,  Zw3146),  clumpy  (e.g.,  A11),  point-like  (e.g.,
Centaurus), disk-like  (Hydra A), and filamentary
(e.g., A2597,  R1504). There is however significant  overlap between
these   classes. For example,   A2597  could be arbitrarily described  
as a hybrid of filamentary,  amorphous, and clumpy structures, 
illustrating the need for a more objective measure of morphology.

We therefore   quantify all  projected FUV morphology by means of scale-
invariant  structural indices, as is done frequently for galaxies  in the
literature.  We adopt the commonly used  Concentration-Asymmetry-Smoothness
($CAS$)   system   described   by \citet{conselice03},  which  posits  that
galaxy  morphology  can  be entirely  quantified by  measuring  the
concentration  of light  ($C$) around a  photocentric point, the azimuthal
asymmetry  of light about this  point  ($A$),  and  the  high spatial
frequency  smoothness  or clumpiness ($S$)  of that light.   The $CAS$ indices
are  useful in that they  (a) are  independent of any  assumption about
galaxy light distribution and (b) correlate  with galaxy processes such as
star formation, mergers, colors, emission line widths, etc. Galaxies of
different Hubble type appropriately stratify within the optical $CAS$ volume,
which  has  been expanded to include other wavelength regimes over the years
(including for extragalactic FUV imaging, see e.g., \citealt{holwerda12}).

In   our  case,   there  is   a    risk   that  any   use  of   a
concentration-of-light parameter $C$  (typically defined by the ratio of
curve-of-growth radii containing 80\% and 20\%  of all light) may be
significantly contaminated by    the    SBC    red     leak,    for    reasons
discussed    in \S\ref{section:redleak}. We therefore make  use of only the
asymmetry and clumpiness parameters $A$ and $S$, which (even without $C$) are
useful in quantifying spatial  anisotropy in FUV  surface brightness.
Following \citet{conselice03}, we  compute the  asymmetry index $A$  by
rotating each FUV image by 180$^\circ$ about  a central point (discussed
below), subtracting this from the original  unrotated  image,  and   then
summing  the  absolute  value intensities from the resulting  residual map.
The resultant value is then normalized by two times the original galaxy flux.
Expressing the above more quantitatively, the asymmetry index is given by
\begin{equation}   \label{equation:asymmetry} A = k\times \frac{\Sigma
\left|~I_{\theta = 0} - I_{\theta = 180}~\right|}{2\times\Sigma\left|
~I_{\theta=0}~ \right|}, \end{equation} where $I_{\theta = 0}$  and
$I_{\theta = 180}$ are  the intensity distributions in the original and
180$^\circ$ rotated images, respectively.  On both the rotated and original
images, the sum is taken over all pixels within matching regions that
encompass all galaxy FUV flux (e.g., an ``all the flux you see'' circular
aperture).  The resulting value of $A$ depends strongly on the pixel about
which   the image is rotated, as is discussed by \citet{conselice03}.   For
sample-wide consistency, we have  chosen to rotate each FUV image around the
pixel that is cospatial with both the radio core and optical photocentroid of
the host galaxy,  such that our computed  $A$ values are at least somewhat
related to the  projected reflection asymmetry of young stars around the AGN.
In a few cases the radio core was not cospatial with the host galaxy optical
photocentroid, but this was typically because of a central dust lane (Hydra A
is one example).  Comparison of both $A$ and $S$ values with those in other
papers is beyond the scope of this work, so in the interests of simplicity we
use  $k$ as a scalar normalization to  set the range spanned by the  $A$
distribution to $0 \le A \le 1$. Sources with lower values  of $A$ are more
azimuthally symmetric about the centre of the galaxy than are sources  with
higher values of $A$.

The ``clumpiness'' parameter $S$ was calculated by summing pixel intensities in 
an unsharp mask of the FUV image, made by 
gaussian smoothing the map with both small and  large kernel sizes, subtracting 
the heavily smoothed map from the lightly smoothed one, and then normalizing the residuals 
by the flux in the original image. 
More specifically, $S$ is given by  
\begin{equation}
\label{equation:clumpiness}
S= j \times \Sigma\frac{\left( I - I_\sigma \right) - B}{I}, 
\end{equation}
where  $I$  and $I_\sigma$  are  the  intensity  distributions in  the
lightly and heavily smoothed images, respectively. All of these sums are
taken within  a matching area, and  all galaxy apertures are made with  
a central hole
that  intentionally  excludes  the  galaxy nucleus  (where an FUV point source  
might artificially weight the $S$  value).
$B$ is the background intensity distribution 
within an off-source ``sky'' aperture. 
 We again normalize by $j$
to set the range of possible $S$  values equal to that of $A$, i.e. $0
\le  S \le  1$.  A  galaxy with  with $S\approx1$  will feature  many high
spatial frequency  clumps, whereas an object with  $S\approx0$ is very
smooth.

We plot $A$ vs.~$S$ in Fig.~\ref{fig:a_vs_s}. 
The two indices strongly correlate, such that 
sources that are more asymmetric also tend to be more clumpy. 
The trend is strong enough that for the remainder of this paper 
we will discuss $A$ and $S$ together by means of 
a single ``FUV anisotropy index'' $A+S$, which 
can inhabit the range from $ 0 \le A+S \le 2$. 
Galaxies with higher $A+S$ are more clumpy, filamentary, and asymmetric, 
whereas galaxies with lower $A+S$ values are more symmetrically amorphous or point-like.

\begin{figure*}
\begin{center}
\includegraphics[scale=0.48]{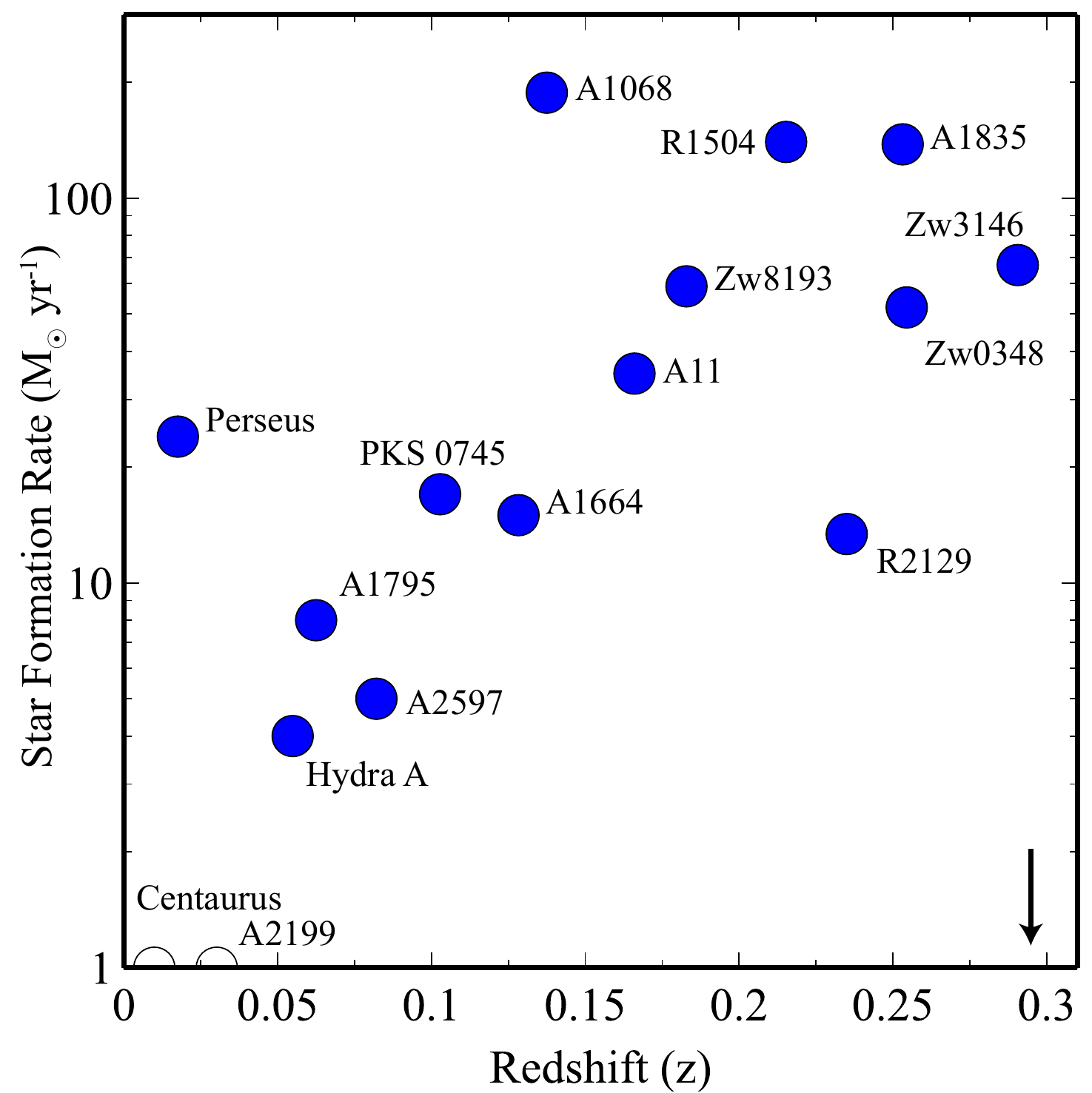}~~~~
\includegraphics[scale=0.48]{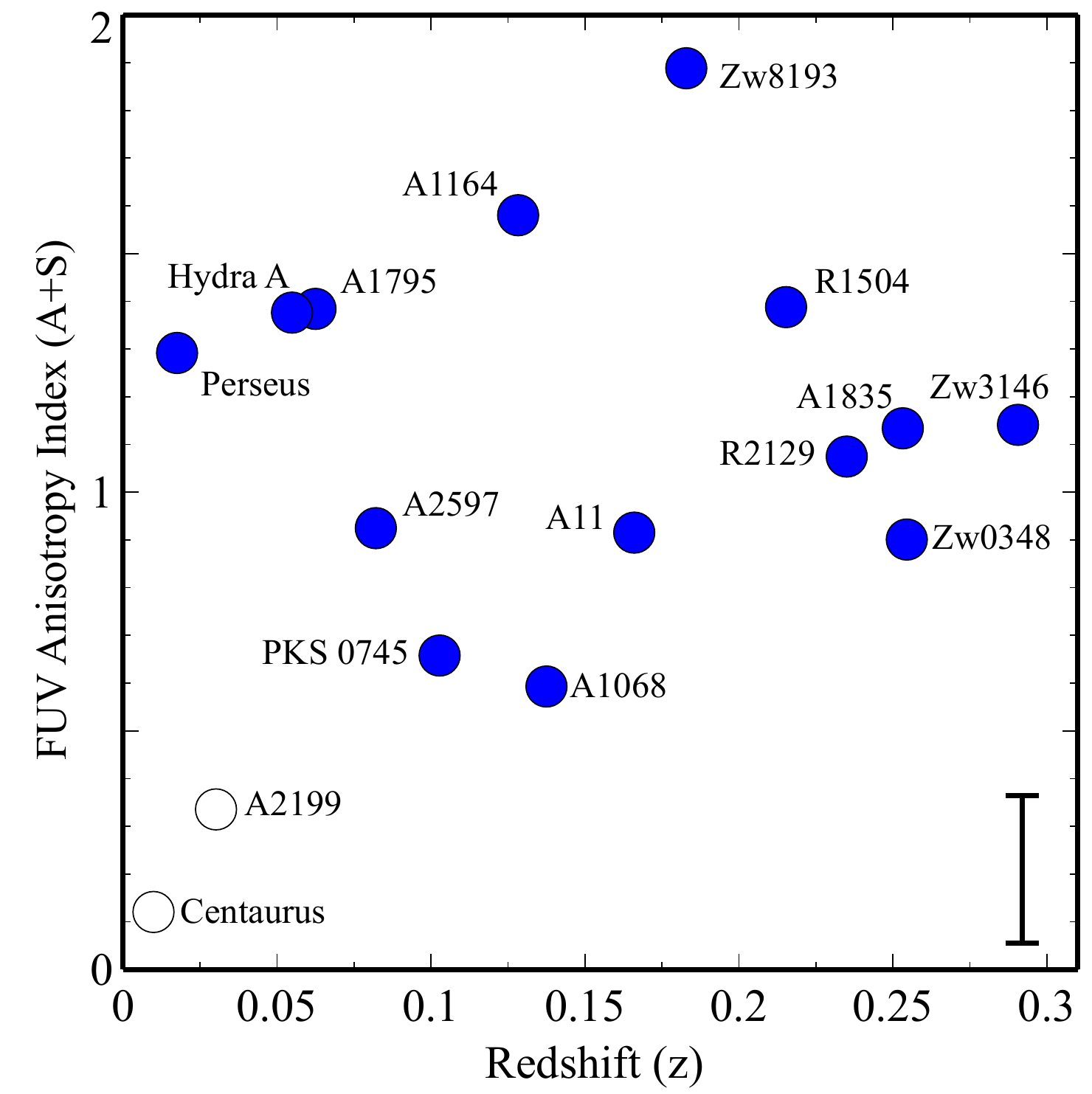}\\
\includegraphics[scale=0.48]{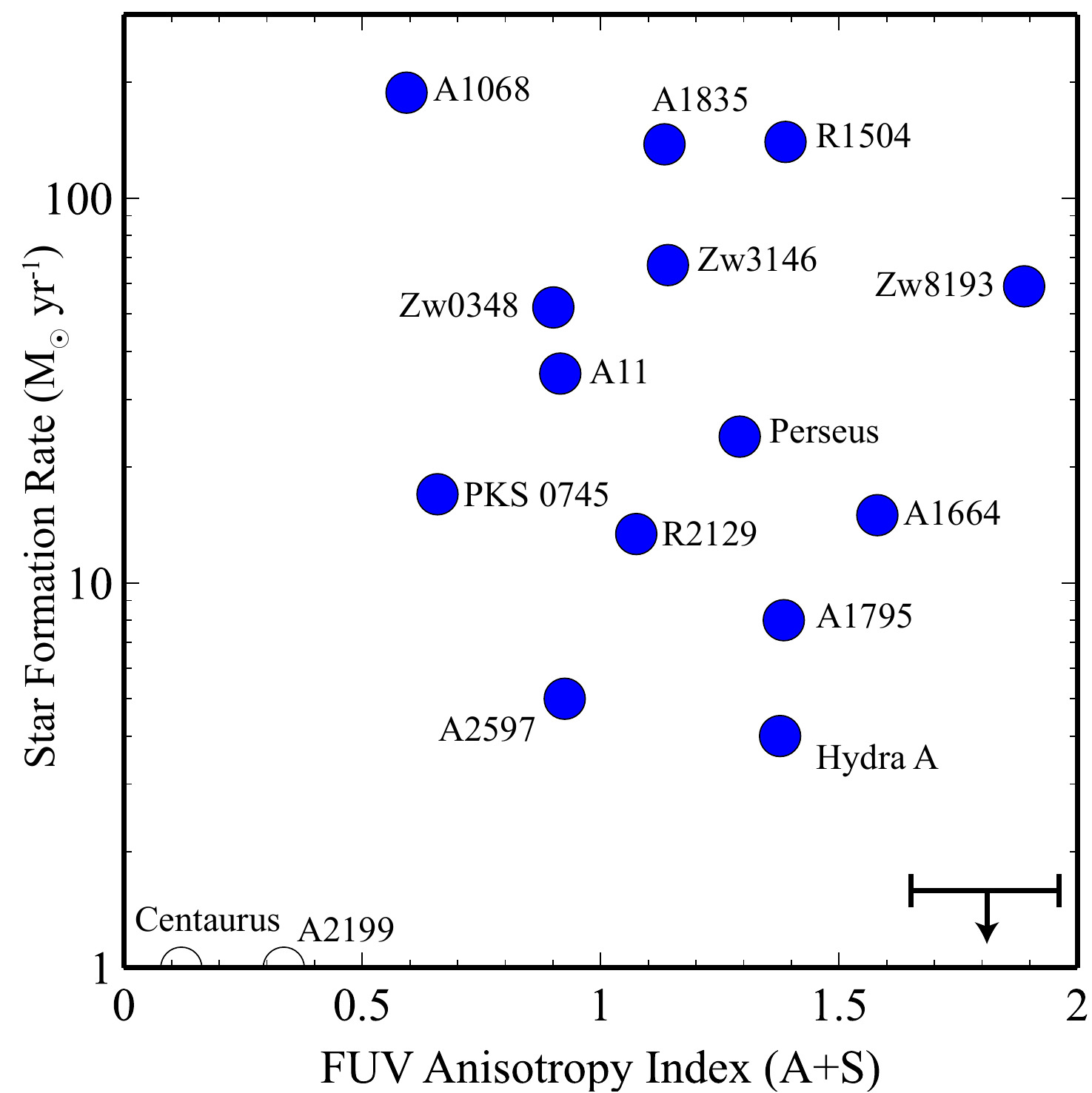}~~~~
\includegraphics[scale=0.48]{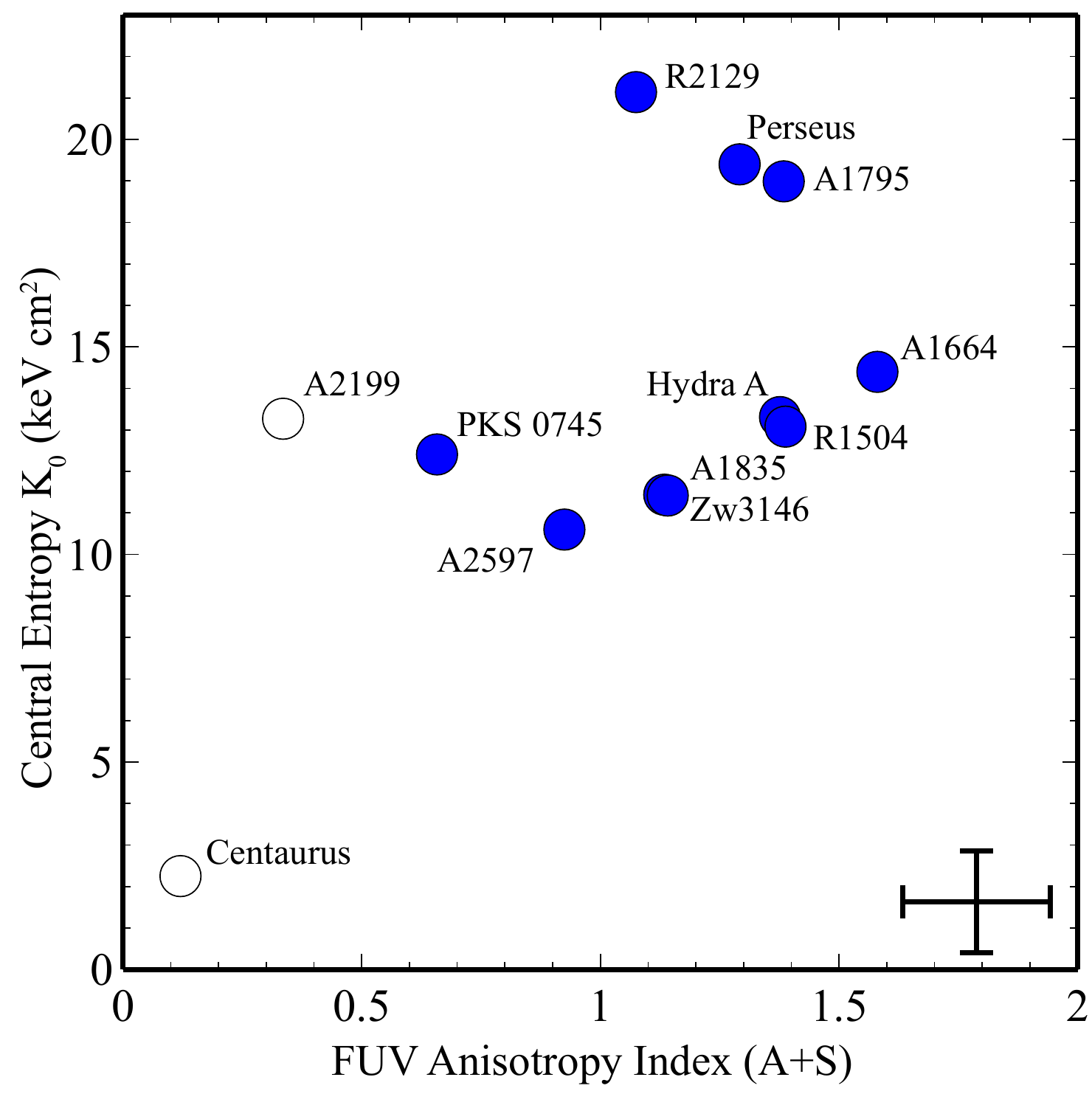}
\end{center}
\caption{({\it top left}) Infrared-estimated star formation rate vs.~redshift for all 
targets in our sample. SFRs estimated by other indicators in other 
wavelength regimes tend to be lower than the IR-based rate, which can 
be considered a rough upper-limit. ({\it top right}) ``FUV anisotropy index'' $\left(A+S\right)$ based on the $CAS$ parameters described by \citet{conselice03}. We define the anisotropy index in \S\ref{section:AS}. A2199 and Centaurus have been ``grayed out'' so as not to give the illusion 
of correlation where there is likely none. These two sources may be highly contaminated by red leak (so their $CAS$ morphology cannot be trusted), and their 
star formation rates are extremely low (perhaps effectively zero).  }
\label{fig:morphologyplots}
\end{figure*}

\subsection{Comparison of FUV morphology with redshift, star formation rate, and 
ICM central entropy}

FUV morphology does not exhibit any obvious redshift dependence, despite the
very strong redshift-luminosity  bias present in our sample (which is assembled
from  flux-limited and therefore Malmquist-biased catalogs).  We demonstrate
this bias in the top-left panel of Fig.~\ref{fig:morphologyplots}, where we
plot the IR-estimated star formation rate vs.~redshift.  Despite
the expected strong upward trend in SFR with redshift associated with the
Malmquist bias, the top right panel of Fig.~2 --- showing FUV anisotropy
index ($A+S$) vs.~redshift --- is effectively a scatter plot (note that, for each of these plots, 
we ``gray out'' red-leak contaminated A2199 and Centaurus, so they do not give 
the illusion of correlation).
Galaxies at higher redshift marginally tend to have a higher FUV anisotropy
value, albeit with very large scatter. The error bar on this plot reflects the
rather large range that $A+S$ can inhabit given slightly different choices of
pixel about which the image is rotated (in the case of $A$) or smoothing
lengths used to make the unsharp mask (in the case of $S$).

We conclude 
that there is no evidence for any correlation between redshift and morphology in our sample.
This is perhaps somewhat surprising, because one naturally expects a trend between redshift and morphology
as $\left(1+z\right)^4$ surface
brightness dimming and angular size scaling should make objects look
increasingly smooth and symmetric as they approach the resolution limit at
higher redshifts.  
To independently test what effect redshift may have on perceived morphology in
our sample, we have  artificially redshifted all of our FUV images to one
common redshift equal to that for our most distant target (Zw3146,
$z=0.2906$). An IRAF script was used to accomplish this, implementing the
technique described by \citet{giavalisco96} (specifically, see their equations
2-7).  While artificially redshifting our targets had only a small effect on
overall asymmetry $A$, it is clear from this test that redshift can have a
strong effect on perceived smoothness $S$.  Our lowest redshift targets around
$z=0.01$, for example, suffer a factor of $\sim 20$ degradation in spatial
resolution, lowering their $A$ value negligibly and their $S$ value moderately
(depending on choice of smoothing scale lengths). 
It is therefore possible that our high redshift targets are intrinsically far
more clumpy than they appear. We therefore note that, while the top right panel of Fig.~\ref{fig:morphologyplots}
provides no evidence for a correlation between $A+S$ and redshift, it cannot be used to rule it out.

We plot IR-estimated star formation rate vs.~FUV anisotropy index in the lower left panel of Fig.~\ref{fig:morphologyplots}, 
finding no correlation. 
We do  observe a weak upward trend of central ICM entropy with FUV
anisotropy index, as is evident from the lower right panel of
Fig.~\ref{fig:morphologyplots}. Here, entropy $S$ (in units of keV cm$^2$) is
defined as $S=kTn_e^{-2/3}$, where $k$ is the Boltzmann constant, $T$ is the
gas temperature, and $n_e$ the electron density. Central entropy values have
been adopted from the ACCEPT  sample \citep{donahue06,cavagnolo09}.  
We again caution against over-interpretation here, particularly because 
calculations of ICM central entropy from the X-ray data can be problematic and strongly tied to data quality  \citep{panagoulia14}.
The plot merely demonstrates that sources with higher central entropy {\it may} have more spatially anisotropic star formation, 
at least within our sample.

\begin{figure*}
\begin{center}
\includegraphics[scale=0.5]{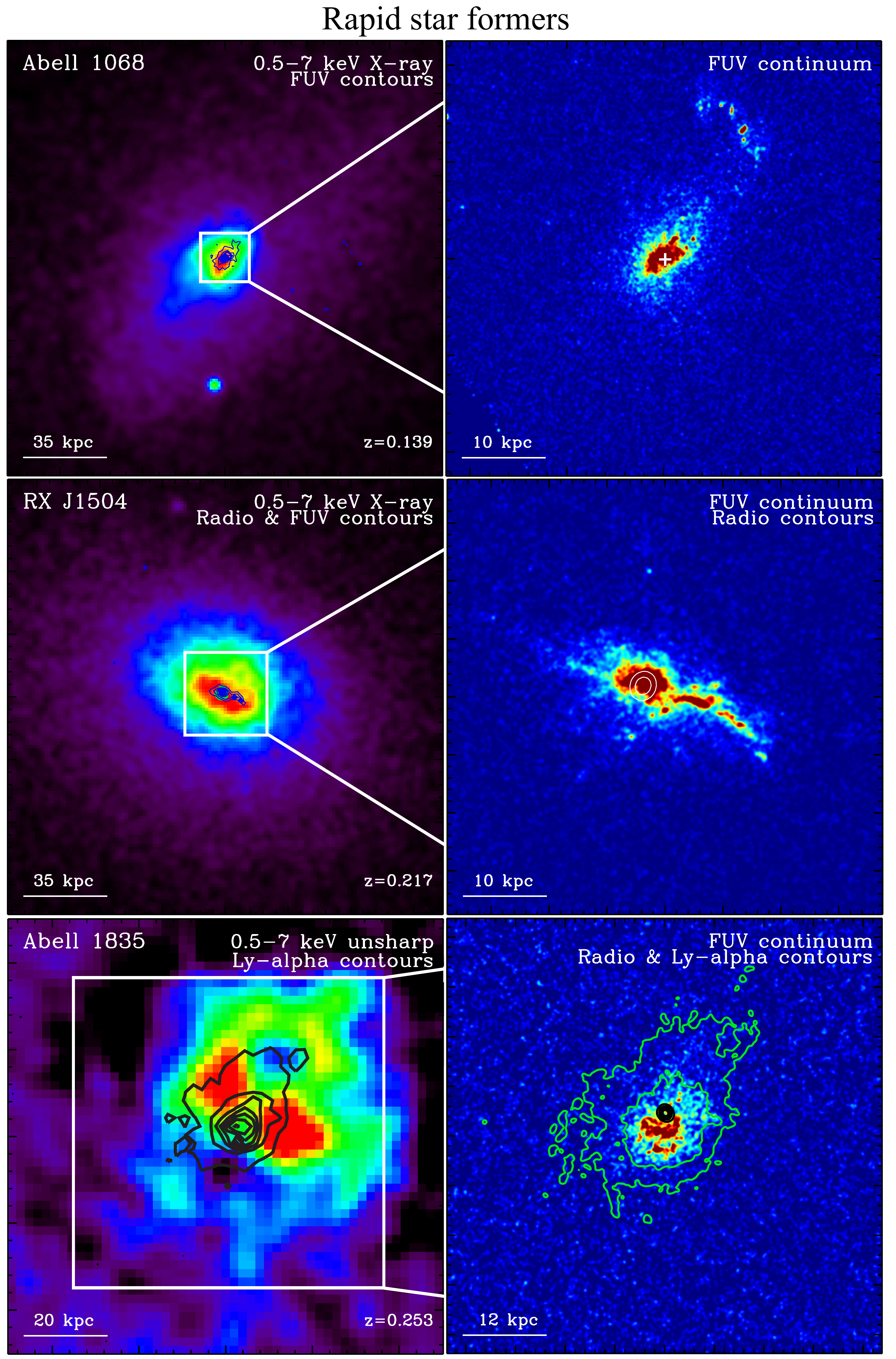}
\end{center}
\vspace*{-2mm}
\caption{The three sources in our sample with the highest star formation rates ($\gae 100$ \Msol\ yr\mone). The left-hand panels show a wide
view of X-ray emission cospatial with the BCG and its outskirts. White boxes are used to indicate the FOV of the right-hand panels, which show FUV continuum emission. Various contour sets are overlaid, and are labeled appropriately in their respective panels.  }
\label{fig:highsfr}
\end{figure*}

\begin{figure*}
\begin{center}
\includegraphics[scale=0.5]{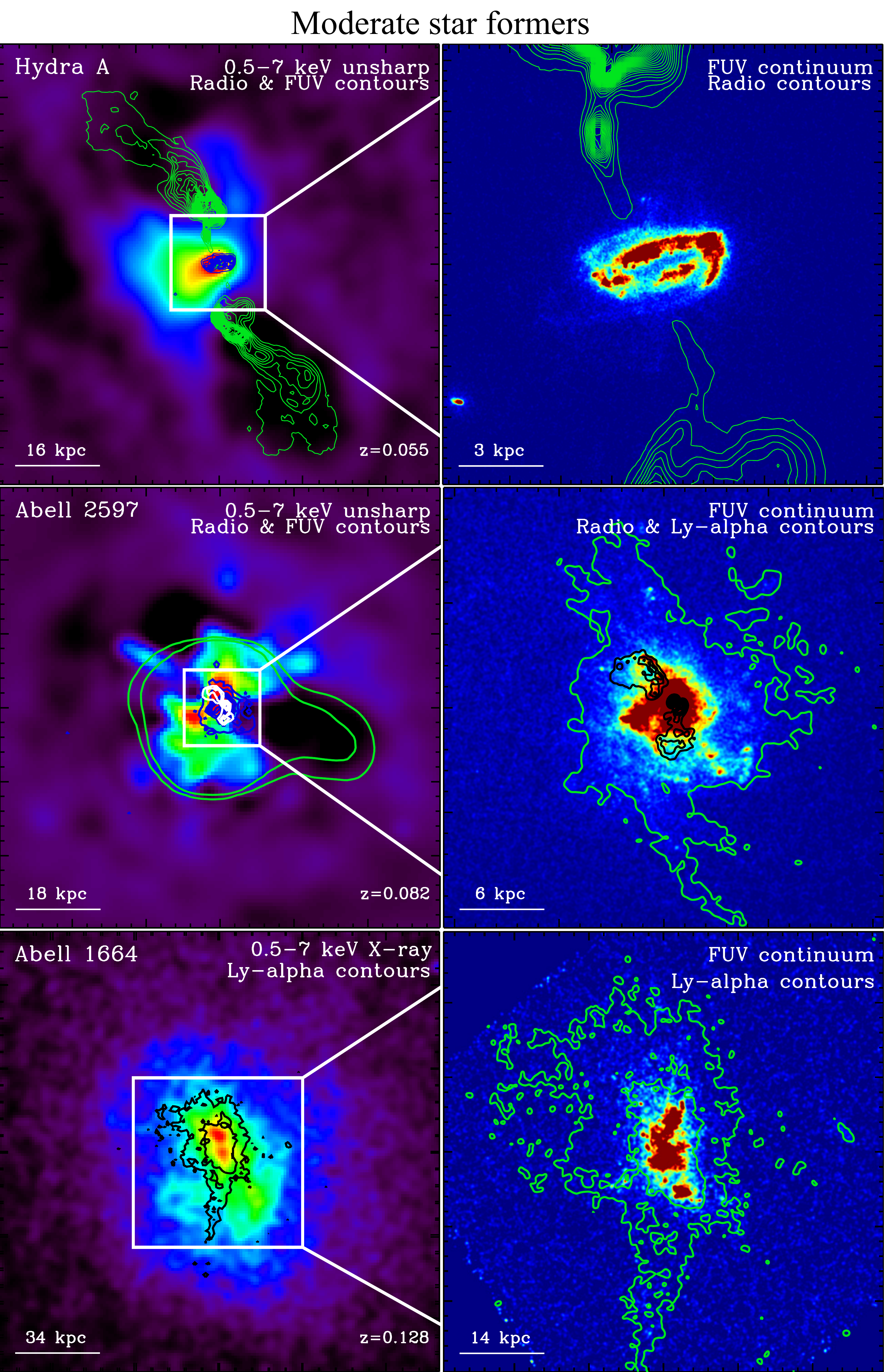}
\end{center}
\vspace*{-2mm}
\caption{A selection of sources in our sample that show moderate star formation rates $\sim 5 < \mathrm{SFR} < 15$ \Msol\ yr\mone. The left-hand panels show a wide
view of X-ray emission cospatial with the BCG and its outskirts. White boxes are used to indicate the FOV of the right-hand panels, which show FUV continuum emission. Various contour sets are overlaid, and are labeled appropriately in their respective panels. }
\label{fig:midsfr}
\end{figure*}

\begin{figure*}
\begin{center}
\includegraphics[scale=0.5]{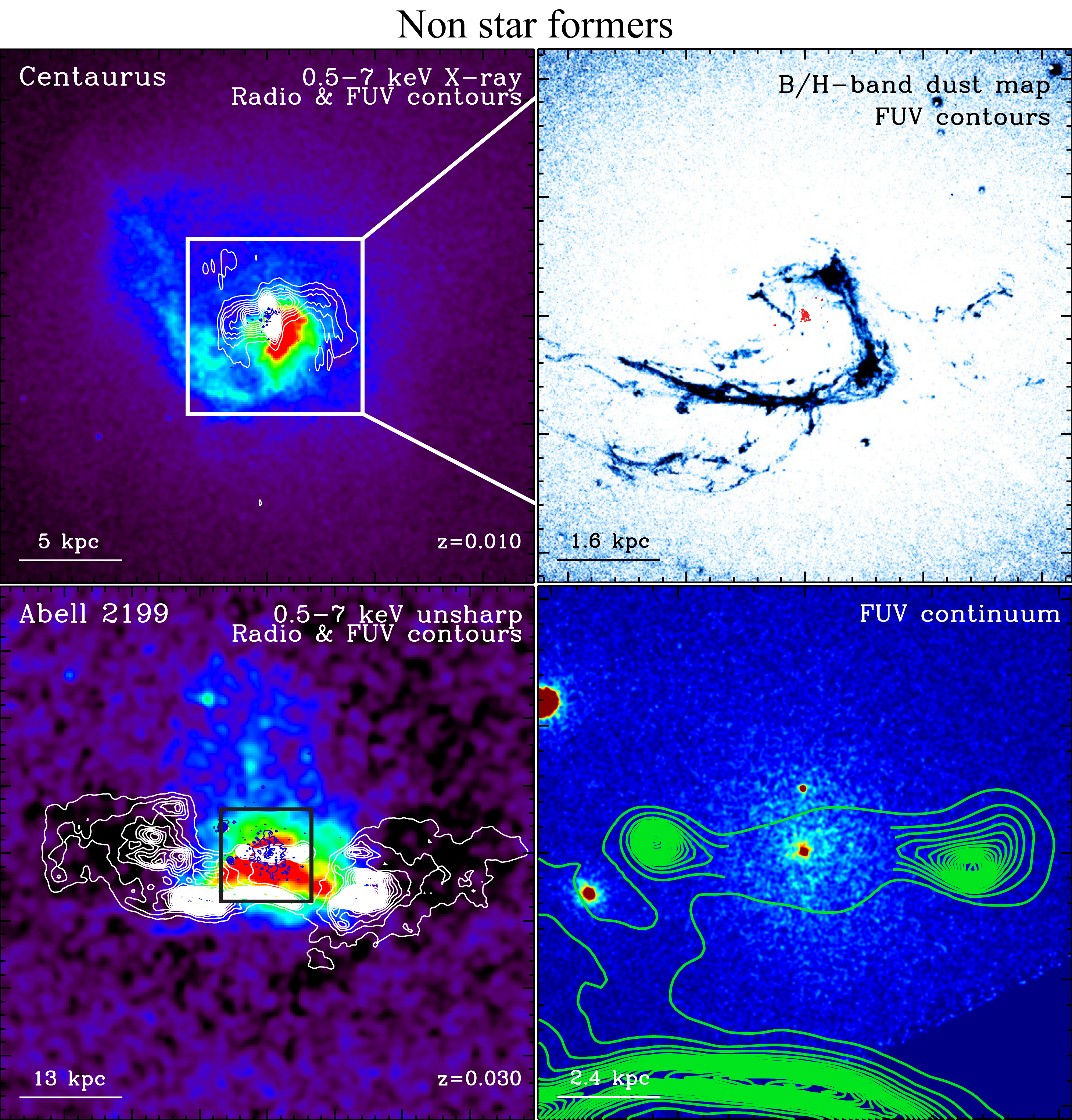}
\end{center}
\vspace*{-2mm}
\caption{The two sources in our sample in which there is effectively 
no ongoing star formation or FUV-bright young stars (or $\ll0.1$ \Msol yr\mone).The left-hand panels show a wide
view of X-ray emission cospatial with the BCG and its outskirts. White boxes are used to indicate the FOV of the right-hand panels. For Centaurus, we show a $B$/$H$-band color map made with {\it HST} data, showing the famous dust lane associated with the source. The (almost nonexistent) FUV emission in the radio core is shown in red contours.  For A2199, the rightmost panel shows FUV continuum emission.  The diffuse emission seen in A2199 may be an artifact of the ACS SBC red leak.   }
\label{fig:lowsfr}
\end{figure*}

\begin{figure*}
\begin{center}
\includegraphics[scale=0.33]{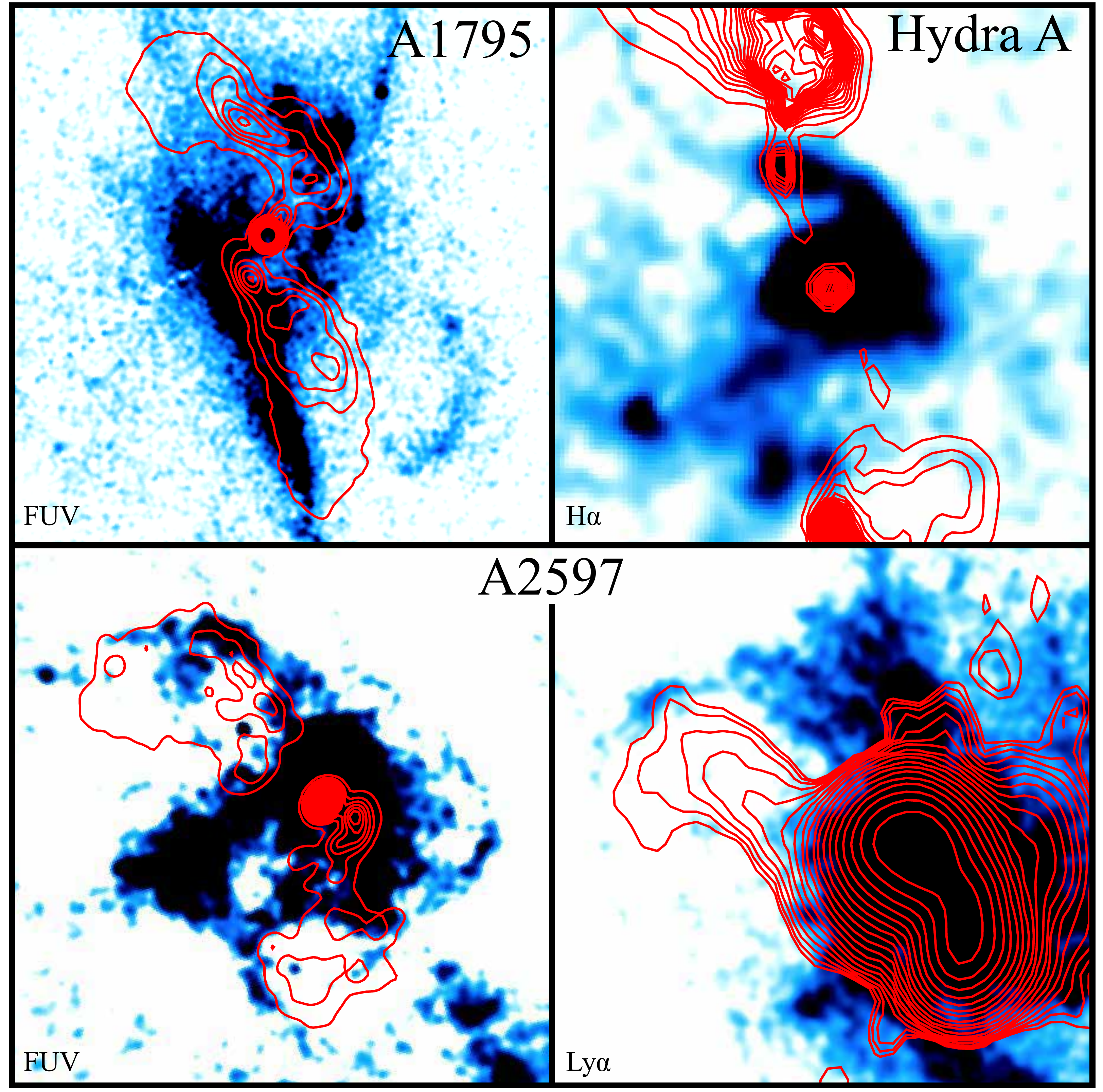}
\end{center}
\vspace*{-2mm}
\caption{The strongest examples of FUV/radio morphological correlation and anticorrelation in our sample. This is perhaps evidence for  
for (a) star forming filaments that have been uplifted, entrained, or swept aside by the propagating radio source or (b) jet-triggered star formation. Note also that PKS 0745 (Fig.~\ref{fig:pks0745_figure}) shows weaker evidence for a similar alignment.  }
\label{fig:jet_triggered_sfr_figure}
\end{figure*}

\begin{figure*}
\begin{center}
\includegraphics[scale=0.355]{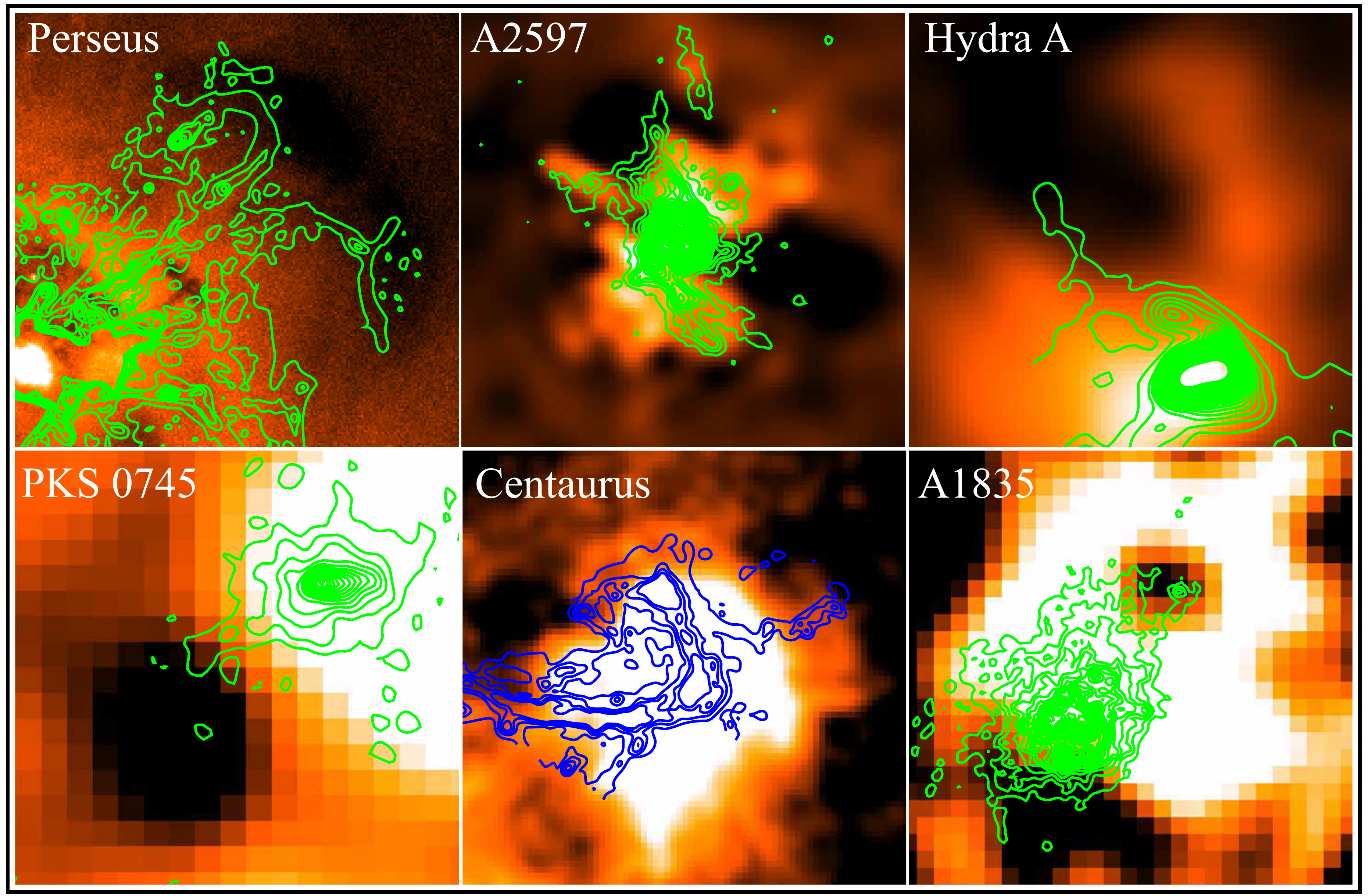}
\end{center}
\vspace*{-2mm}
\caption{Six sources for which there is some evidence of kpc-scale 
filaments extending in projection towards and around X-ray cavities. FUV contours are overlaid in green on X-ray unsharp masks. Centaurus has no discernable star formation, but its dust contours do possess filaments that extend toward cavities. We show these dust contours in blue. For Hydra A we show H$\alpha$ contours as these better show 
the faint filament that follows the northern cavity. While not shown, FUV emission at low 
surface brightness in Hydra A is fully cospatial with the H$\alpha$ filaments shown above.   }
\label{fig:filamentcavity}
\end{figure*}

\subsection{Star forming filaments aligned with radio jets and lobes}

\label{section:jet_triggered_sf}

Fig.~\ref{fig:highsfr} through Fig.~\ref{fig:lowsfr} shows selected X-ray, FUV, Ly$\alpha$, and radio 
overlay figures for a subset of our sample. 
The three figures are presented in order of highest to lowest star formation rate, respectively. 
In these panels one will find several clear spatial correlations between FUV emission, radio 
emission, and/or X-ray emission. We discuss these correlations in the next two sections. 

In Fig.~\ref{fig:jet_triggered_sfr_figure} we highlight four
examples of  strong morphological alignment between FUV continuum /
line emission and radio jets or lobes (shown in red contours).  These  include
A1795, Hydra A, and A2597. While A1795 and
A2597 are  known and well-studied  examples (e.g., \citealt{odea04}),  such
alignment has not previously been noted for Hydra A. Moreover, Fig.~\ref{fig:pks0745_figure}
shows some evidence of alignment in PKS 0745, whose ``spike-like'' FUV filament is aligned 
with the axis about which the radio source appears to kink or fold over.  

There are now too many examples of clumpy/filamentary star formation alinging with radio jets and lobes 
for this cospatiality to credibly be pure coincidence or a projection effect (see e.g., Cen A --- \citealt{crockett12,hamer15,santoro15}; Minkowski's Object
/ NGC 541 --- \citealt{vanbreugel85}; 3C 285 --- \citealt{vanbreugel93}; 4C
41.47; \citealt{bicknell00}; see also 3C 305, 3C 321, 3C 171, and 3C 277.3).
These filaments 
may have been dynamically entrained, uplifted, or swept aside by the radio source, or may have formed {\it in situ} along the working surface 
of the radio lobe in an example of {\it positive} AGN feedback.
It has long been  predicted that star formation may be triggered by shock-
induced  cloud collapse  as the propagating radio plasma  entrains and
displaces cold gas  phases (see e.g., the      shock/jet-induced      star
formation     models      by
\citealt{elmegreen78,voit88,deyoung89,mcnamara93}).   Jet-induced star
formation has for many years been considered as a plausible  explanation for
the high redshift alignment effect \citep{rees89,daly90}.

Whatever the case, examples  like these demonstrate that, 
at least for some time, star formation can survive
(and  may  indeed  be  triggered  by)  dynamical  interaction  with  a
propagating radio source. 
If jet-triggered star formation is indeed a real effect, and is responsible for
the alignment seen for the four targets in our sample, we can roughly estimate
whether or not such an effect has a significant or negligible impact on the
global star formation rate in the galaxy.  If we take A1795, our most dramatic
example, and assume that {\it all} FUV emission associated with the
``P''-shaped filament cospatial with the radio lobes is directly induced by
propagation of the jet, then up to 50\% of all star formation in the galaxy
could be jet-triggered.  The upper limit percentages for the other sources in
our sample are much lower, such that even if jet-triggered star formation is
indeed real, it probably does not play the dominant role in driving star
formation in the galaxy.  We estimate that the effect may play a role at the
few percent level at best. 
Regardless, the apparently competing roles of radio mechanical feedback
simultaneously quenching {\it and} triggering star formation can be reconciled
with one another.  Even if the propagating radio source does not (immediately)
inhibit or truncate star formation directly, it may still work to starve it of gas for {\it future} star formation by excavating cavities and
driving sound waves in the hot gas, preventing it from cooling and forming stars.

\begin{figure*}
\begin{center}
\includegraphics[scale=0.28]{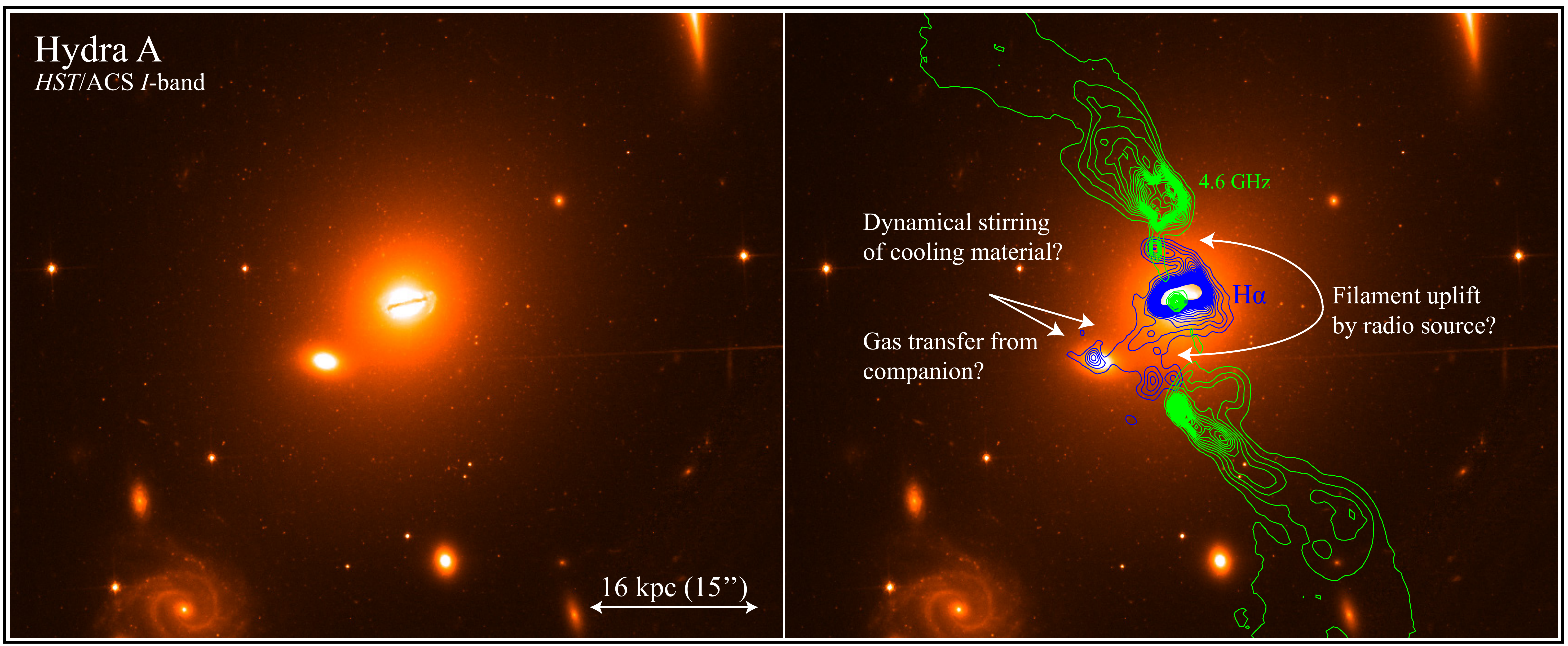}
\end{center}
\vspace*{-3mm}
\caption{The new {\it HST} $I$-band optical image of the Hydra A BCG and its surrounding environment. On the right-hand panel we overlay 
both the VLA 4.6 GHz radio contours and the MMTF narrowband H$\alpha$ contours. The H$\alpha$ distribution shows evidence for dynamical 
interaction with {\it both} the radio source (note the apparently uplifted filaments northward and southward of the nucleus) as well as
the small companion galaxy $\sim 10\arcsec$ to the southeast. While this companion is unlikely to to provide a substantial cold gas mass
to the BCG via merger-driven flow, it may be acting to dynamically ``stir'' the low entropy gas already present in the BCG. The companion 
is cospatial with a bright knot of both H$\alpha$ and FUV continuum emission.}
\label{fig:hydraA_companion}
\end{figure*}

\subsection{Spatial correlations and anti-correlations of star forming filaments with X-ray cavities}

As we demonstrate in Fig.~\ref{fig:filamentcavity}, six sources in our sample
possess one or more kpc-scale narrow filaments that, in projection, extend
toward, into,  or wrap around the edges of kpc-scale X-ray cavities.  These
include Perseus, A2597, Hydra A, PKS 0745, Centaurus, and A1835.  Two
additional sources (A1664 and A1068) show weaker evidence (due perhaps to the
unavailability of deeper X-ray imaging) of the same effect.  Most of those
filaments that extend toward and into cavities are FUV bright and forming stars, 
while Perseus and Centaurus show only dusty, H$\alpha$ bright
filaments that lack cospatial FUV continuum.

This may be evidence for buoyant uplift of the filament by the cavity as it
rises amid the ICM, as has been discussed (typically in the context of Perseus)
by many previous authors (e.g., \citealt{fabian03,hatch06,mcdonald11a,canning10,canning14}). The
spatial associations are certainly compelling, and filament uplift by cavities
may indeed be an important and even common effect. It is, however, unlikely to
be the only effect driving the morphology of all narrow filaments ubiquitously
observed in CC BCGs, as there are many examples of filaments with no obvious
association with either a cavity (or radio source, for that matter).
Ever-deeper observations of X-ray cool cores do however tend to reveal ever
more numerous X-ray cavities, so we cannot necessarily rule out the unlikely
possibility that {\it all} filaments in CC BCGs have at some point been
uplifted by a cavity. We find this unlikely, however, as filament kinematics
(at least for the small sample that has been fully mapped with an IFU) are
generally inconsistent with expectations if they are indeed dragged outwards
(see, for example, the northern filament in Perseus, which shows a smooth
velocity gradient of a few 100 km sec\mone, consistent with a laminar flow;
\citealt{hatch06}).  Alternatively, those filaments that wrap (in projection)
{\it around} X-ray cavities may have formed {\it in situ} in the cavity's
compressed shell, though it is entirely unknown whether or not direct cooling 
from the X-ray to molecular phase is even possible absent dust (e.g., \citealt{fabian94}). 
We expand on this below, in our discussion of FUV morphology in the contex of ICM cooling and AGN heating.

\begin{figure*}
\begin{center}
\includegraphics[scale=0.5]{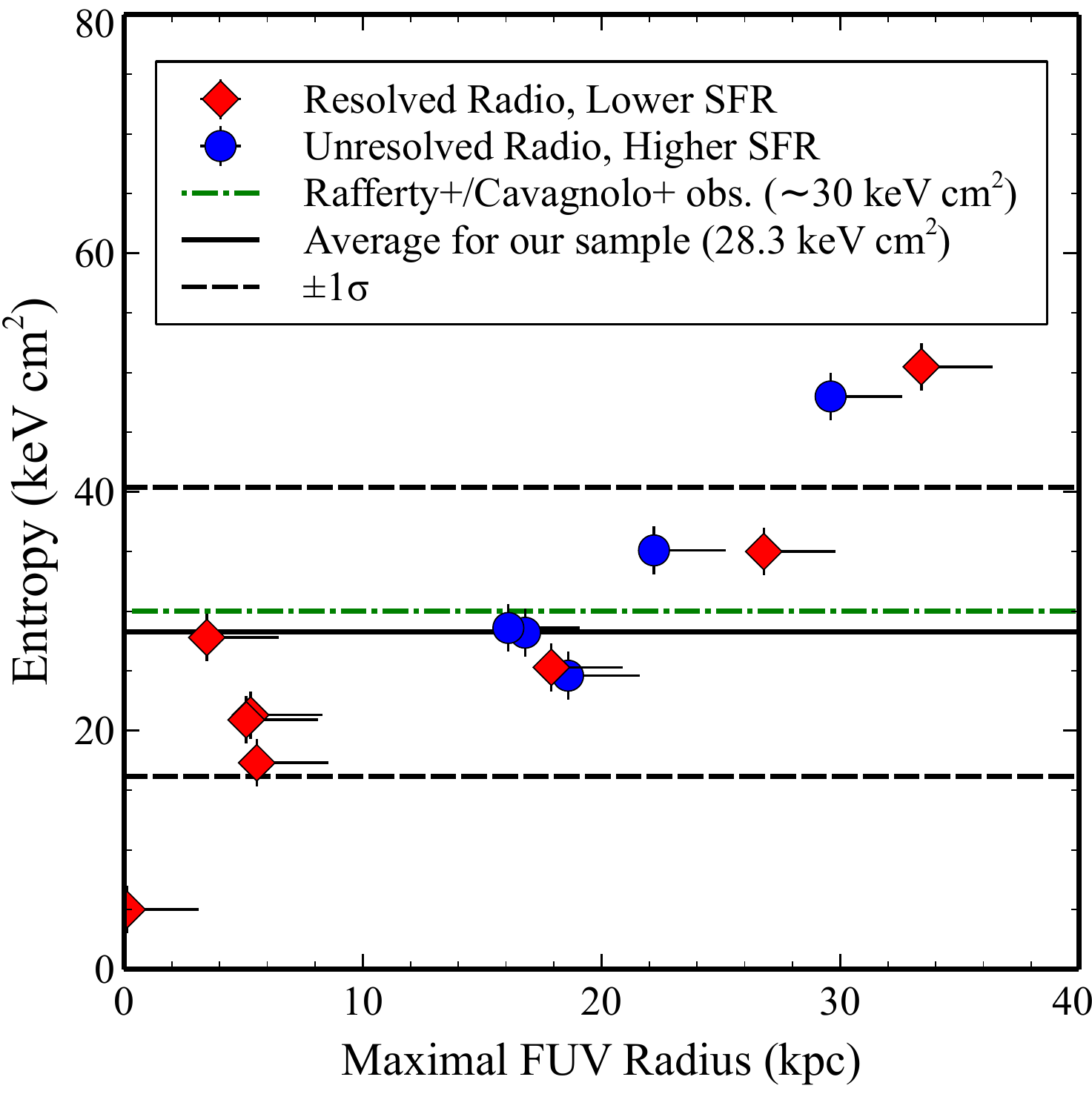}~~~~
\includegraphics[scale=0.5]{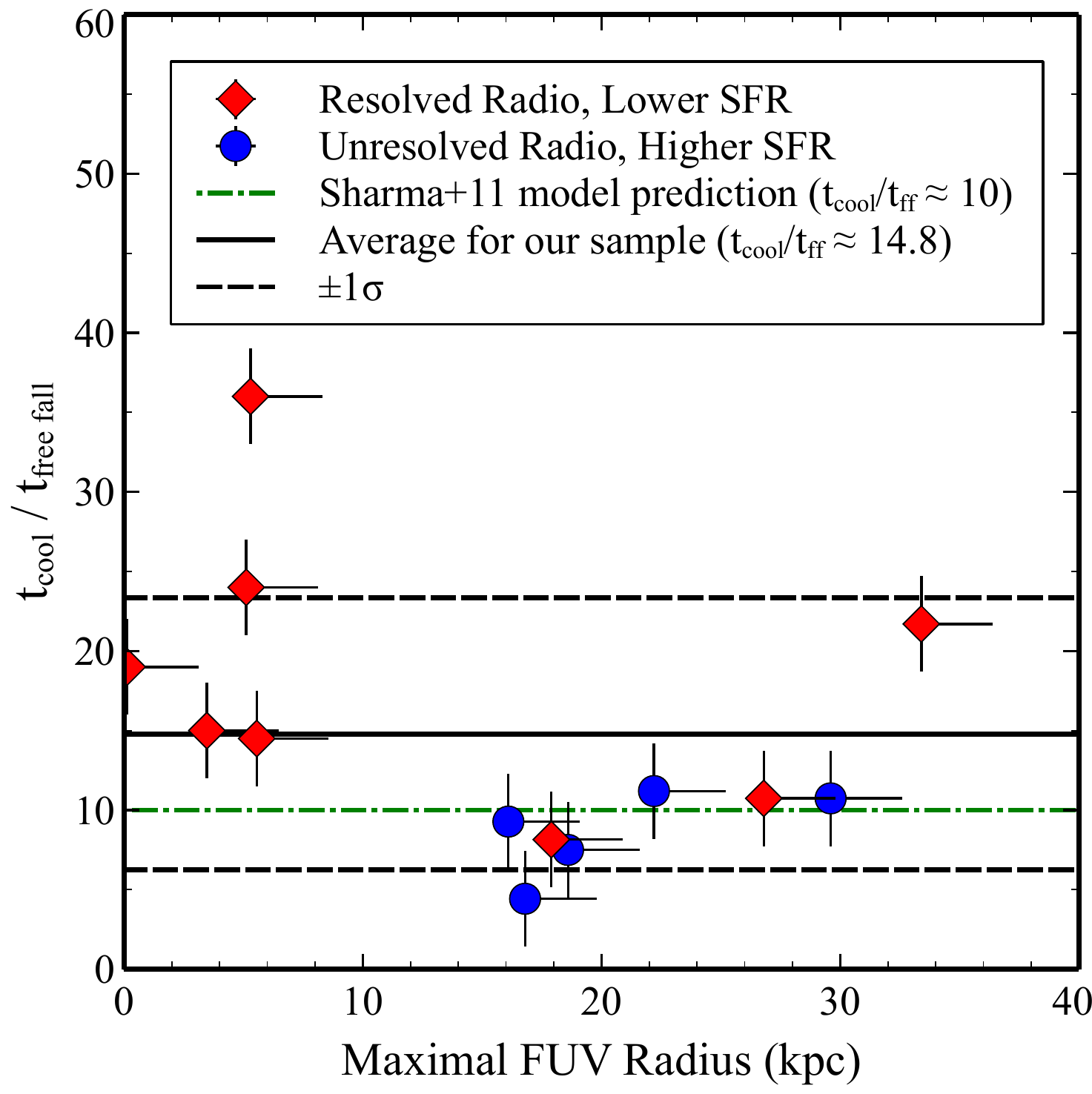}\\
\end{center}
\caption{Here we compare the maximal FUV radius with the X-ray entropy
  (left) and cooling-to-dynamical time ratios (right) measured at that
  same cluster-centric  radius.  Assuming (a) that  the star formation
  in our sources is indeed powered  by a cooling flow and (b) there is
  no  non-detected  star formation  beyond  the  largest measured  FUV
  radius,  the maximal radius  within which  FUV emission  is observed
  serves as a rough observable  tracer of the radial threshold for the
  onset of  cooling flow powered  star formation.  Within  our sample,
  the average  maximal radius (and its $\pm  1\sigma$ interval) within
  which FUV continuum emission is detected is $14.7 \pm 10.14$ kpc. At
  this average  radius, the average (and $\pm  1\sigma$) cooling time,
  entropy, and cooling-to-dynamical time ratio is $1.09 \pm 0.68$ Gyr,
  $28.27\pm12.12$  keV  cm$^2$, and  $14.78  \pm 8.57$,  respectively.
  These average values and their $\pm 1\sigma$ intervals are marked by
  the black solid and dashed lines (respectively) on both panels.   Blue  and   red  points   are   used  to
  differentiate  between those sources  with unresolved  radio sources
  (and higher SFRs)  and resolved radio sources (and  lower SFRs).  As
  entropy  rises with  radius, the  points  on the  leftmost panel  of
  course  also rise  (i.e.. larger  local entropies  will be  found at
  larger  maximal FUV  radii). Both  results are  found to  be roughly
  consistent  (although  with   large  scatter)  with  the  respective
  predictions by  \citet{cavagnolo08} and \citet{sharma11},  which are
  marked by the green dashed line on both plots.}
\label{fig:thresholds}
\end{figure*}

\begin{figure*}
\begin{center}
\includegraphics[scale=0.7]{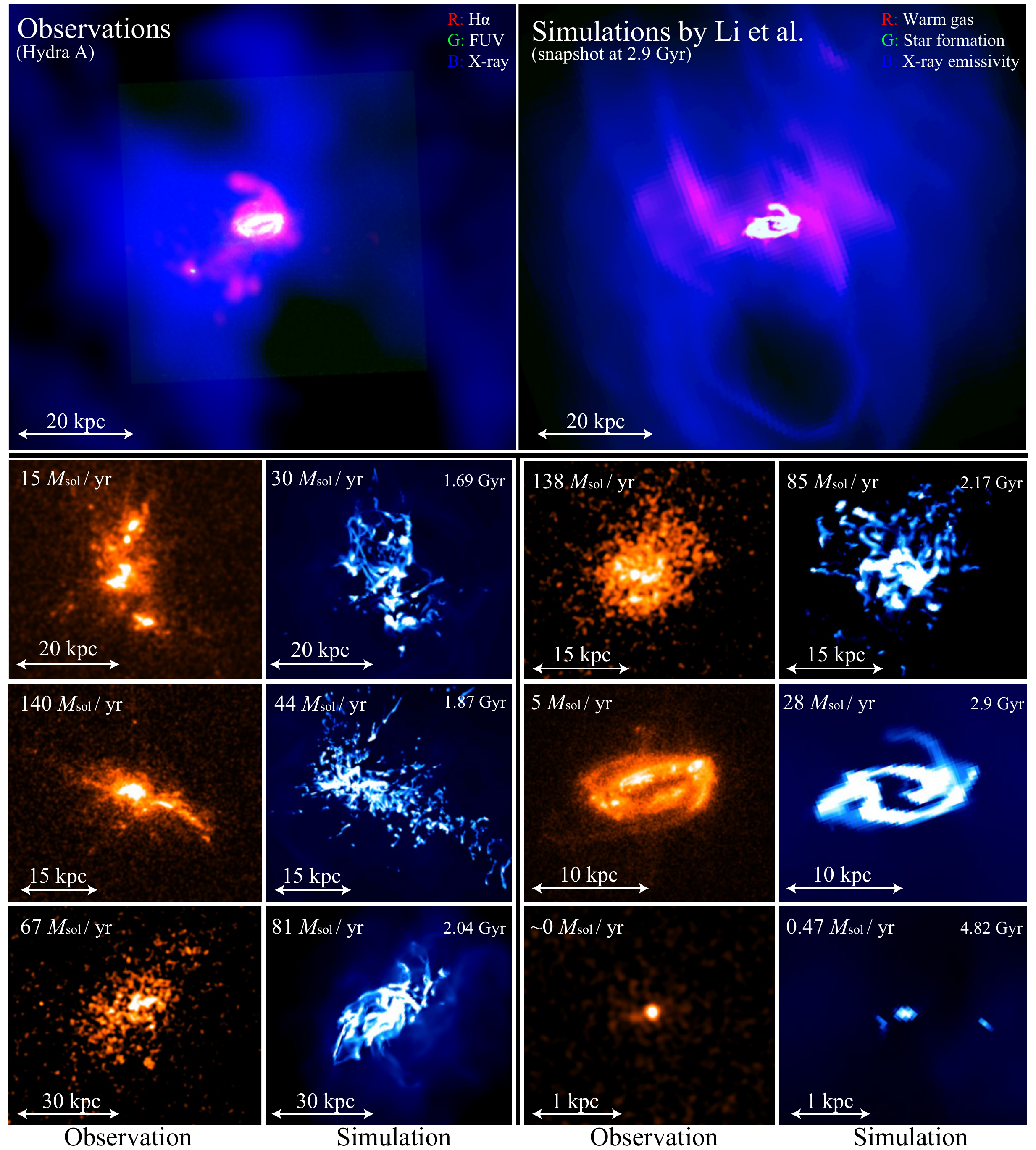}
\end{center}
\vspace*{-3mm}
\caption{Observations compared with simulations by Y.~Li and collaborators (e.g., \citealt{li14a,li14b,li15}). 
({\it Top panels}) A multiwavelength (X-ray, FUV, and H$\alpha$) composite of Hydra A, compared with a single snapshot of the \citet{li15} simulation at 2.9 Gyr.
The RGB channels are set to roughly simulate H$\alpha$, FUV, and X-ray emmissivities, respectively, though two ``cheats'' have been used: the green channel 
shows cold, dense gas where star formation appears in the simulation --- the assumption, then, is that FUV emission from these young stars would roughly show the
same morphology. The red channel does not explicitly show simulated H$\alpha$, but rather intermediate temperature gas whose morphology closely matches
a rough scaling to simulate collisionally ionized (but not photoionized) H$\alpha$. It should therefore be treated as a very rough approximation.
({\it Bottom panels}) Selected FUV images from our sample are shown in orange, and projected density-weighted density snapshots from the simulation are shown in blue. 
The bright knots and filaments show the high density low temperature clouds that are forming stars in the simulation 
(see \citealt{li15} for details).  
Our sample's redshift range 
spans $\sim3$ Gyr of cosmic history. The simulation is capable of producing star forming structures similar in axis ratio, physical extent, and star formation rate within the same $\sim 3$ Gyr temporal slice of the simulated cluster's evolution.}
\label{fig:simulations}
\end{figure*}

\section{Discussion}

The exquisitely complex and highly diverse range of far ultraviolet morphologies 
prestented in this paper reflect the dynamical response of low entropy gas
to a highly energetic, chaotic environment. Besides the graviational potential of their host galaxies, these star forming nebulae reside amid
AGN-driven jets, bubbles, sound waves, bulk ICM motions, and stellar feedback, and few structures observed in our FUV dataset are likely to be 
long-lived. 
Hydra A's rotating, star forming disk \citep{hamer14} reveals cold gas that is largely in dynamical equilibrium,
though it still features narrow filaments that have likely been lifted outward by the radio jet (Figs.~\ref{fig:jet_triggered_sfr_figure} \& \ref{fig:hydra_figure}), or dynamically 
stirred by a small companion galaxy (Fig.~\ref{fig:hydraA_companion}). A1795 (Fig.~\ref{fig:a1795_figure}) not only features radio lobes frosted
with young stars, but a $\sim 20$ kpc southern tail deposited perhaps by a cooling wake 
that lags behind the BCG \citep{mcdonald09}. Centaurus (Fig.~\ref{fig:centaurus_figure}) features no discernable star formation whatsoever, 
but a spectacular winding dust lane whose shape mirrors that of the larger scale X-ray spiral.

In many ways, conclusions drawn from a small collection of highly complex
individual galaxies are doomed to ambiguity.  If each source is so chaotic and
time-varying, what can we  learn in a ``big picture'' context? The answer may
be that our data are snapshots of a prototypical cool core BCG at different stages
of an AGN outburst cycle.
Throughout the 
past $\sim 3$ Gyr of cosmic history that our sample's redshift range spans, perhaps 
each galaxy has spent (or will spend) some time resembling each of the others
as the tug-of-war between ICM cooling and AGN heating cycles and varies. 
Indeed, AGN can vary in power and switch 
on and off over a manifold range of timescales, and the associated balance of AGN heating and ICM cooling 
can vary still more.

Recent numerical  work   by    \citet{sharma11,mccourt12},   and
\citet{gaspari12} has  shown that  thermal instabilities in  a cooling
flow can produce a  multiphase and star  forming ISM when
the  ratio  of  the   cooling  time  $t_\mathrm{cool}$  to  the  local
gravitational      free-fall     timescale      $t_\mathrm{ff}$     is
$t_{\mathrm{cool}}/t_{\mathrm{ff}}\lae  10$ (see also \citealt{pizzolato05,pizzolato10}). This theoretical framework has since been expanded 
in a series of papers by Voit and collaborators, who propose a precipitation-regulated 
AGN feedback model applicable not only to BCGs and giant ellipticals \citep{voit15,voit15b,voit15c}, 
but perhaps galaxies in general (Voit et al.~2015c, in press).  
In the model, cold clouds 
precipitate out of the ambient hot medium via thermal instability wherever $t_{\mathrm{cool}} \lae 10 t_{\mathrm{ff}}$. 
Cold chaotic accretion \citep{gaspari13,gaspari14} from the now ``raining'' ambient hot atmosphere
boosts black hole feeding to $\sim 100$ times the Bondi rate, powering jets 
that can stimuluate further precipitation by dragging low entropy gas to higher 
altitudes, where the cooling-to-dynamical time ratio will lessen. 
At the same time, jet heating works to {\it raise} the local cooling time, resulting in 
the system's self-regulation at nearly $t_{\mathrm{cool}} \approx 10 t_{\mathrm{ff}}$ \citep{voit15,voit15b,voit15c}.

The maximal radius within which FUV emission is detected in our images
can be  used in  a very rough comparison with this theoretical prediction.  
Assuming (a) that the star formation in our sources
is indeed powered  by a precipitation-based cooling flow and (b)  there is no non-detected
star formation  beyond the largest measured FUV  radius, this ``maximal FUV radius''
serves as  a rough observable tracer  of the radial  threshold for the
onset of  cooling flow powered  star formation. Models of precipitation-regulated feedback predict 
that this radius should coincide with $t_{\mathrm{cool}} \approx 10 t_{\mathrm{ff}}$. 
Of  course, assumption
(b)  may be  not reasonable,  as  deeper FUV  observations may  reveal
larger maximal FUV radii for a significant fraction of our sample. 
One must also consider that these filaments may have been uplifted by jets 
or buoyant cavities in some cases. 
The maximal FUV radius should therefore be treated as a lower limit in this 
caveat-laden test.

To derive $t_{\mathrm{ff}}$, 
we adopt the spectrally   deprojected  X-ray
emissivity, cooling time, and electron density  profiles from  work on  the
ACCEPT  sample by \citet{donahue06}  and \citet{cavagnolo09}. To these, we fit 
third-order polynomials in log space, and then 
analytically differentiate to obtain the gravitational free fall time
$t_\mathrm{ff}$.   The  presence  of  the  BCG was  accounted  for  by
enforcing a minimum value  of the gravitational acceleration $g$ equal
to that of an isothermal sphere with a velocity dispersion of $250$ km
s\mone\ (a correction that is only important at radii $\lae 10$ kpc).

The result is shown in Fig.~\ref{fig:thresholds}, where we compare the maximal FUV radius with the
X-ray entropy  and cooling-to-dynamical  time ratios measured  at that
same   cluster-centric  radius.  
Within our sample, the average maximal radius (and its $\pm 1\sigma$ interval) within which FUV continuum emission is detected 
is $14.7 \pm 10.14$ kpc. At this average radius, the average (and $\pm 1\sigma$) cooling time, entropy, 
and cooling-to-dynamical time ratio is $1.09 \pm 0.68$ Gyr, $28.27\pm12.12$ keV cm$^2$, and $14.78 \pm 8.57$, respectively. 
These average values and their $\pm 1\sigma$ intervals are 
marked by the black solid and dashed lines (respectively) 
on the two panels in Fig.~\ref{fig:thresholds}. Blue and red points 
are used to differentiate between those sources with unresolved radio 
sources (and higher SFRs) and resolved radio sources (and lower SFRs). 
As entropy rises with radius, the points on the leftmost panel 
of course also rise (i.e., larger local entropies will be found 
at larger maximal FUV radii).

We find that the observed {\it average} ``star formation onset'' threshold  of
$S=28$ keV cm$^2$ is  very close to the empirical (rough) threshold of
$S\approx30$ keV cm$^2$ from \citet{cavagnolo08}, which we mark with the
green dashed line on the left panel of Fig.~\ref{fig:thresholds}.  The average
observed $t_{\mathrm{cool}}/t_{\mathrm{ff}}$ ratio of  $14.78$ is close to the
predicted threshold of $t_{\mathrm{cool}}/t_{\mathrm{ff}}\lae 10$ by
\citet{sharma12b}, which we mark with the green dashed line on the rightmost
panel of  Fig.~\ref{fig:thresholds}. Sources with higher star formation rates
cluster more strongly around $t_{\mathrm{cool}} \approx 10 t_{\mathrm{ff}}$
than do sources with lower star formation rates. Heeding the strong caveats noted above, 
this may be roughly  
consistent (or at least not obviously inconsistent) with theoretical predictions by \citet{sharma12b}, \citet{mccourt12}, and \citet{voit15b}.

\citet{li14a,li14b} and \citet{li15} have published new adaptive mesh hydrodynamical 
simulations of BCGs consistent with the precipitation-regulated AGN feedback framework.  
In Fig.~\ref{fig:simulations}, we show selected snapshots of the cold, star forming gas 
produced by 
the standard run in \citet{li15} for a single BCG, and compare these with 
our FUV images. 
The simulated cooling flow begins at $t\sim300$ Myr (roughly the central cooling
time of Perseus), which ignites AGN feedback. The jets trigger more
ICM to cool into filamentary structures, causing more cold gas to rain into the nucleus. This fuels both star
formation and black hole accretion, leading to a major AGN outburst. AGN feedback heats up
the core, reducing the cooling rate and causing $t_{\mathrm{cool}}$ to increase; star
formation gradually consumes the cold gas until it vanishes, which turns off
AGN feedback and allows the ICM to cool again. The cluster experiences three
such cycles within 6.5 Gyr. 

Most of the images shown in Fig.~\ref{fig:simulations} 
are from the second cycle that
starts around $t=1.6$ Gyr. 
Within this cycle, the simulation produces star forming structures that are remarkably 
similar to our FUV images not only in morphology, axis ratio, and physical extent, but also in star formation rate.
That is, when the simulated morphology matches that of one of our FUV images, its associated star formation rate 
also roughly matches.  
These structures are reproduced by the simulation in the same $\sim 3$ Gyr temporal slice
spanned by the redshift range of our sample. Their simulated multiphase (hot, warm, and cold)
gas morphology is also remarkably similar to the multiwavelength morphologies we have presented here. 
This is illustrated in the topmost panels of Fig.~\ref{fig:simulations}, in which the RGB multiwavelength 
composite of Hydra A in X-ray, H$\alpha$, and FUV is nearly indistinguishable from the simulation snapshot
at 2.9 Gyr, which comes complete with a rotating, star forming disk.

While surely  other stochastic events such as mergers play
an additional role in sculpting the morphology of star forming filaments in CC BCGs, 
the Li et al.~simulations show that
they need not be invoked to explain the FUV morphologies we observe (the same was true for a similar comparison made in \citealt{donahue15}). 
Instead, nearly all FUV morphologies shown in this paper appear in a simulated BCG whose 
evolution is driven by precipitation-regulated AGN feedback. The FUV images
presented in this paper may be snapshots at successive stages of an ICM cooling / AGN heating cycle.

\section{Summary \& Concluding Remarks}

We have analysed the far ultraviolet morphology of star forming clouds
and filaments in 16 low-redshift ($z<0.29$) cool core 
brightest cluster galaxies.  X-ray,  Ly$\alpha$,
H$\alpha$, broadband optical/IR, and radio maps were compared with the
FUV  emission,  providing a  high  spatial  resolution  atlas of  star
formation locales  relative to  the ambient hot  and warm  ionised gas
phases, as  well as  the old stellar  population and  radio-bright AGN
outflows.  
The main results of this paper are summarised as follows. 

\begin{itemize}

\item Nearly half of the sample possesses kpc-scale narrow filaments that, in
  projection, extend toward, into, and  around  radio lobes and/or X-ray cavities. Most (but not all) of these 
  filaments are FUV-bright and forming stars, and we suggest that they
  have  either been  uplifted by  the radio lobe or buoyant X-ray  cavity, or  have
  formed {\it  in situ}  by jet-triggered star formation or rapid cooling  in the  cavity's compressed
  shell.

\item The maximal projected radius to which FUV emission is observed to extend
corresponds to a cooling-to-freefall time of $t_{\mathrm{cool}}/t_{\mathrm{ff}}\sim 10$ for the 
majority of the sample. Sources with higher star formation rates cluster more strongly about this 
ratio than do sources with lower star formation rates. This may be roughly  
consistent (or at least not inconsistent) with theoretical predictions by \citet{sharma12b}, \citet{mccourt12}, and \citet{voit15b}.
We nevertheless stress that maximal FUV radius is not the ideal tracer for the onset of purported ICM precipitation, 
as one must consider imaging depth, extinction, and other morphological drivers such as filament uplift.

\item The diverse range of morphology, axis ratio, spatial extent, and star formation rate in our FUV sample 
is almost entirely recovered in a
single simulation by \citet{li15}, demonstrating that galaxy-scale stochastic events such as mergers need not be invoked to
explain the complex FUV morphologies we observe. Instead, we suggest that our 
images represent 
snapshots of a prototypical cool core BCG at many stages of its evolution.  

\end{itemize}

\section*{Acknowledgments} 
The authors thank  Drs.~Richard Bower, Tim Davis,
Bill Forman, Nina Hatch, Claudia Lagos,  Robert Laing, Jason Spyromilio, and
Aurora Simionescu for thoughtful  discussions.  G.~R.~T.~acknowledges support
from a NASA Einstein Fellowship under award number PF-150128, as well as a
European  Southern Observatory  (ESO)   Fellowship  partially  funded   by
the  European Community's Seventh Framework  Programme (/FP7/2007-2013/) under
grant agreement number 229517.   Support for this work was provided by the
National Aeronautics  and Space Administration through Einstein Postdoctoral
Fellowship  Award Number PF-150128 issued by the Chandra X-ray Observatory
Center, which is operated by the Smithsonian Astrophysical  Observatory for
and on behalf of NASA under contract NAS8-03060. Y.~L.~acknowledges financial
support from NSF grants AST-0908390,  AST-1008134, AST-1210890, AST-1008454,
NASA grants NNX12AH41G, NNX12AC98G, and ATP12-ATP12-0017, as well as
computational resources from NASA, NSF XSEDE and Columbia University.  T.~E.~C.~was 
partially supported by NASA through Chandra award G06-7115B issued by the
{\it Chandra} X-ray  Observatory Center for and on behalf of NASA under
contract NAS8-39073.  Basic research into radio astronomy at the Naval
Research Laboratory is supported by 6.1 Base funds.  This paper is based on
observations made with the NASA/ESA {\it  Hubble Space   Telescope}, obtained
at the Space Telescope Science Institute, which is  operated  by  the
Association  of Universities  for  Research  in Astronomy, Inc., under NASA
contract 5-26555.  This research made use of Astropy, a community-developed core Python package for Astronomy \citep{astropypaper}.
We have also made extensive use of  the NASA Astrophysics  Data
System bibliographic  services and the NASA/IPAC  Extragalactic Database,
operated by  the Jet Propulsion Laboratory,  California Institute of
Technology, under  contract with NASA.

\appendix

\section{Multiwavelength overlay figures}

\begin{figure*}
\begin{center}
\includegraphics[scale=0.39]{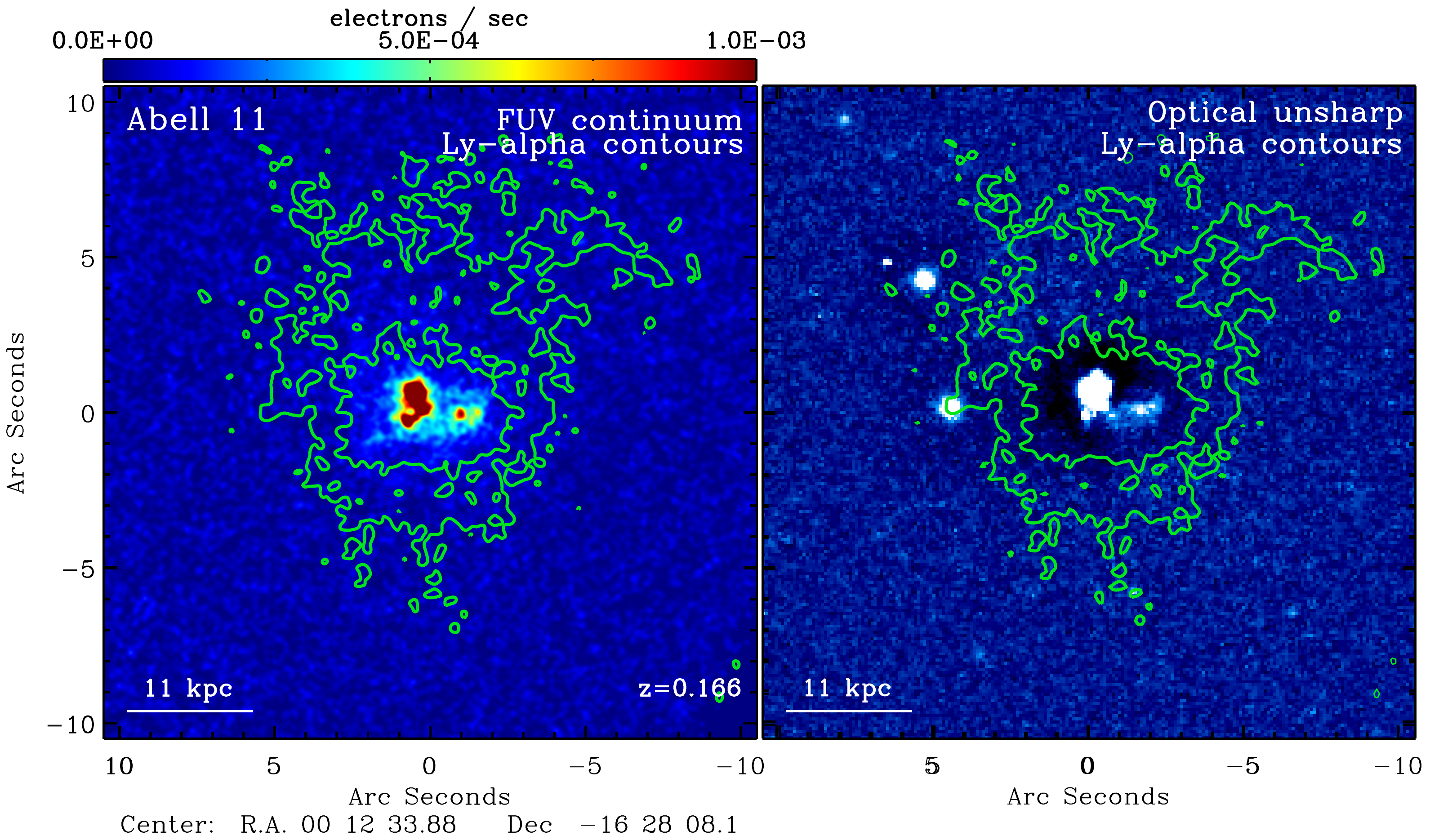}
\end{center}
\caption{The Abell 11 BCG ($z=0.1660$). 
The FUV continuum image and broadband optical unsharp mask 
are shown in the left and right panels, respectively. Ly$\alpha$ contours 
are overlaid in green. The radio source is unresolved. 
The FUV colour bar can be scaled to a flux density by the inverse sensitivity 
$4.392\times 10^{-17}$ ergs cm\mtwo\ \AA\mone\ electron\mone.  
The centroids of both panels are aligned, with east left and north up.  }
\label{fig:a11_figure}
\end{figure*}

\begin{figure*}
\begin{center}
\includegraphics[scale=0.41]{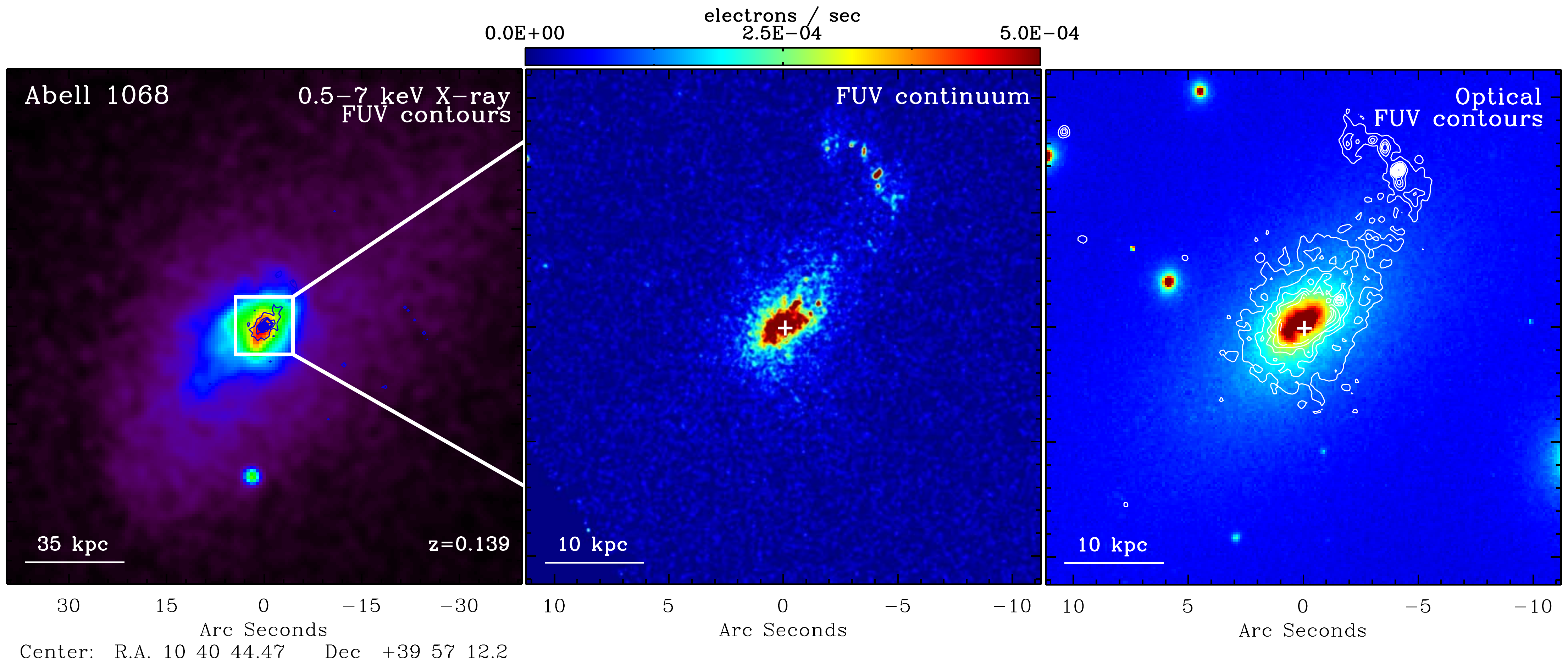}
\includegraphics[scale=0.49]{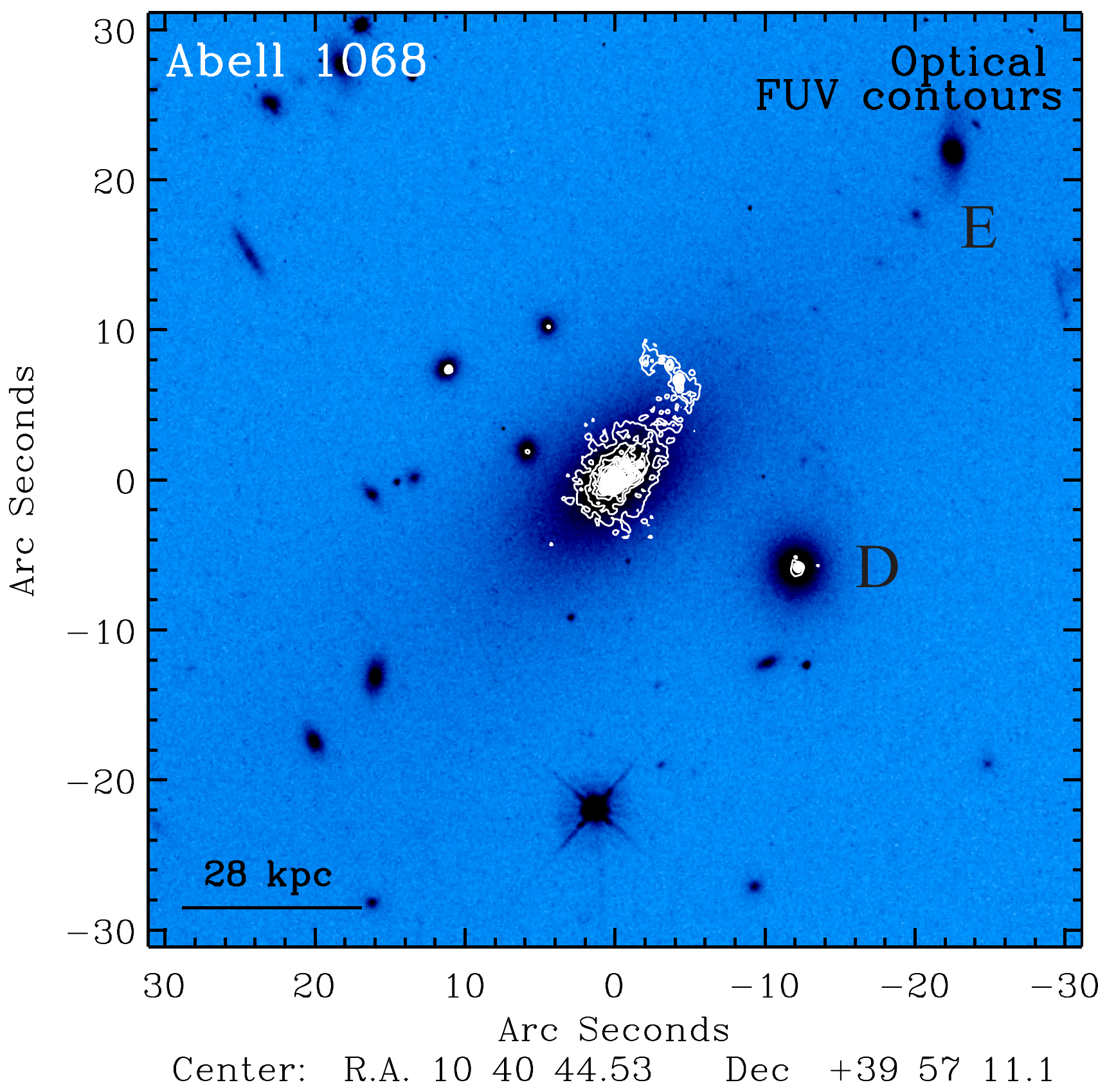}
\end{center}
\caption{A multiwavelength view of the Abell 1068 BCG ($z=0.1375$). X-ray, FUV continuum, and broadband optical 
images are shown in the top left, centre, and right panels, respectively. Contours of constant FUV continuum 
surface brightness are overlaid in blue on the X-ray panel, and in white on the optical panel. The white cross
on the FUV panel marks the location of the unresolved radio source. 
The FUV colour bar can be scaled to a flux density by the inverse sensitivity 
$4.392\times 10^{-17}$ ergs cm\mtwo\ \AA\mone\ electron\mone.  
The white box on the X-ray panel marks the FOV of the two rightmost panels.
The bottom panel shows the same broadband optical image in a different colour scale, a wider FOV, and with FUV contours shown in white. The 
nearby companions are labeled ``D'' and ``E'' to correspond to the notation used by \citet{mcnamara04} in their 
discussion of these companions as possible gas donors. The centroids of all panels are aligned, with east left and north up.  }
\label{fig:a1068_figure}
\end{figure*}

\begin{figure*}
\begin{center}
\includegraphics[scale=0.41]{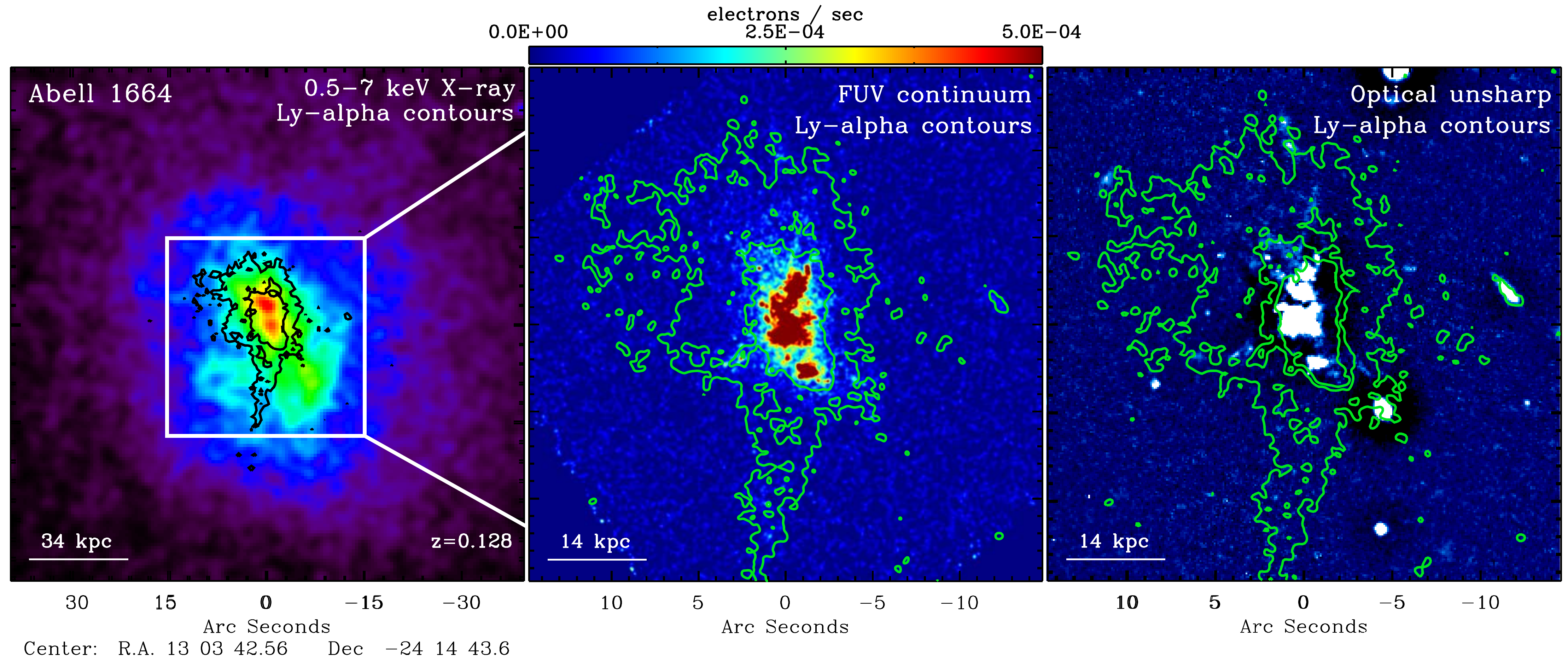}
\end{center}
\caption{The Abell 1664 BCG ($z=0.1283$).  The X-ray, FUV continuum, and 
broadband optical unsharp mask images  
are shown in the left, centre, and right panels, respectively. Ly$\alpha$ contours 
are overlaid in black (on the X-ray panel) and green (on the FUV and optical panels). The radio source is unresolved. 
The FUV colour bar can be scaled to a flux density by the inverse sensitivity 
$4.392\times 10^{-17}$ ergs cm\mtwo\ \AA\mone\ electron\mone.  
The centroids of all panels are aligned, with east left and north up.    }
\label{fig:a1664_figure}
\end{figure*}

\begin{figure*}
\begin{center}
\includegraphics[scale=0.41]{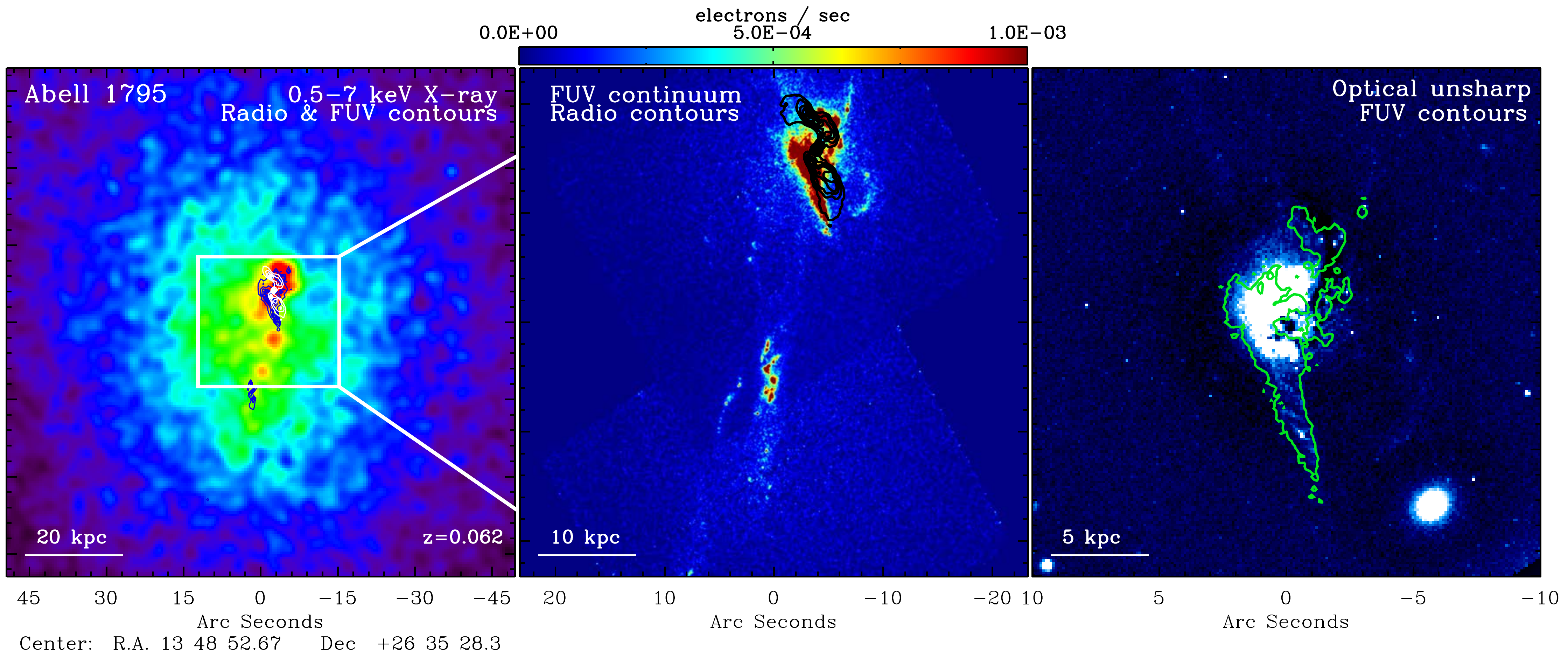}
\end{center}
\caption{The Abell 1795 BCG ($z=0.0625$).   The X-ray, FUV continuum, and 
broadband optical unsharp mask images  
are shown in the left, centre, and right panels, respectively. FUV contours 
are overlaid in blue and green on the X-ray and optical panels, respectively. 
The radio source is shown in white and black contours on the X-ray and FUV panels, 
respectively. Note the remarkable spatial correspondence between the FUV continuum 
emission and the radio source. 
The FUV colour bar can be scaled to a flux density by the inverse sensitivity 
$2.173\times 10^{-17}$ ergs cm\mtwo\ \AA\mone\ electron\mone.  
The centroids of all panels are aligned, with east left and north up.     }
\label{fig:a1795_figure}
\end{figure*}

\begin{figure*}
\begin{center}
\includegraphics[scale=0.41]{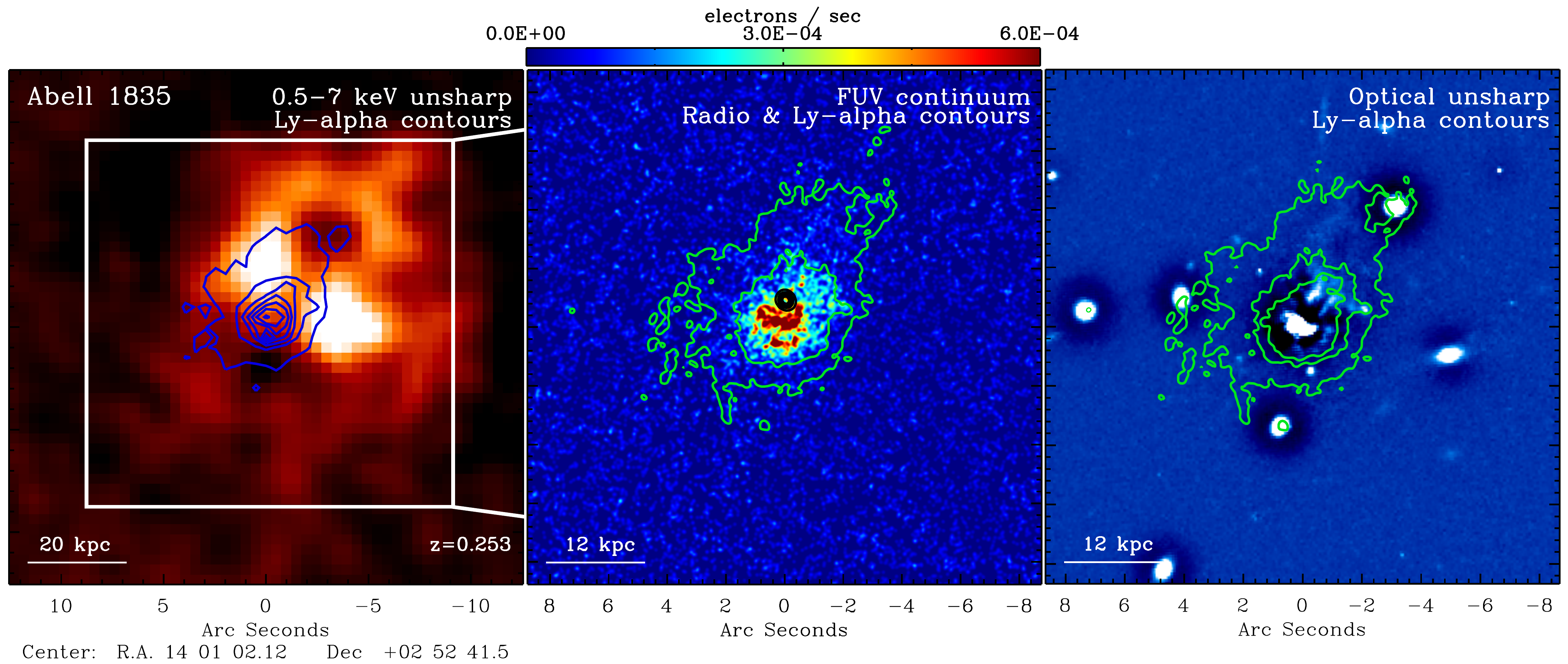}
\end{center}
\caption{The Abell 1835 BCG ($z=0.2532$).  The X-ray unsharp, FUV continuum, and 
broadband optical unsharp mask images  
are shown in the left, centre, and right panels, respectively. Ly$\alpha$ contours 
are overlaid in blue on the X-ray panel and in green on the FUV and optical panels. 
The unresolved radio source is overlaid in black contours on the FUV panel. 
The FUV colour bar can be scaled to a flux density by the inverse sensitivity 
$1.360\times 10^{-16}$ ergs cm\mtwo\ \AA\mone\ electron\mone.  
The centroids of all panels are aligned, with east left and north up.     }
\label{fig:a1835_figure}
\end{figure*}

\begin{figure*}
\begin{center}
\includegraphics[scale=0.41]{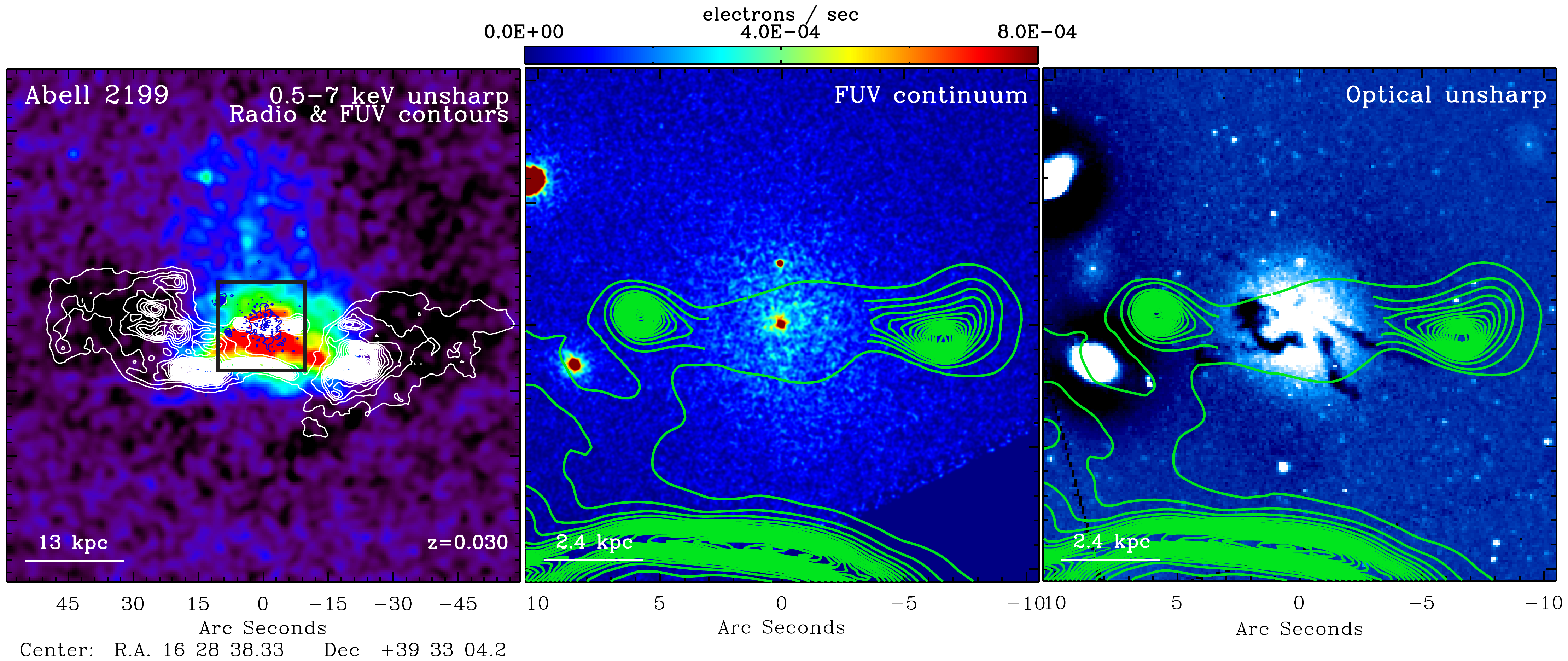}
\includegraphics[scale=0.49]{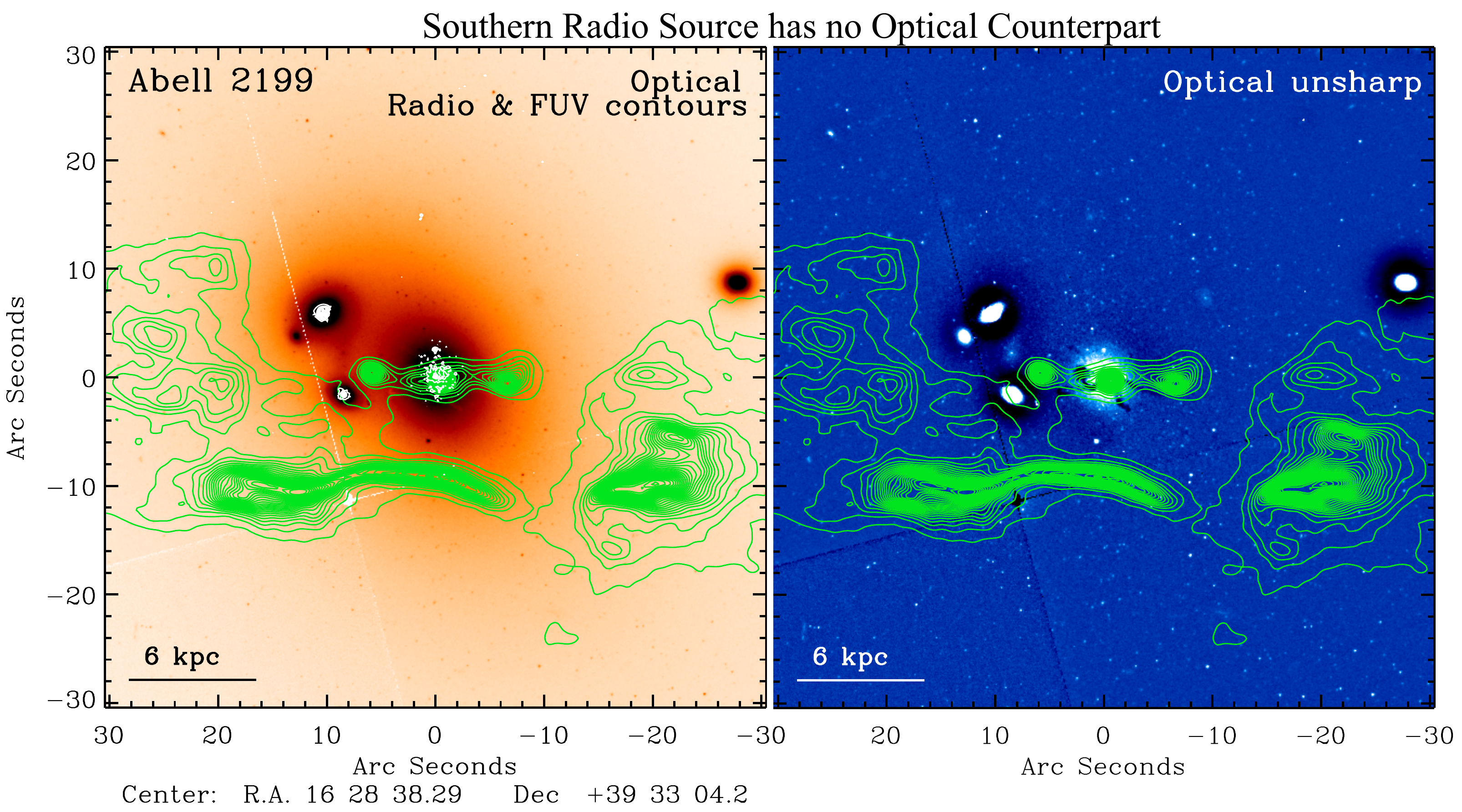}
\end{center}
\caption{A multiwavelength view of the Abell 2199 BCG ($z=0.0302$) and its remarkable radio source. 
An X-ray unsharp mask, FUV continuum map, and optical unsharp mask  
are shown in the top left, centre, and right panels, respectively. 
The double-double FR~I radio source 3C 338 is shown in white contours on the X-ray panel, and in green 
contours on all other panels. 
Note the  $\sim 20$ kpc-scale X-ray cavities cospatial with the radio lobes.
FUV continuum contours are overlaid in blue and white on the X-ray and bottom-left 
optical panel, respectively.  
The FUV colour bar can be scaled to a flux density by the inverse sensitivity 
$2.713\times 10^{-17}$ ergs cm\mtwo\ \AA\mone\ electron\mone.  
The black box on the X-ray panel marks the FOV of the two rightmost panels.
The bottom panel shows wider FOVs of the broadband optical (and optical unsharp mask) images, and clearly 
demonstrates that the southern component of the radio source has no optical counterpart. 
As discussed in \citet{nulsen13}, the radio source has either restarted while the 
host galaxy has moved north (in projection) at very high peculiar velocity (unlikely), or sloshing X-ray 
gas has pushed the relic radio source south.  
The centroids of all panels are aligned, with east left and north up.    }
\label{fig:a2199_figure}
\end{figure*}

\begin{figure*}
\begin{center}
\includegraphics[scale=0.41]{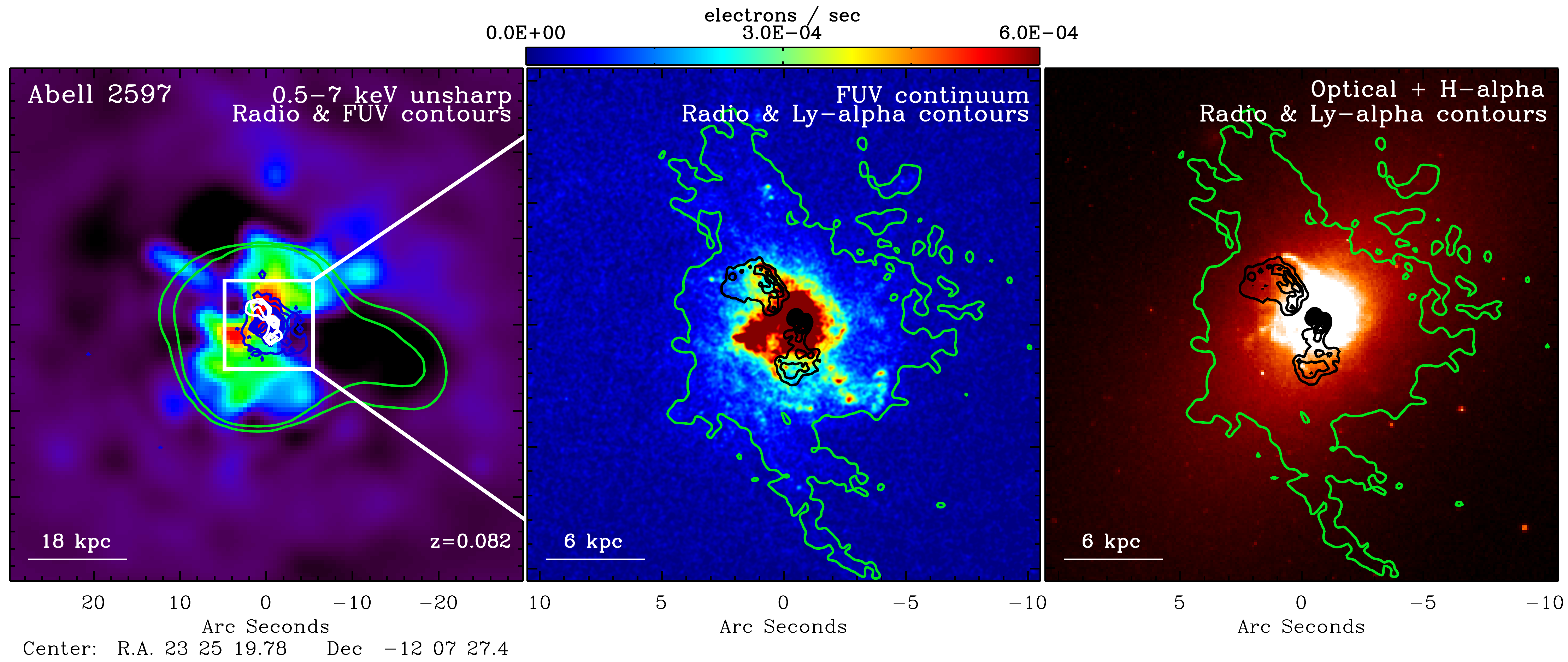}
\end{center}
\caption{The Abell 2597 BCG ($z=0.0821$).  The X-ray unsharp, FUV continuum, and 
broadband optical images  
are shown in the left, centre, and right panels, respectively. 
330 MHz and 8.4 GHz radio contours are shown in green and white (respectively) 
on the X-ray panel, and 8.4 GHz contours are shown in black on the FUV and optical panels. 
Ly$\alpha$ contours 
are overlaid in blue on the X-ray panel and in green on the FUV and optical panels.  
The FUV colour bar can be scaled to a flux density by the inverse sensitivity 
$4.392\times 10^{-17}$ ergs cm\mtwo\ \AA\mone\ electron\mone.  The multiwavelength data for A2597 are discussed at length in \citet{tremblay12a,tremblay12b}.
The centroids of all panels are aligned, with east left and north up.   }
\label{fig:a2597_figure}
\end{figure*}

\begin{figure*}
\begin{center}
\includegraphics[scale=0.41]{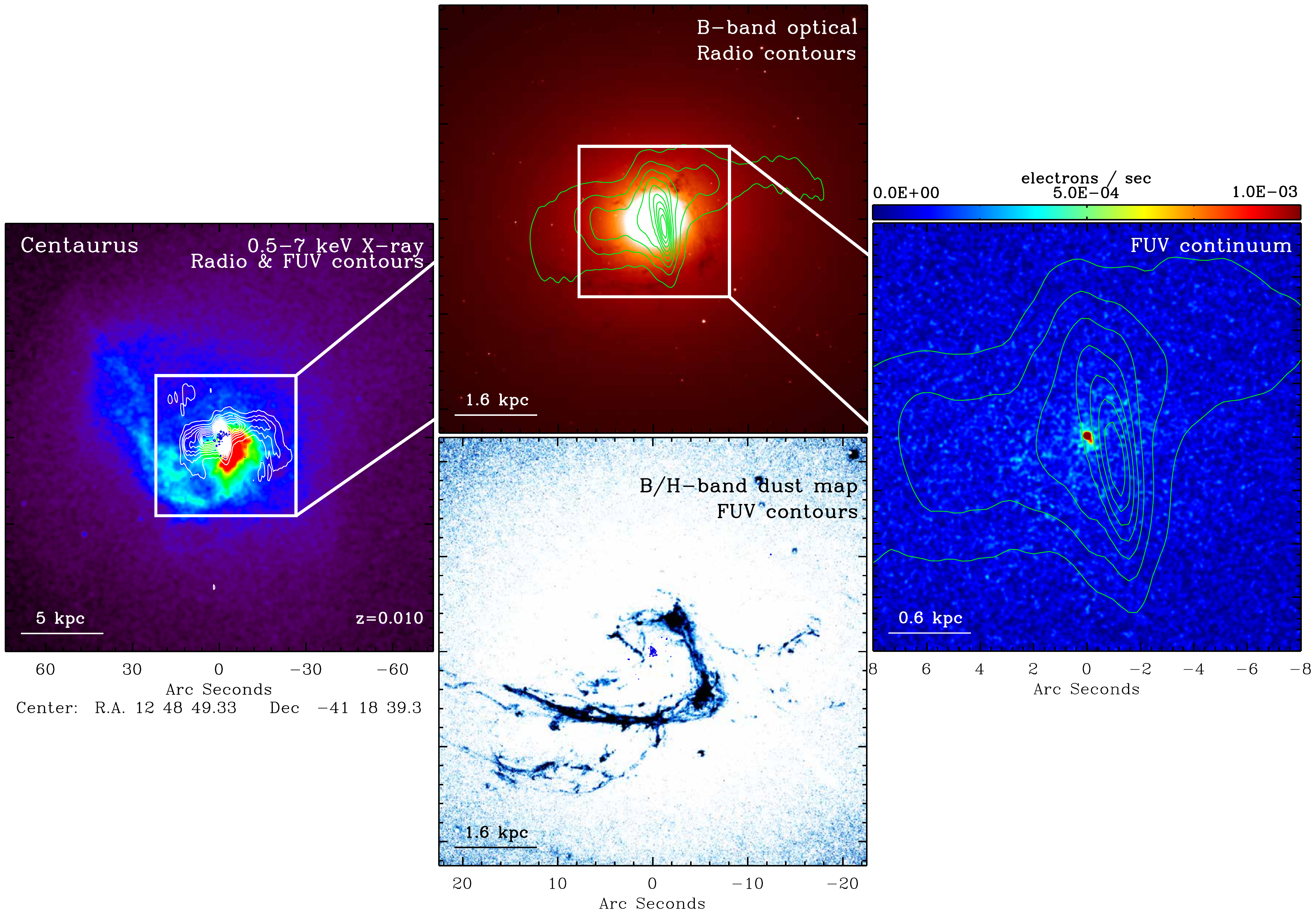}
\end{center}
\caption{NGC 4696, the BCG of the Centaurus Cluster ($z=0.0099$). The X-ray surface brightness map is shown in the leftmost panel, with radio and FUV contours overlaid in white and blue contours, respectively. The centre-top panel shows the $B$-band 
optical map with radio contours overlaid in green. The centre-bottom panel shows a $B$/$H$-band colour map that highlights the dramatic (and well-known) 5 kpc-scale dust lane. FUV contours are overlaid in blue. The rightmost panel shows the FUV continuum data, with radio contours overlaid in green. The FUV colour bar can be scaled to a flux density by the inverse sensitivity 
$4.392\times 10^{-17}$ ergs cm\mtwo\ \AA\mone\ electron\mone.  
The centroids of all panels are aligned, with east left and north up.    }
\label{fig:centaurus_figure}
\end{figure*}

\begin{figure*}
\begin{center}
\includegraphics[scale=0.41]{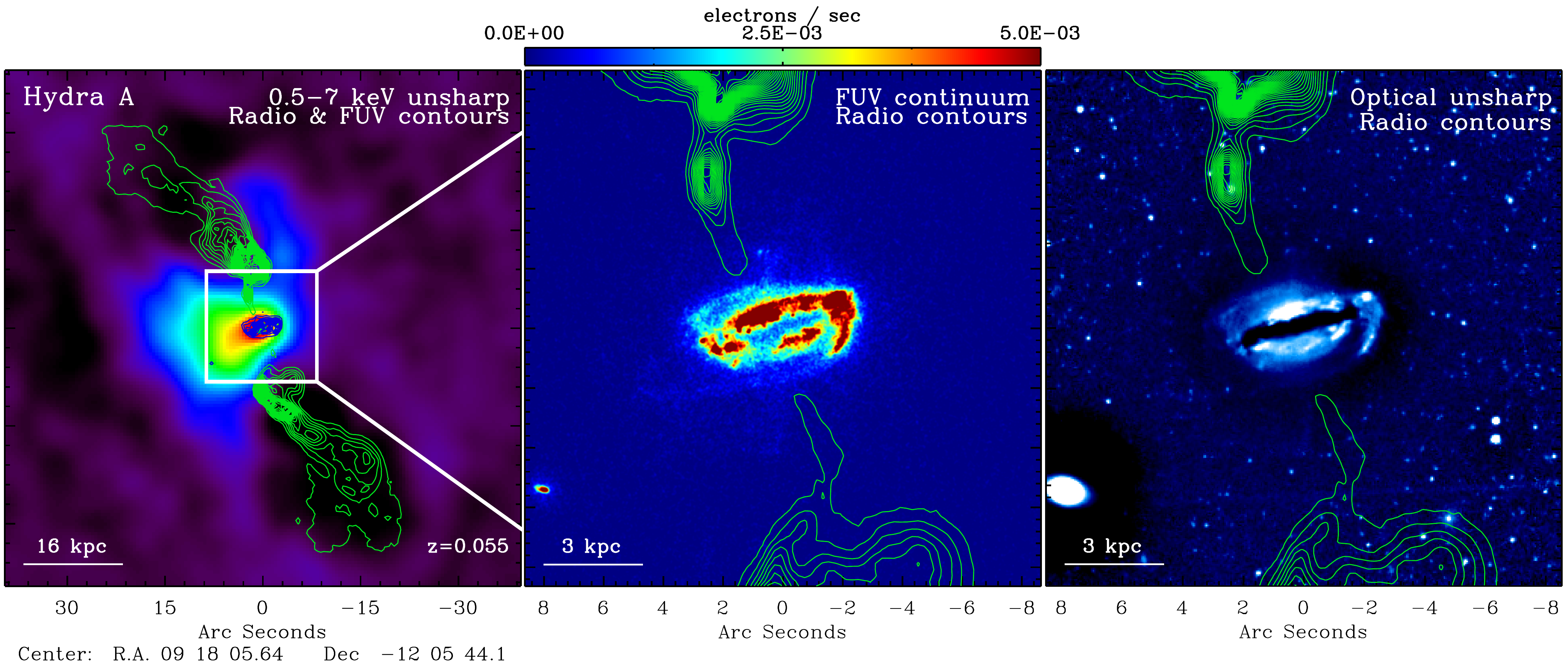}
\end{center}
\caption{A multiwavelength view of the Hydra A (Abell 780) BCG ($z=0.0549$), one of the most mechanically 
powerful radio sources in the known Universe \citep{wise07}. 
An X-ray unsharp mask, FUV continuum map, and optical unsharp mask  
are shown in the left, centre, and right panels, respectively. 
The FR~I radio source 3C~218 is shown in green contours on all panels.
Note the  $\sim 50$ kpc-scale X-ray cavities cospatial with the radio lobes.
The FUV colour bar can be scaled to a flux density by the inverse sensitivity 
$2.713\times 10^{-17}$ ergs cm\mtwo\ \AA\mone\ electron\mone.  
The white box on the X-ray panel marks the FOV of the two rightmost panels.
The centroids of all panels are aligned, with east left and north up. }
\label{fig:hydra_figure}
\end{figure*}

\begin{figure*}
\begin{center}
\includegraphics[scale=0.41]{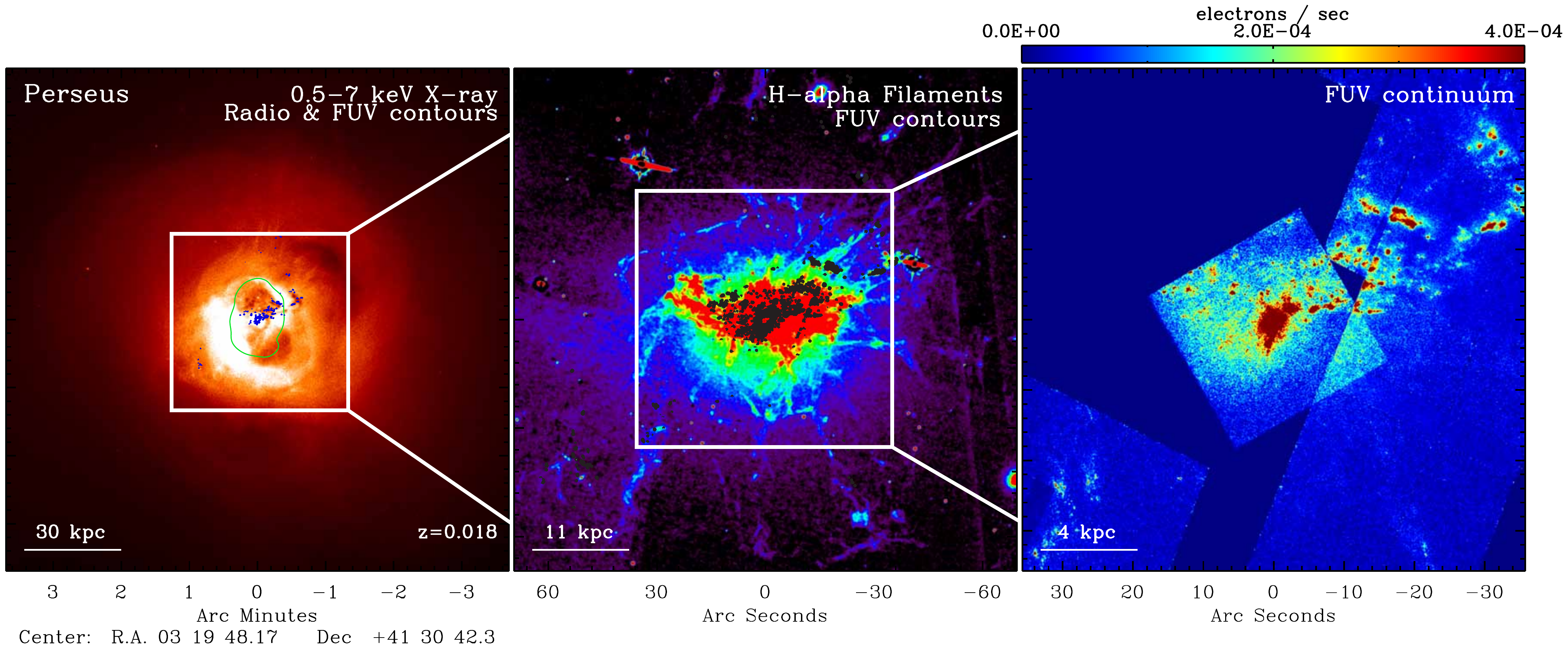}
\end{center}
\caption{NGC 1275, the BCG of the Perseus Cluster ($z=0.0176$).  The X-ray, 
H$\alpha$+[\ion{N}{ii}], and FUV continuum maps are shown in the left, centre, and right panels, respectively. 
Radio contours are shown in green on the X-ray panel, and FUV contours are shown in 
blue and black on the X-ray and optical panels, respectively. The white box on each 
panel shows the FOV of the panel to the right. Some of the FUV emission may be attributable
 to an unrelated galaxy that is superimposed on the line of sight.  
The FUV colour bar can be scaled to a flux density by the inverse sensitivity 
$2.713\times 10^{-17}$ ergs cm\mtwo\ \AA\mone\ electron\mone.  
The centroids of all panels are aligned, with east left and north up.  }
\label{fig:perseus_figure}
\end{figure*}

\begin{figure*}
\begin{center}
\includegraphics[scale=0.41]{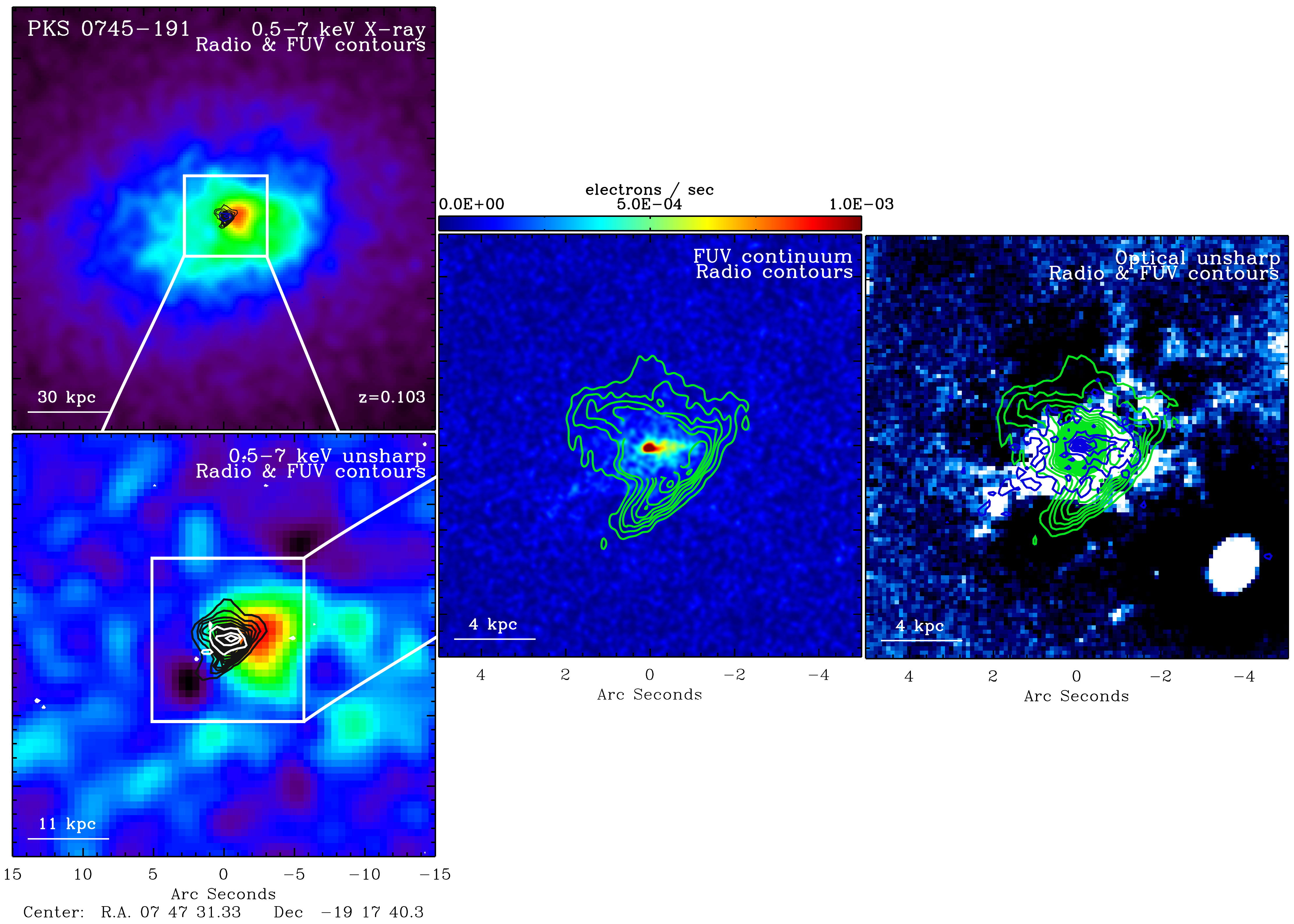}
\end{center}
\caption{A multiwavelength view of the PKS 0745-191 BCG ($z=0.1028$).
The two leftmost panels show an X-ray surface brightness map at top and 
a X-ray unsharp mask at bottom (whose zoomed-in FOV is marked by the white box 
on the top panel). Radio and FUV contours are overlaid in black and blue 
on the top panel, and in black and white on the bottom panel, respectively. 
The center and rightmost panel shows the FUV continuum map and an optical unsharp mask 
image, with radio and FUV contours overlaid in green and blue, respectively. 
The FUV colour bar can be scaled to a flux density by the inverse sensitivity 
$2.713\times 10^{-17}$ ergs cm\mtwo\ \AA\mone\ electron\mone.  
Note that the FUV continuum follows the ``spine'' of the ``bird-shaped'' radio source. 
The white box on the bottom-most X-ray panel marks the FOV of the two rightmost panels.
The centroids of all panels are aligned, with east left and north up.  }
\label{fig:pks0745_figure}
\end{figure*}

\begin{figure*}
\begin{center}
\includegraphics[scale=0.41]{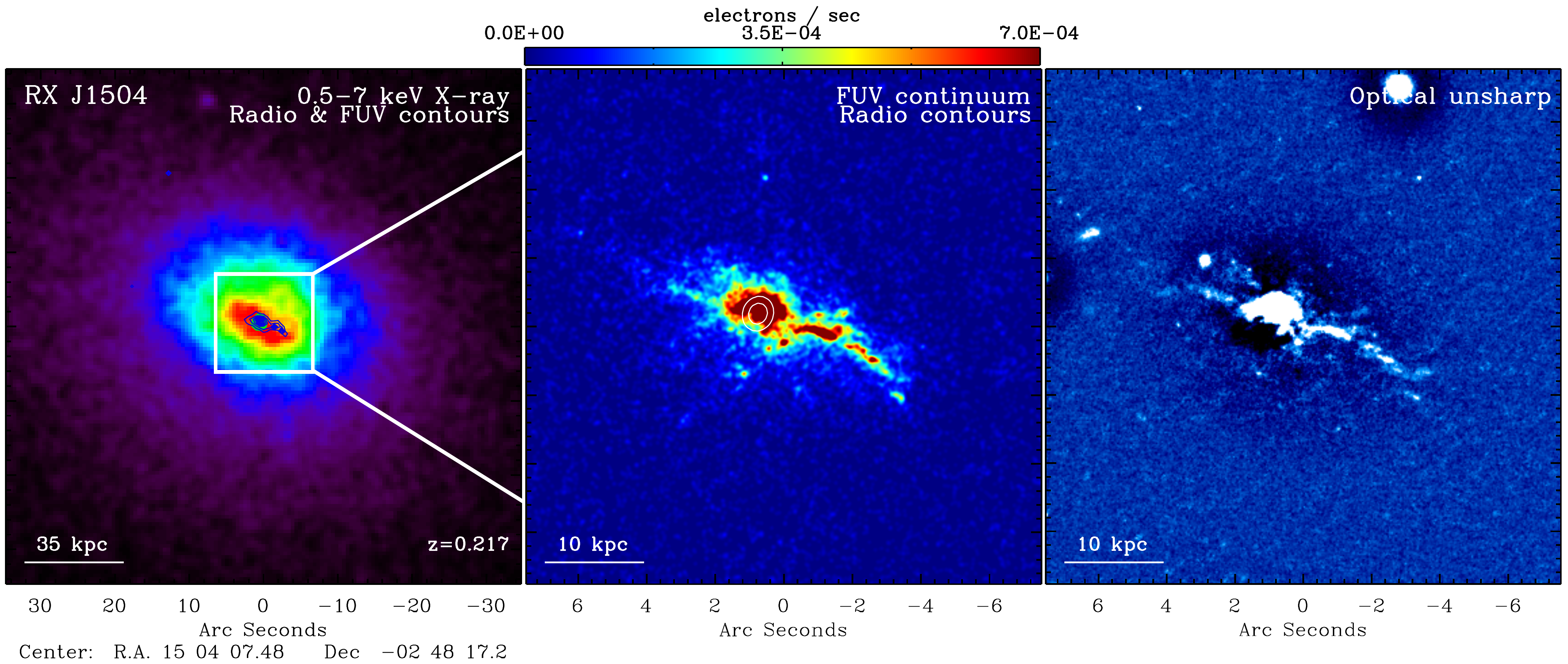}
\end{center}
\caption{A multiwavelength view of the RX J1504.1-0248 BCG ($z=0.2153$).
X-ray, FUV continuum, and optical unsharp mask images are shown in the   
the left, centre, and right panels, respectively. 
FUV continuum contours are shown in blue on the X-ray panel, 
and the unresolved radio source is shown in white contours on the FUV panel. 
Note the $\sim 20$ kpc stellar filament (FUV and broadband optical) extending along the BCG major 
axis.
The FUV colour bar can be scaled to a flux density by the inverse sensitivity 
$1.360\times 10^{-16}$ ergs cm\mtwo\ \AA\mone\ electron\mone.  
The white box on the X-ray panel marks the FOV of the two rightmost panels.
The centroids of all panels are aligned, with east left and north up.}
\label{fig:rxj1504_figure}
\end{figure*}

\begin{figure*}
\begin{center}
\includegraphics[scale=0.41]{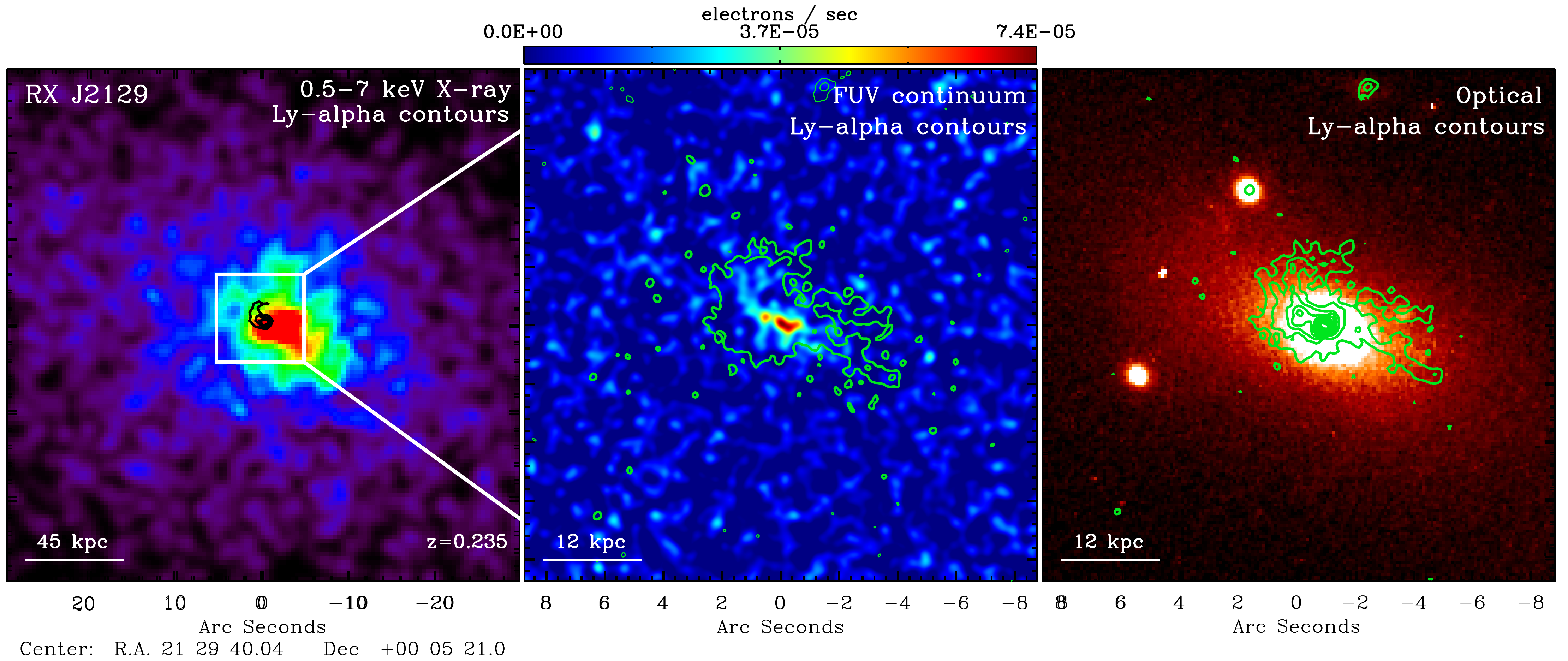}
\end{center}
\caption{The RX J2129.6+0005 BCG ($z=0.235$).   The X-ray, FUV continuum, and 
broadband optical images  
are shown in the left, centre, and right panels, respectively. Ly$\alpha$ contours 
are overlaid in black on the X-ray panel, and in green on the FUV and optical panels. 
The radio source is unresolved. The white box on the X-ray panel marks the FOV of the FUV and optical panels.  The FUV colour bar can be scaled to a flux density by the inverse sensitivity 
$1.360\times 10^{-16}$ ergs cm\mtwo\ \AA\mone\ electron\mone.  
The centroids of all panels are aligned, with east left and north up.   }
\label{fig:rxj2129_figure}
\end{figure*}

\begin{figure*}
\begin{center}
\includegraphics[scale=0.41]{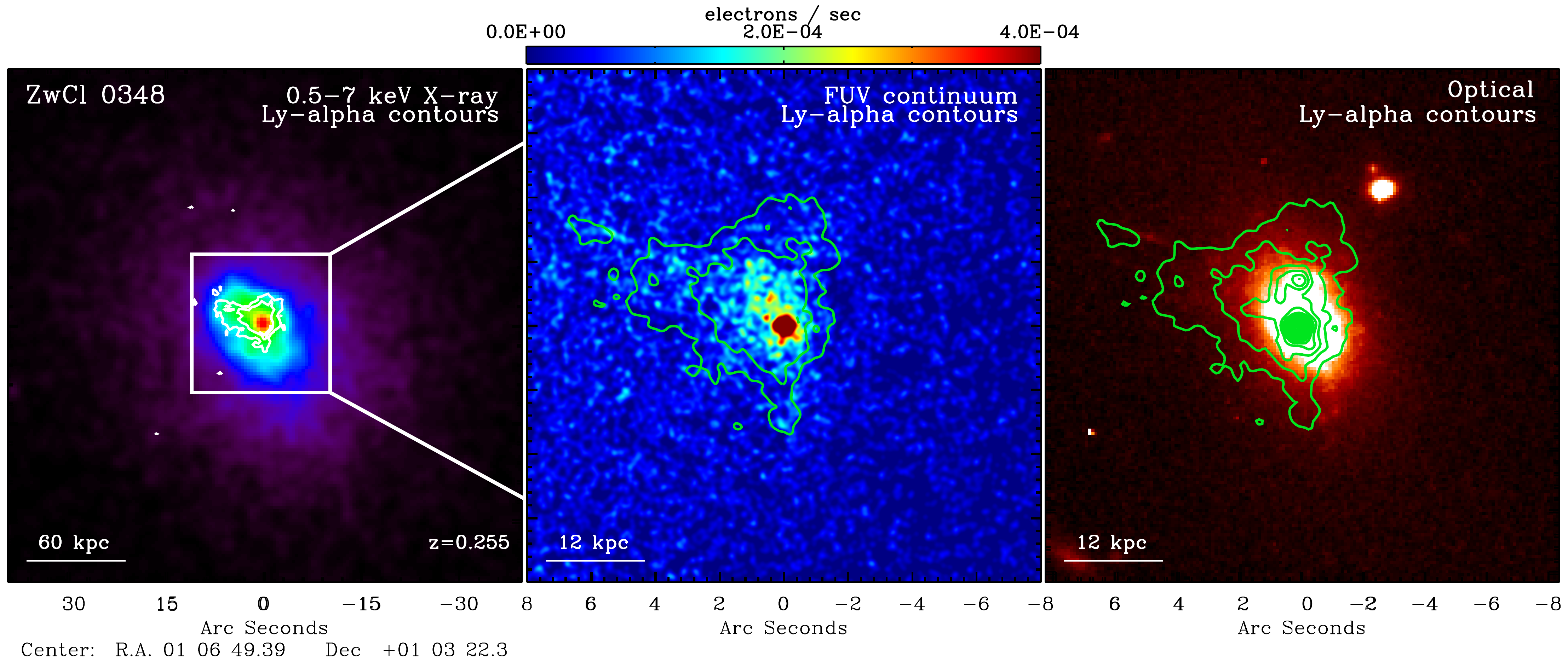}
\end{center}
\caption{The ZwCl 0348 BCG ($z=0.255$).  The X-ray, FUV continuum, and 
broadband optical images  
are shown in the left, centre, and right panels, respectively. Ly$\alpha$ contours 
are overlaid in white on the X-ray panel, and in green on the FUV and optical panels. 
The radio source is unresolved. The white box on the X-ray panel marks the FOV of the FUV and optical panels.  The FUV colour bar can be scaled to a flux density by the inverse sensitivity 
$1.360\times 10^{-16}$ ergs cm\mtwo\ \AA\mone\ electron\mone.  
The centroids of all panels are aligned, with east left and north up.  }
\label{fig:zw0348_figure}
\end{figure*}

\begin{figure*}
\begin{center}
\includegraphics[scale=0.41]{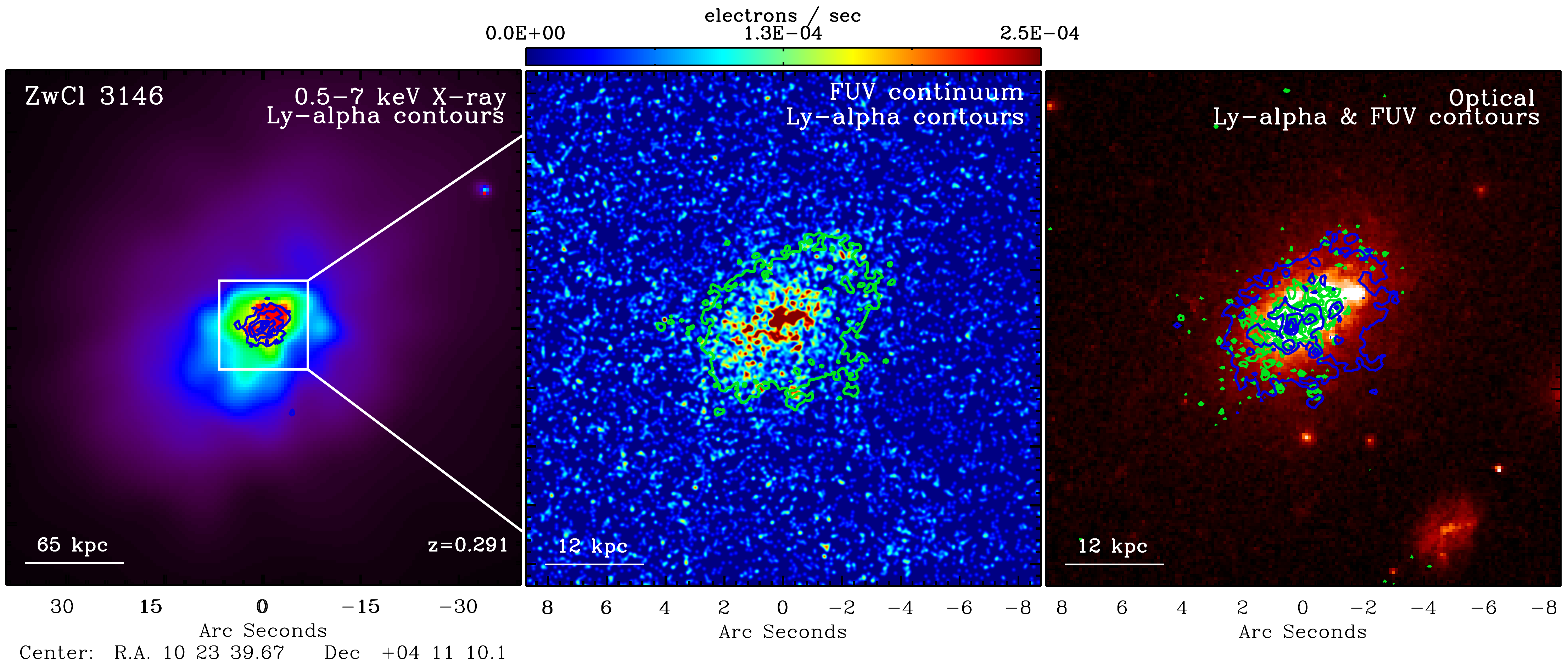}
\end{center}
\caption{The ZwCl 3146 BCG ($z=0.291$). The X-ray, FUV continuum, and 
broadband optical images  
are shown in the left, centre, and right panels, respectively. Ly$\alpha$ contours 
are overlaid in blue on the X-ray panel, in green on the FUV panel, and in blue on the optical panel. We also overlay FUV contours are in green on the optical panel.  
The radio source is unresolved. The white box on the X-ray panel marks the FOV of the FUV and optical panels.  The FUV colour bar can be scaled to a flux density by the inverse sensitivity 
$1.360\times 10^{-16}$ ergs cm\mtwo\ \AA\mone\ electron\mone.  
The centroids of all panels are aligned, with east left and north up.  }
\label{fig:zw3146_figure}
\end{figure*}

\begin{figure*}
\begin{center}
\includegraphics[scale=0.46]{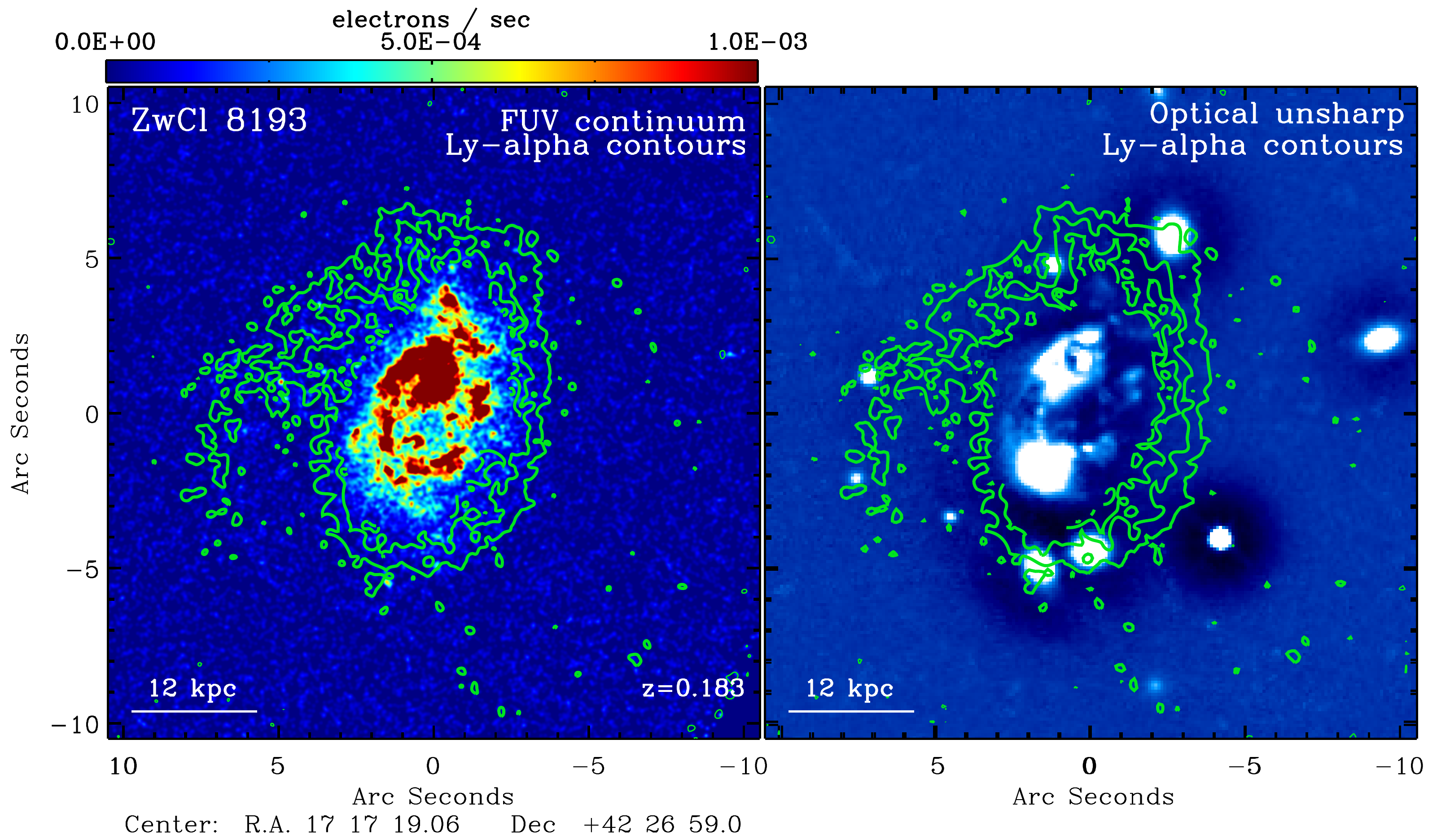}
\end{center}
\caption{The Zw8193 BCG ($z=0.1829$) may be undergoing a (minor?) merger. FUV continuum and a broadband optical 
unsharp mask are shown in the left and right panels, respectively. Ly$\alpha$ contours are overlaid in green (innermost contours have been removed to aid viewing). Note the highly disturbed morphology of the optical counterpart, 
as well as its apparent double nucleus. Brighter FUV emission 
associated with ongoing star formation is cospatial with the northernmost 
optical nucleus. Clumpy tendrils wind counter-clockwise from this bright northern FUV knot, suggestive of non-negligible net angular momentum perhaps stemming from the merger.  
The FUV colour bar can be scaled to a flux density by the inverse sensitivity 
$4.392\times 10^{-17}$ ergs cm\mtwo\ \AA\mone\ electron\mone.  
The centroids of both panels are aligned, with east left and north up. The radio source in Zw8193 is unresolved.}
\label{fig:zw8193_figure}
\end{figure*}

\clearpage

\hfill

\bibliographystyle{mn2e}
\bibliography{masterrefs}

\label{lastpage}

\end{document}